\documentclass[fleqn,usenatbib]{mnras}
\usepackage{newtxtext,newtxmath}
\usepackage[T1]{fontenc}
\usepackage{ae,aecompl}
\usepackage{relsize}
\usepackage{dblfloatfix}
\usepackage[dvipsnames]{xcolor}
\usepackage{graphicx}	
\usepackage{amsmath}	
\usepackage{upgreek}
\usepackage{bm}
\usepackage{ulem}

\graphicspath{{./}{figures/}}
\newcommand{\Disco}{{\texttt{Disco}}}
\newcommand{\tensorSymbol}[1]{{\overset{\leftrightarrow}{#1}}}

\title[Circumbinary Disc Thickness and Viscosity]{A Survey of Disc Thickness and Viscosity in Circumbinary Accretion: Binary Evolution, Variability, and Disc Morphology}

\author[A. J. Dittmann \& G. Ryan]{
Alexander J. Dittmann,$^{1}$\thanks{E-mail: \href{mailto:dittman@astro.umd.edu}{dittmann@astro.umd.edu}}
Geoffrey Ryan$^{2}$
\\
$^{1}${Department of Astronomy and Joint Space-Science Institute, University of Maryland, College Park, MD 20742-2421}\\
$^{2}${Perimeter Institute for Theoretical Physics, 31 Caroline St. N., Waterloo, ON, N2L 2Y5, Canada}
}
\date{Accepted XXX. Received YYY; in original form ZZZ}
\pubyear{2022}

\begin{document}
\label{firstpage}
\pagerange{\pageref{firstpage}--\pageref{lastpage}}
\maketitle

\begin{abstract}
Much of the parameter space relevant to the evolution of astrophysical circumbinary accretion discs remains unexplored.
We have carried out a suite of circumbinary disc simulations surveying both disc thickness and kinematic viscosity, using both constant-$\nu$ and constant-$\alpha$ prescriptions. We focus primarily on disc aspect ratios between $0.1$ and $0.033$, and on viscosities between $\nu=0.0005$ and $\nu=0.008$ (in units of binary semi-major axis and orbital frequency), and specialise to circular equal-mass binaries. Both factors strongly influence the evolution of the binary semi-major axis: at $\nu=0.0005,$ inspirals occur at aspect ratios $\lesssim0.059$, while at $\nu=0.004$ inspirals occur only at aspect ratios $\lesssim0.04$. Inspirals occur largely because of the increasingly strong negative torque on the binary by streams of material which lag the binary, with negligible contributions from resonant torques excited in the circumbinary disc. 
We find that reductions in accretion rate occur when simulations are initialised too far from the eventual quasi-steady state driven by interaction with the binary, rather than being intrinsically linked to the disc aspect ratio.
We find not only that the cavity size increases as viscosity is decreased, but that thinner circumbinary discs become more eccentric. Our results suggest that supermassive black hole binaries 
should be driven, more rapidly than previous estimates, from $\sim$parsec separations to distances where gravitational waves drive their inspiral, potentially reducing the number of binaries observable by pulsar timing arrays.  
\end{abstract}

\begin{keywords}
accretion, accretion discs -- hydrodynamics -- black hole physics -- binaries: general -- stars: pre-main-sequence
\end{keywords}

\section{Introduction} 
As most massive galaxies are thought to harbor supermassive black holes (SMBHs) in their centers \citep{2013ARA&A..51..511K}, the orbital evolution of those black holes following galaxy mergers is of significant interest \citep[e.g.][]{2000ApJ...532L..29G,1980Natur.287..307B}. Gravitational interactions between SMBH binaries and their circumbinary discs can potentially facilitate black hole mergers in binaries too far separated to inspiral efficiently via gravitational wave emission \citep[e.g.][]{2009ApJ...700.1952H}, solving the so-called `final parsec problem' \citep{2003ApJ...596..860M,2003AIPC..686..201M}, although other solutions have been proposed \citep[e.g.][]{2013ApJ...773..100K}. This paradigm has recently been called into question by \citet{2020ApJ...889..114M} who suggest, based on circumbinary disc simulations at aspect ratio $(H/r)=0.1$ simulations, that SMBH binaries will be driven outwards, potentially stalling before their evolution reaches the gravitational wave (GW)-driven regime. 

Whether or not SMBH binaries stall at large $(\sim$pc) radii has important implications for both optical and gravitational wave observations. For example, if many binaries stall at large radii, only a select few would have orbital periods short enough to be reliably detected in optical surveys, although more recent candidates with low estimated false alarm probabilities suggest that binary stalling may not be common \citep[e.g.][but see also \citet{2021arXiv211002982F}]{{2015MNRAS.453.1562G},{2016MNRAS.461.3145V},{2016MNRAS.463.2145C},{2018ApJ...859L..12L},{2020MNRAS.499.2245C},{2021MNRAS.500.4025L}}.
However, if circumbinary discs drive rapid inspirals up to GW frequencies of $\sim10^{-8}$ Hz, they may limit the number of SMBH binaries contributing to the stochastic GW signals probed by pulsar timing arrays \citep[e.g.][]{2011MNRAS.411.1467K,2020ApJ...905L..34A}.

Apart from SMBH binaries, circumbinary discs can also occur in a variety of other astrophysical contexts, such as from disc fragmentation during star formation \citep[e.g.][]{{1986ApJS...62..519B},{2010ApJ...710.1375K}} or the fallback of ejecta from post-common envelope binaries \citep{2011MNRAS.417.1466K}. Unlike SMBH binaries, resolved images of stellar binaries and their discs can realistically be captured, sometimes in great detail \citep[e.g.][]{2020A&A...639A..62K}. However, in order to extract the physical properties of both the binary and disc from observations, a wide parameter space must be probed through simulations, including binary mass ratio $q\equiv M_2/M_1\leq1$, binary eccentricity $(e)$, binary inclination relative to the circumbinary disc, disc viscosity ($\nu$), and disc scale height ($H$). 

Classical models of binary-disc interaction suggest that the binaries are driven together, losing angular momentum to the circumbinary disc through resonantly excited density waves \citep{1991ApJ...370L..35A,1994ApJ...421..651A}.
Additionally, negative torques from the circumbinary disc on the binary have been identified in some simulations \citep[e.g.][]{2008ApJ...672...83M,2017MNRAS.469.4258T,2020ApJ...900...43T,2020A&A...641A..64H}, although the cavity locations found in simulations tend to disagree with estimates of cavity sizes from balancing gravitational and viscous torques in linear theory \citep[e.g.][]{2008ApJ...672...83M,2015MNRAS.452.2396M}. 

However, the circumbinary disc is not the only source of torque on the binary. Considering binaries with $q\gtrsim0.1$, the time-variable gravitational potential of the binary clears out a low-surface density cavity near the binary and the inner circumbinary disc becomes eccentric. Near the circumbinary disc periapse, material can be captured by the binary or flung away from it through the cavity \citep[e.g.][]{2021arXiv211104721T}. Additionally, accretion discs (`minidiscs') form around each binary component. Over time, an overdense `lump' forms on the cavity walls, modulating accretion onto the binary strongly at the orbital period of the lump. \citep[e.g.][]{2008ApJ...672...83M,2012A&A...545A.127R,2012ApJ...749..118S,2015ApJ...807..131S,2013MNRAS.436.2997D,2017MNRAS.466.1170M, 2019ApJ...879...76B,2021arXiv210312100N}

Thus, in addition to the torque from the circumbinary disc, the binary can experience a net gravitational torque from the minidiscs, as well as a torque from the streams of material in the cavity and the torque associated with the accretion of material onto each object. These effects have been studied extensively, and have been found to dictate the evolution of binaries over a wide range of parameter space. Discs with aspect ratio $H/r=0.1$, or equivalently Mach number $\mathcal{M}\equiv\Omega r /c_s=10$, have been studied most thoroughly, in which case circular binaries with not-too-unequal masses $(q=M_2/M_1\gtrsim0.1)$ have been shown to outspiral \citep{2017MNRAS.466.1170M,2017MNRAS.469.4258T,2020ApJ...889..114M,2019ApJ...871...84M,2019ApJ...875...66M,2020ApJ...901...25D,2021ApJ...921...71D}, a conclusion which holds in both 2D and 3D simulations \citep[at $q=1$;][]{2019ApJ...875...66M}, and is robust to simulation boundary conditions \citep{2020ApJ...889..114M}, viscosity and rate of mass removal by sink particles \citep{2020ApJ...901...25D}, and sink mass removal algorithm \citep{2021ApJ...921...71D}. Eccentric equal-mass binaries, on the other hand, can expand or shrink depending on the eccentricity of the binary \citep{2017MNRAS.466.1170M,2021arXiv210309251D}. 

Thicker, $\mathcal{M}\sim10$, discs may be appropriate for stellar binaries \citep[e.g.][]{1999ApJ...516..335W}, but accretion discs around SMBHs are typically expected to be significantly thinner, with $\mathcal{M}\sim10^2-10^3$ \citep[e.g.][]{2001ApJ...559..680H}. The $\mathcal{M}>10$ regime has been explored more sparsely
\citep[e.g.][]{2015MNRAS.446L..36F,2016MNRAS.459.2379D,2016MNRAS.460.1243R,2020ApJ...900...43T,2020A&A...641A..64H, 2020MNRAS.499.3362R}, with only a subset of these studies exploring orbital evolution as a function of Mach number \citep{2020ApJ...900...43T,2020A&A...641A..64H}. Both studies found that inspirals occur at higher Mach number and simulated finite accretion discs, allowing the discs to spread viscously, without a steady supply of gas from large radii. \citet{2020ApJ...900...43T} noted that the angular momentum change in the binary due to gravitational torques per unit accreted mass in the cavity region and outer disc became increasingly negative at higher Mach number, and that the time-averaged peak surface density around the cavity wall increased with Mach number. \citet{2020A&A...641A..64H} argued that for discs with $H/r \lesssim0.1$ resonant torques are strong enough to overcome viscous torques and impede the accretion flow, leading to inspirals. However, we find that any accretion rate reduction observed at high Mach numbers in our simulations is due to inconsistency between initial conditions and net torque through the disc due to interaction with the binary. 

Both \citet{2020ApJ...900...43T} and \citet{2020A&A...641A..64H} primarily varied the Mach number in their simulations, although \citet{2020A&A...641A..64H} used an $\alpha$-viscosity model, setting the kinematic viscosity as $\nu=\alpha c_s H$, while \citet{2020ApJ...900...43T} primarily focused on constant-$\nu$ simulations, but found good agreement with a limited set of simulations holding $\alpha$ fixed. However, both studies hinted at important viscosity dependence to the evolution of accreting binaries.
The convergence tests in \citet{2020ApJ...900...43T} show that their measured torques become more negative as resolution improves - at lower numerical viscosity.
\citet{2020A&A...641A..64H} used a particle-based code \citep{2018PASA...35...31P}, finding that poor resolution (higher numerical viscosity) can change an inspiraling binary to an outspiraling one. We present a thorough study of both viscosity and Mach number using numerical methods that are generally less susceptible to numerical viscosity: the moving-mesh finite volume code \Disco{} using cylindrical geometry \citep{2016ApJS..226....2D} (using a corrected viscosity implementation described in \citet{2021ApJ...921...71D}), which is better-suited to probe this parameter space. 

Our numerical methods and diagnostic tools are detailed in Section \ref{methods}. We present comparisons with previous works in Section \ref{sec:prelim} and describe the results of convergence tests, indicating that our results are not dependent on numerical resolution and that the numerical viscosity in our simulations is small compared to the physical Navier-Stokes viscosity employed. We present our results on binary orbital evolution in Section \ref{sec:binOrb}, showing not just that binaries with high-$\mathcal{M}$ low-$\nu$ discs inspiral rapidly, but that viscosity strongly affects the rate of inspiral, or whether inspiral occurs at all. We show that strong negative torques arise from streams of gas which lag behind the binary, and that the density contrast between the accretion streams and cavity minimum is reduced with increasing viscosity and disc aspect ratio, leading to less-negative torques. In Sections \ref{sec:alpha} and \ref{sec:global} we check the robustness of our results to our viscosity model and equation of state respectively. We discuss in section \ref{sec:diskOrb} how the orbital characteristics of the circumbinary disc evolve with Mach number and viscosity, showing that higher-Mach number discs are more eccentric, and that lower-viscosity discs have larger cavities. We discuss the implications of our results for gravitational wave and optical observations of SMBH binaries in Section \ref{sec:SMBHB}, after which we summarise our results in Section \ref{sec:conclude}. Measurements from our simulations relevant to binary orbital evolution and the transport of mass and angular momentum through the disc are collected in Appendix \ref{app:summary}.

\section{Methods}\label{methods}
This study uses the moving-mesh code \Disco{} \citep{2016ApJS..226....2D,2021ApJ...921...71D}
to solve the 2D ($r,~\phi$) equations of isothermal vertically-integrated viscous hydrodynamics:

\begin{align}
\partial_t\Sigma + \bm{\nabla}\!\cdot\!(\Sigma\mathbf{v}) &= S_\Sigma \label{eq:continuity} \\
\partial_t(\Sigma\mathbf{v}) + \bm{\nabla}\!\cdot\!(\Sigma\mathbf{v}\mathbf{v}+\Pi\tensorSymbol{I} - 2 \Sigma \nu \tensorSymbol{\sigma})
&= -\Sigma\bm{\nabla}\Phi + \mathbf{S}_{p} \label{eq:momentum} \\
\Pi&=c_s^2(\mathbf{x})\Sigma,
\end{align}
where $\Pi$ is the vertically-integrated pressure, $\Sigma$ is the disc surface density, $c_s$ is the sound speed, $\mathbf{v}$ is the fluid velocity vector, $\Phi$ is the gravitational potential, $\tensorSymbol{I}$ is the identity matrix, $S_\Sigma$ is a mass sink term, $\mathbf{S}_{p} $ is a momentum sink term, $\tensorSymbol{\sigma}$ is the velocity shear tensor, and $\nu$ is the kinematic viscosity. 

We primarily use a `locally isothermal' equation of state, setting the sound speed based on the local gravitational potential
\begin{equation}\label{lociso}
c_s^2(\mathbf{x}) = -\Phi(\mathbf{x})\mathcal{M}^{-2},
\end{equation}
where $\mathcal{M}$ is the fiducial Mach number and fixed for each simulation. We also carried out a limited set of `globally isothermal' simulations, setting
\begin{equation}\label{globiso}
c_s^2=a_b^2\Omega_b^2\mathcal{M_*}^{-2},
\end{equation}
where we have used $\mathcal{M}_*$ to denote a characteristic Mach number in globally-isothermal simulations.

We use a softened gravitational potential, where the total potential is given by $\Phi=\sum_i \Phi_i$
and $\Phi_i$ is the gravitational potential of a point mass of mass $M_i$ at position $\mathbf{x}_i$ given by
\begin{equation}
\Phi_i = \frac{-GM_i}{\sqrt{|\textbf{x}-\textbf{x}_i|^2+\epsilon_g^2}},
\end{equation}
where $\epsilon_g$ is the gravitational softening length. We fix the gravitating point masses on circular orbits, calculating their positions analytically each time step, assuming that binary evolution occurs on much longer timescales than the orbital period of the binary.

Accretion discs are frequently modelled using an $\alpha-$viscosity \citep{1973A&A....24..337S}, where $\nu=\alpha c_s H$, and $H$ is the disc scale height which we calculate using
\begin{equation}
H = \frac{c_s}{\sqrt{\sum_i GM_i(|\textbf{x}-\textbf{x}_i|^2+\epsilon_g^2)^{-3/2}}}.
\end{equation}
However, in this formulation changes in Mach number result in changes in the viscosity as well, making it challenging to isolate effects related to gas pressure from those associated with viscosity. Thus, we primarily hold $\nu$ constant in our simulations to better differentiate between pressure and viscous effects, but also perform a limited number of tests using an $\alpha$-viscosity. 

The surface density sink term is given by 
\begin{equation}\label{surfSink}
S_\Sigma = -\gamma \Omega_b \Sigma \sum_i s_i(|\textbf{x}-\textbf{x}_i|),
\end{equation}
where $s_i$ is a function specifying the sink profile for each particle, $\Omega_b$ is the angular frequency of the binary, and $\gamma$ is the sink rate, such that $(\Omega_b\gamma)^{-1}$ is the characteristic timescale for mass removal ($t_s$). We vary $\gamma$ so that ratio between $t_s$ and the viscous inflow timescale $(t_\nu)$ at the characteristic sink radius $(r_s)$ is constant between simulations.
In this work we always use a sink profile of the form
\begin{equation}\label{GaussSquared}
s_i = \exp\left(-\frac{|\mathbf{x}-\textbf{x}_i|^b}{Ar_s^b}  \right),
\end{equation}
where $A$ and $b$ are positive constants. In almost all cases we use $A=1~,b=4$, but to make direct comparisons with \citet{2020ApJ...900...43T} we perform a limited set of simulations using $A=2, b=2$.

We use a torque-controlled momentum sink term in \Disco{}, given by
\begin{align} 
    \mathbf{S}_{p} &= -\gamma \Omega_b \Sigma \sum_i s_i(|\textbf{x}-\textbf{x}_i|)\mathbf{v}^*_i \label{eq:tf1} \\
    \mathbf{v}^*_i &= \left(\mathbf{v} - \mathbf{v}_i\right)\cdot \left(\hat{\mathbf{r}}_i \hat{\mathbf{r}}_i + \delta\ \! \hat{\bm{\phi}}_i \hat{\bm{\phi}}_i \right) + \mathbf{v}_i, \label{eq:tf2}
\end{align}
where $\mathbf{v}_i$ is the velocity of the sink particle,  $\hat{\mathbf{r}}_i$ and $\hat{\bm{\phi}}_i$ are the unit basis vectors at position $\mathbf{x}$ of a polar coordinate system centered on the sink particle, and $\delta$ is a dimensionless control parameter. In almost all cases we set $\delta=0$, which leads to `torque-free' sinks, which prevent the sink term from exerting a torque on the fluid in the frame of the particle.
Torque-free sinks emulate accretion onto an unresolved point mass are able to reproduce analytic steady-state disc profiles in single-object discs \citep{2020ApJ...892L..29D,2021ApJ...921...71D},
and reduce the degree to which the rate of gas removal affects the inferred evolution of the binary and time-averaged minidisc structure (see \citet{2021ApJ...921...71D} for further details). We also performed a limited set of simulations using `standard' sinks, setting $\delta=1$ which change the angular momentum of the gas relative to the sink particle but keep the fluid velocity constant, to illustrate the impact of our sink choices and make comparisons with previous studies such as \citet{2020ApJ...900...43T}. 

\subsection{Diagnostics}\label{sec:diag}
We regularly store various fluid quantities and their averages in time and azimuth, e.g. snapshots of $\Sigma(r,\phi,t_i)$, $\mathbf{v}(r,\phi,t_i)$, and  $\langle\Sigma\rangle(r,t_i)$ at times $t_i$, typically once per orbit, where 
\begin{equation}
\langle\Sigma\rangle(r, t_i)\equiv\frac{1}{2\pi}\frac{1}{t_i-t_{i-1}}\int_{t_{i-1}}^{t_i}dt\int_0^{2\pi}\Sigma(r,\phi) d\phi.
\end{equation}
We calculate other quantities such as the gravitational torque density in the disc $\langle\Sigma\partial_\phi\Phi\rangle$, mass-weighted radial velocity $\langle v_r\rangle=\langle\Sigma v_r\rangle/\langle\Sigma\rangle$, and the mass-weighted eccentricity vector of each annulus $\langle\mathbf{e}\rangle=\langle \mathbf{e}\Sigma\rangle/\langle\Sigma\rangle$, where
\begin{equation}
\mathbf{e}\equiv\left(\frac{\mathbf{v}\cdot\mathbf{v}}{GM}-\frac{1}{r}\right)\mathbf{r}-\frac{\mathbf{r}\cdot\mathbf{v}}{GM}\mathbf{v},
\end{equation}
and where $M=\sum_iM_i$, given here in terms of the binary semi-major axis $(a_b)$ and angular frequency $(\Omega_b)$ by $\mu=\Omega_b^2a_b^3$. After calculating $\langle\mathbf{e}\rangle$, we calculate the average scalar eccentricity as $e=\sqrt{\langle\mathbf{e}\rangle\cdot\langle\mathbf{e}\rangle}$ and argument of periapsis as $\omega\equiv\tan^{-1}{(\langle \mathbf{e}\rangle_x/\langle \mathbf{e}\rangle_y)}$.

We also calculate various quantities related to the sinks, such as the accretion rate onto each particle
\begin{equation}
    \dot{M}_i = - \int\! dA\ S_{\Sigma, i}\ ,
\end{equation}
where $S_{\Sigma, i} = -\gamma \Omega_b \Sigma s_i(|\mathbf{x}-\mathbf{x}_i|)$.  Similarly, we define the $\mathbf{S}_{p, i} = S_{\Sigma_i} \mathbf{v}_i^*$ by Equation \eqref{eq:tf1}.  The sink profile function $s_i(x)$ sharply truncates the contributions from distances more than a few $r_s$ from the sink particle, although the integral formally includes the entire domain. 
We also record the accretion torque $\dot{J}_a$, the rate of angular momentum delivered through the sink terms.  
The total accretion torque on each sink particle is:
\begin{align}
    \dot{J}_{a,i} &= - \int \! dA\ \mathbf{x} \times \mathbf{S}_{p, i} = - \int \! dA\ S_{\Sigma, i}\ \mathbf{x} \times \mathbf{v}^*_{i}\ .
\end{align}
The accretion torque contains three independent contributions, corresponding to the three terms in $\mathbf{v}^*_i$.  The first is proportional to $\hat{\mathbf{r}}_i$ and corresponds to the direct absorption of linear momentum from the gas, the second is proportional to $\hat{\bm{\phi}}_i$ and contributes dominantly to the spin of the sink particle (the `spin torque', $\dot{J}_{a,s}$), and the third is proportional to the sink particle velocity $\mathbf{v}_i$ and corresponds to the direct accretion of mass to the sink. The total torque, the spin torque, and the component that contributes to the orbit of the binary ($\dot{J}_{a,o}$) are related by 
\begin{equation}
\dot{J}_a = \dot{J}_{a,o}+\dot{J}_{a,s}, 
\end{equation}
where 
\begin{align}
    \dot{J}_{a,s,i} &= - \int \! dA\ \left(\mathbf{x} - \mathbf{x}_i\right) \times \left(\mathbf{S}_{p, i} - \mathbf{v}_i S_{\Sigma, i} \right)\\
    &= - \int \! dA\ \delta \ S_{\Sigma, i}\left|\mathbf{x} - \mathbf{x}_i\right| \left(\mathbf{v} - \mathbf{v}_i\right)_{\hat{\phi}_i}. \label{eq:spinTorque}
\end{align}

Apart from a limited number of simulations to gauge the impacts of torque-free sinks, we use a sink prescription that sets the term contributing to the spin of the sink particles to zero.
This is especially justified for SMBH ($M_\bullet \sim 10^8~\rm M_\odot$) binary systems separated by a few parsecs, in which case the radius of the innermost stable circular orbit is $\sim10^{-5}$ times the orbital separation and size of the minidiscs. The grid scale in our simulations is typically $\sim 10^{-2}$ times the binary semi-major axis, so measuring the accreted spin torque using standard sinks would overestimate the torque by orders of magnitude. We note that torque-free sinks also reproduce analytic steady-state accretion disc surface density profiles while standard sinks cause anomalous deficits, but see \citet{2020ApJ...892L..29D,2021ApJ...921...71D} for further discussion. 

Sink particles may also acquire orbital momentum through gravitational interactions. These are also computed by integrating the corresponding source terms over the simulation domain:
\begin{equation}
    \dot{J}_{g,i} =  -\int \! dA\ \mathbf{x} \times \left(-\Sigma \bm{\nabla} \Phi_i\right) =  \int \! dA\ \Sigma \partial_\phi \Phi_i.
\end{equation}
We also define a time-averaged and $\phi$-integrated linear gravitational torque density $\tau_g\equiv2\pi r\langle \Sigma\partial_\phi\Phi \rangle$.

The total torque on the system of sink particles is the sum of the torques on each component:
\begin{equation}
    \dot{J} = \sum_i \dot{J}_{g, i} + \dot{J}_{a,i} .
\end{equation}
The rate of change of the total \textit{orbital} angular momentum, the orbital torque $\dot{J}_{\mathrm{orb}}$, is: 
\begin{equation}
    \dot{J}_{\mathrm{orb}} = \sum_i \dot{J}_{g, i} + \dot{J}_{a,o,i} .
\end{equation}

We calculate the orbital evolution of the binary from the torques measured in our simulations using the equation for the binary angular momentum
\begin{equation}
J_b=\frac{M_1M_2}{M}\sqrt{GMa_b\left(1-e_b^2\right)},
\end{equation}
where $M$ is the total mass of the binary and $e_b$ is the eccentricity of the binary. We specialise to circular binaries in this work, for which the torque on the binary can be expressed as 
\begin{equation} \label{eq:evolEqn1}
\frac{\dot{J}_{\rm orb}}{J_b} = \frac{\dot{M}_1}{M_1} + \frac{\dot{M}_2}{M_2} - \frac{1}{2}\frac{\dot{M}}{M} + \frac{1}{2}\frac{\dot{a}_b}{a_b}.
\end{equation}
The above expression assumes that the binary remains circular, which is supported by previous studies which have found that binary eccentricity is damped for $e\lesssim 0.08$ \citep{2019ApJ...871...84M,2020arXiv201009707Z}.
For binaries of arbitrary mass ratio, the evolution of the binary semi-major axis is given by \begin{equation}
\frac{d\log{a_b}}{d\log{M}} = 1+ 2\left[\frac{l_0}{l_b}-\frac{\dot{M}_1}{\dot{M}}(1+q)-     \frac{\dot{M}_2}{\dot{M}}\frac{1+q}{q}\right], \label{eq:evolEqn2}
\end{equation}
where $q=M_2/M_1$ is the mass ratio, $l_0=\langle\dot{J}_{\rm orb}\rangle/\langle\dot{M}\rangle=\langle(\dot{J}_{a,o}\rangle+\langle\dot{J}_g\rangle)/\langle\dot{M}\rangle$ is the $\textit{orbital}$ angular momentum change in the binary per unit mass, and $l_b$ is the specific angular momentum of the binary $\Omega_ba^2_bq/(1+q)^2$.  For equal-mass binaries $q=1$ and $\langle\dot{M}_1/\dot{M}\rangle=\langle\dot{M}_2/\dot{M}\rangle=1/2$ by symmetry.   Furthermore, the accreted orbital angular momentum can be modelled as $\langle\dot{J}_{a,o}\rangle=\dot{M}_1l_1 + \dot{M}_2l_2$, where $l_1=\Omega_ba^2_bq^2/(1+q)^2$ and $l_2=\Omega_ba^2_b/(1+q)^2$ are the specific angular momenta of the primary and secondary respectively \cite{2021ApJ...921...71D}. Thus, for equal-mass binaries, such as those studied in this work,

\begin{equation} \label{eq:orbJg}
\frac{d\log{a_b}}{d\log{M}}=\frac{8\dot{J}_g}{\dot{M}a_b^2\Omega_b} - 1.
\end{equation}
We therefore focus on $d\log{a_b}/d\log{M}$ and $\dot{J}_g/\dot{M}$ when studying the effects of Mach number and viscosity on binary evolution.

We note that in the relevant astrophysical discs the timescale of orbital evolution is typically many orders of magnitude longer than the binary orbital period.  Hence, while the sink particles track their accrued mass and angular momentum, the assumption that the binary moves on a fixed orbit is typically well-justified in our simulations.

\subsection{Simulation setup}
The computational domain used in this study extends from $r=0$ to $r=30a_b$, and the polar coordinate system is centred on the barycenter of the binary. We discretise this domain radially into 512 zones, which we found to produce converged results when compared to simulations using 786 radial zones (see Section \ref{sec:prelim}). Grid spacing was linear from $r=0$ to $r=a_b$, and logarithmic from $r\geq a_b$ to $r=30a_b$. In each annulus, the number of zones in $\phi$ was chosen so that the cell aspect ratio was as close to 1 as possible. We use two additional outer annuli for specifying boundary conditions, which we hold fixed at their initial values throughout the simulation. We utilise a moving mesh, where cells in each annulus move in $\phi$ with the average angular velocity of the gas within that annulus. This technique is able to better capture contact discontinuities and significantly reduce numerical viscosity compared to a fixed mesh for flows with significant azimuthal velocity \citep{2016ApJS..226....2D}, and reduces numerical noise associated with re-meshing in fully moving-mesh codes \citep[e.g.][]{2010MNRAS.401..791S}.

We used axisymmetric initial conditions, with an outer disc in an approximate viscous steady state, but with a cavity excised around the binary to be filled by accreting gas. The corresponding surface density profile was
\begin{equation}\label{eq:initSig}
\Sigma(r)=\Sigma_0\exp\left[ -\left(r/r_e \right)^{-\xi} \right]\left[1-l_i\sqrt{\frac{a_b}{r}} \right],
\end{equation}
where we set $r_e=2.5a_b$ and $\xi=30$. Here, 
\begin{equation}\label{eq:sig0}
\Sigma_0=\frac{\dot{M}_0}{3\pi\nu}, 
\end{equation}
where $\nu$ is the kinematic viscosity, $\dot{M}_0$ is the accretion rate at the outer edge of the disc, and $l_i$ is an initial guess for the steady-state value of $\dot{J}/\dot{M}$ onto the binary (and thus through the disc) \citep{1974MNRAS.168..603L}. We typically use $l_i=0$. However, it is well-known that starting simulations with a value of $l_i$ that is too small compared to the stead-state value of $\dot{J}/\dot{M}$ results in anomalously high accretion rates, which can be remedied by performing additional iterated simulations setting $l_i$ equal to the values of $\dot{J}/\dot{M}$ measured in the previous simulation \citep[e.g.][]{2017MNRAS.466.1170M,2020ApJ...891..108D}. We demonstrate in Section \ref{sec:acc} that a value of $l_i$ that is too large results in anomalously low measured accretion rates. We ignore the self-gravity of the disc, so the scale of $\Sigma$ and $\dot{M}_0$ can be adjusted without loss of generality, although this assumption is expected to break down in discs which are sufficiently massive, particularly at large radii \citep[e.g.][]{1980SvAL....6..357K,1989ApJ...341..685S,2020MNRAS.493.3732D}.

The initial angular velocity profile of the disc was given by 
\begin{equation}
\Omega^2(r) = \Omega_k^2(\mathcal{R})\left(1 + \frac{3a_b^2}{4\mathcal{R}^2}\frac{q}{(1+q)^2} \right) +  \frac{1}{\mathcal{R}\Sigma}\frac{d\Pi}{dr},
\end{equation}
where both pressure gradients and the orbit-averaged quadrupole moment of a circular binary cause deviations in angular velocity from the Keplerian value, and $\mathcal{R}=\sqrt{r^2+\epsilon_g^2}$. The initial angular velocity profile inside the cavity is rapidly erased by the binary. The initial radial velocity was set approximately to the viscous rate,
\begin{equation}
v_r = -\frac{3\nu}{2\mathcal{R}},
\end{equation}
resulting in a constant inflow of matter due to our fixed outer boundary condition, keeping the fluid variables fixed to their initial values.

All simulations were second-order in space and time, using piecewise linear spatial reconstruction with a generalised van Leer slope limiter \citep{{1979JCoPh..32..101V},{2000JCoPh.160..241K}}, setting $\theta=1.5$, a Harten-Lax-Van Leer-Contact approximate Riemann solver \citep{1994ShWav...4...25T} and 2nd-order total variation diminishing Runge-Kutta time stepping \citep{1998MaCom..67...73G}
with a typical Courant–Friedrichs–Lewy safety factor of 0.5, which we lower to $\sim0.2$ in some of our lowest-viscosity simulations. We set the sink rate $\gamma$ so that the ratio of the viscous timescale at the sink radius to the sink timescale is $t_\nu/t_s\sim r_s^2\gamma\Omega_b(2\nu)^{-1}=5/3$ at the sink radius \citep{2021ApJ...921...71D}. Unless otherwise noted, we set $\epsilon_g=r_s=0.05a_b$ in our simulations, leading to $\gamma=4/3$ and $t_s = 3/4 \Omega_b^{-1} $ for $\nu=10^{-3}$. Each grid annulus was set to move with the $\phi$-averaged azimuthal fluid velocity in that annulus. Our $\nu=0.0005,~\nu=0.0001,$ and $\nu\geq0.002$ simulations were $4000,~2000,$ and $1000$ binary orbital periods in duration respectively, a few viscous timescales at $r\sim3~a_b$ and at least two cavity precession periods.

\section{Preliminary Tests} \label{sec:prelim}
We begin by comparing our results with those found by other authors and confirming that our simulations use sufficiently high resolution. We also demonstrate that at higher Mach numbers, a sink prescription that does not malignantly alter the global flow is crucial. 

\begin{figure}
\centering
\includegraphics[width=\columnwidth]{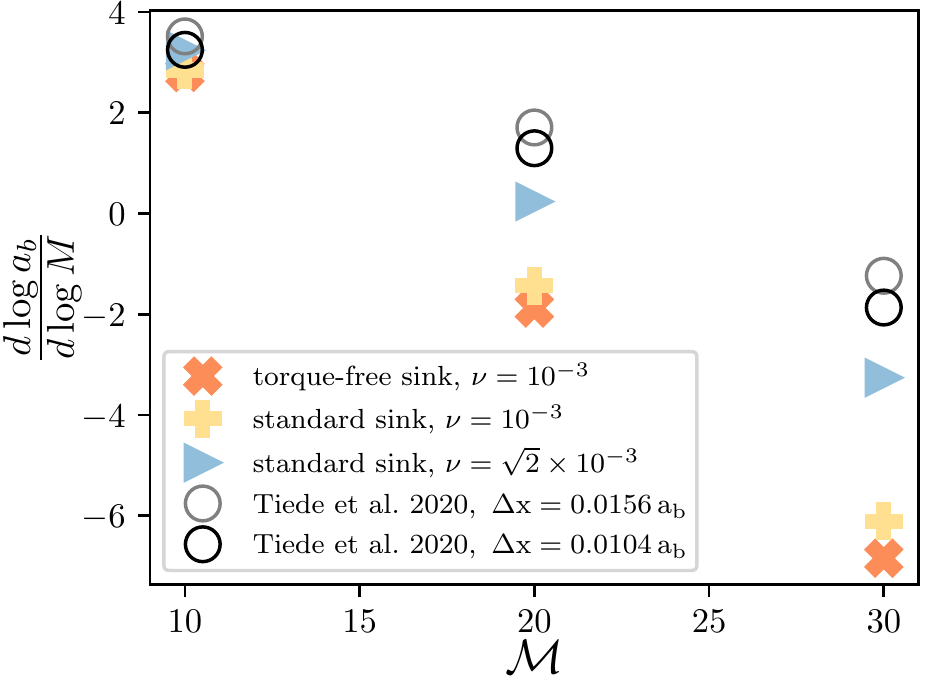}
\caption{The rate of change of the binary semi-major axis. Gray and black circles plot respectively the low- and high-resolution results from \citet{2020ApJ...900...43T} averaged between the 300th and 500th binary orbits. Solid symbols plot the results of our simulations, averaged over the final 500 binary orbits of each 2000-orbit simulation. Orange `x' symbols indicate results at $\nu=0.001$ using torque-free sinks; yellow  `+' symbols indicate results using $\nu=0.001$ and `standard' sinks, which exert a torque on the gas during accretion; blue triangles indicate results using $\nu=\sqrt{2}\times10^{-3}$, and setting $A=2,~b=2$ in Equation (\ref{GaussSquared}) and $\gamma=8$ in Equations (\ref{surfSink}) and (\ref{eq:tf2}), choices made to replicate the sink treatment used in \citet{2020ApJ...900...43T},
unlike all other simulations in this work which set $A=1$ and $b=4$.
We note that in this figure, and only this figure, we have deliberately and incorrectly included the spin component of the accretion torque in the calculation of $d\log{a}_b/d\log{M}$ in order to make better comparison with \citet{2020ApJ...900...43T}, although it does not affect our results using torque-free sinks. }
\label{fig:comparison}
\end{figure}

Previously, \citet{2020ApJ...900...43T} reported that the critical Mach number governing binary orbital evolution, below which binaries outspiral and above which binaries inspiral, was $\sim25$ at $\nu=\sqrt{2}\times10^{-3}$. We present an attempt to confirm their results in Figure \ref{fig:comparison}, along with results using $\nu=0.001$ for two different sink prescriptions. Our simulations using \Disco{} clearly confirm the overall trend shown by \citet{2020ApJ...900...43T}, which used the Cartesian \texttt{Mara3} code \citep{2012ApJ...744...32Z}. Both simulations agree at $\mathcal{M}=10$, although our simulations produce marginally smaller values of $d\log{a}/d\log{M}$ at $\mathcal{M}=20, 30$ and find a correspondingly lower critical Mach number of $\mathcal{M}\approx21.5$.

Note that at a given Mach number, simulations using lower values of $\nu$ find lower values of $d\log{a}/d\log{M}$, which also betrays the results of our investigation into the role of disc viscosity in binary evolution in Section \ref{sec:binOrb}.  \citet{2020ApJ...900...43T} also found
systematically lower values of $d\log{a}_b/d\log{M}$ as they increased  their resolution from $\Delta x = 0.0156a_b$ to $\Delta x = 0.0104a_b$, decreasing their numerical viscosity.
Our lower values of  $d\log{a}_b/d\log{M}$ may result from lower numerical viscosity in our simulations resulting from our use of a moving mesh, a cylindrical coordinate system, and a smaller cell size $\Delta r= 0.0086a_b$, but may also result from \citet{2020ApJ...900...43T} considering finite viscously-spreading discs while we consider infinite discs.  

\begin{figure*}
\centering
\includegraphics[width=\linewidth]{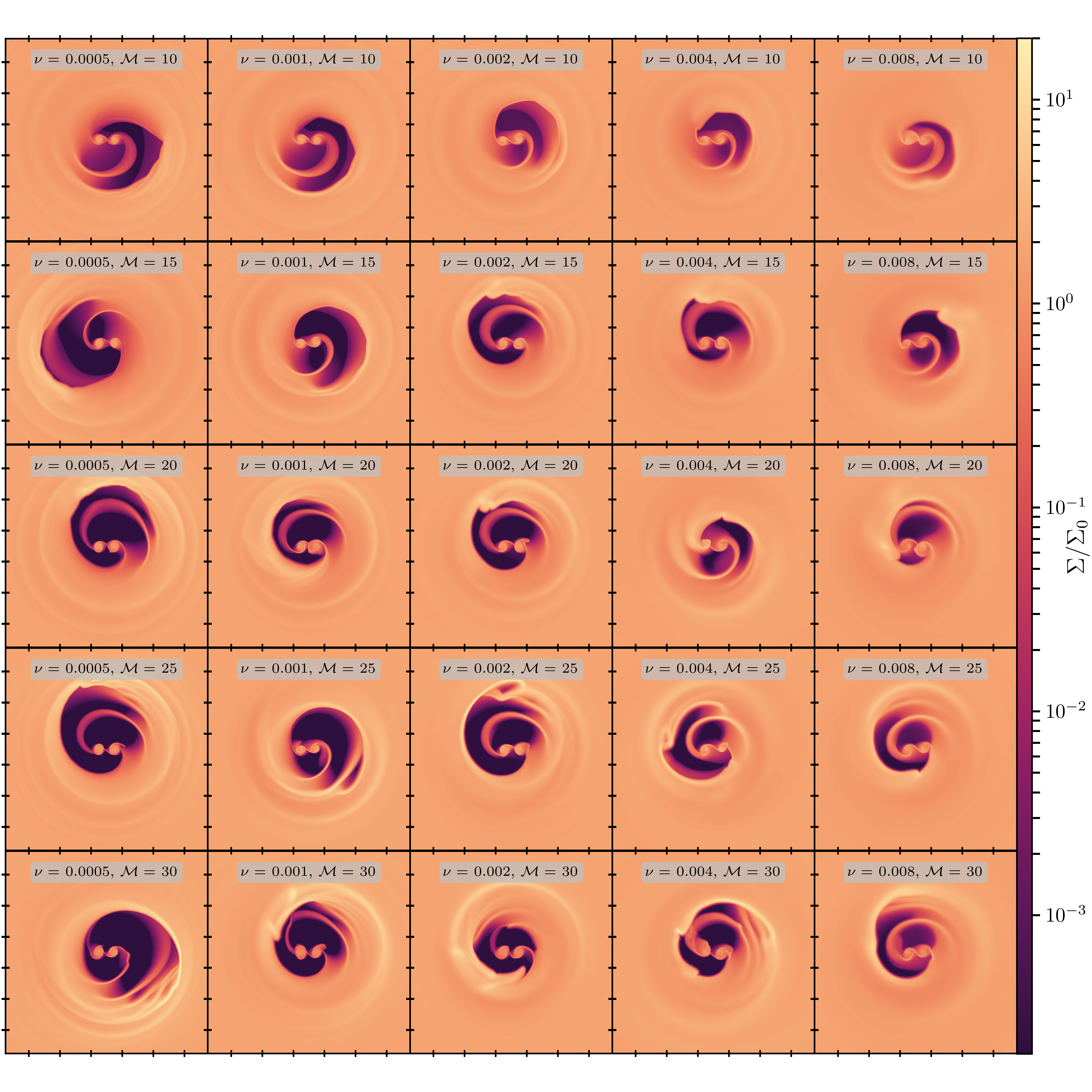}
\caption{The final surface density distributions in our main suite of simulations, normalized by $\Sigma_0$, and visualized using a logarithmic colour scale. Mach numbers range from $\mathcal{M}=10$ in the top row to $\mathcal{M}=30$ in the bottom row, while viscosities range from $\nu=0.0005$ in the leftmost column to $\nu=0.008$ in the rightmost column. Axis ticks are spaced $2a_b$ apart. We note the overdense lumps, which become more prominent at higher Mach numbers, how the cavity size changes with viscosity, and how the cavity becomes more evacuated at high Mach numbers and low viscosities.}
\label{fig:sigGrid}
\end{figure*}

In addition to the importance of viscosity, both physical and numerical, Figure \ref{fig:comparison} shows how at high Mach numbers, sink prescriptions that exert unphysical torques on accreting gas (yellow `+' symbols) result in anomalously large values of $d\log{a_b}/d\log{M}$ by $\sim 0.5 -1$ at $\mathcal{M}=\{20,30\}$ and $\nu=10^{-3}$. Similarly large changes in $d\log{a_b}/d\log{M}$ were found at lower mass ratios $(q\leq0.3)$ by \citep{2021ApJ...921...71D}, but at higher Mach numbers the sink treatment becomes more critical even for equal-mass binaries. We also carried out a small set of simulations at $\mathcal{M}=\{10, 20, 30\}$, $\nu=10^{-3}$ using a numerical resolution of $N_r=786$, with a corresponding smallest cell size of $\Delta r=0.00574a_b$, scaling $\epsilon_g$ and $r_s$ proportionally to $\Delta r$, and measured changes in $d\log{a_b}/d\log{M}\lesssim 1\%$, demonstrating that our choice of resolution is sufficient for this problem.

\section{Binary Orbital Evolution} \label{sec:binOrb}
Herein 
we detail the evolution of binaries and how it is intrinsically linked to the dynamics of gas in the circumbinary disc, cavity, and minidiscs. We focus on viscosities $\nu=\{0.0005,~0.001,~0.002,~0.004,~0.008\}$ and Mach numbers $\mathcal{M}=\{10,12.5,15,17.5,20,25,30\}$. Many key details that govern the evolution of these binaries, and their dependence on Mach number and viscosity, are immediately clear from examining late-time (after initial transients die out and the cavity becomes eccentric) simulation snapshots, such as those shown in Figure \ref{fig:sigGrid}.

First, we see that at high viscosities and low Mach numbers, the cavity is very `full', which is to say that the surface density contrast between the minimum-$\Sigma$ point in the cavity 
and circumbinary disc is small, whereas at high Mach number and low viscosity the surface density contrast between the circumbinary disc and cavity minimum is many orders of magnitude. In addition to the contrast between the cavity minimum and circumbinary disc, the surface density contrast between overdensity along the cavity wall (the `lump') and the circumbinary disc grows with Mach number, although its dependence on viscosity is not as obvious. Similar changes in disc morphology with Mach number have been seen at constant viscosity \citep[e.g.][]{2016MNRAS.460.1243R,2020ApJ...900...43T}.

It is also visually apparent that the radial extent of the cavity decreases as viscosity increases. This scaling follows from a linear analysis, balancing torques from waves excited at Lindblad resonances with the viscous torque in the disc \citep{1994ApJ...421..651A,2015MNRAS.452.2396M,2015ApJ...800...96L}, although these analyses commonly under-predict cavity sizes when compared with numerical simulations \citep[e.g.][]{2008ApJ...672...83M,2020MNRAS.499.3362R}, and should be taken at most as a general guideline \citep{2015MNRAS.452.2396M}. Especially at low viscosity, the cavity extends to a larger radius at high Mach number. However, it is not obvious by eye whether this is because the cavity has a larger semi-major axis, or if it becomes more eccentric at high Mach numbers. This question is explored in Section \ref{sec:diskOrb}. 

A rough picture of the propagation of density waves through the discs is also provided by Figure \ref{fig:sigGrid}. For example, at $\nu=0.0005$, density waves are visibly more tightly wound and propagate shorter distances into the disc at $\mathcal{M}=\{25,30\}$ than at $\mathcal{M}=\{10,15,20\}$, suggesting that they damp more quickly. However, at higher viscosities, e.g. $\nu={0.004, 0.008}$, waves tend to have lower amplitudes and do not propagate as far into the disc, although tighter winding at high Mach number can still be observed.

\subsection{Fiducial models}\label{sec:fiducial}
\begin{figure}
\centering
\includegraphics[width=\linewidth]{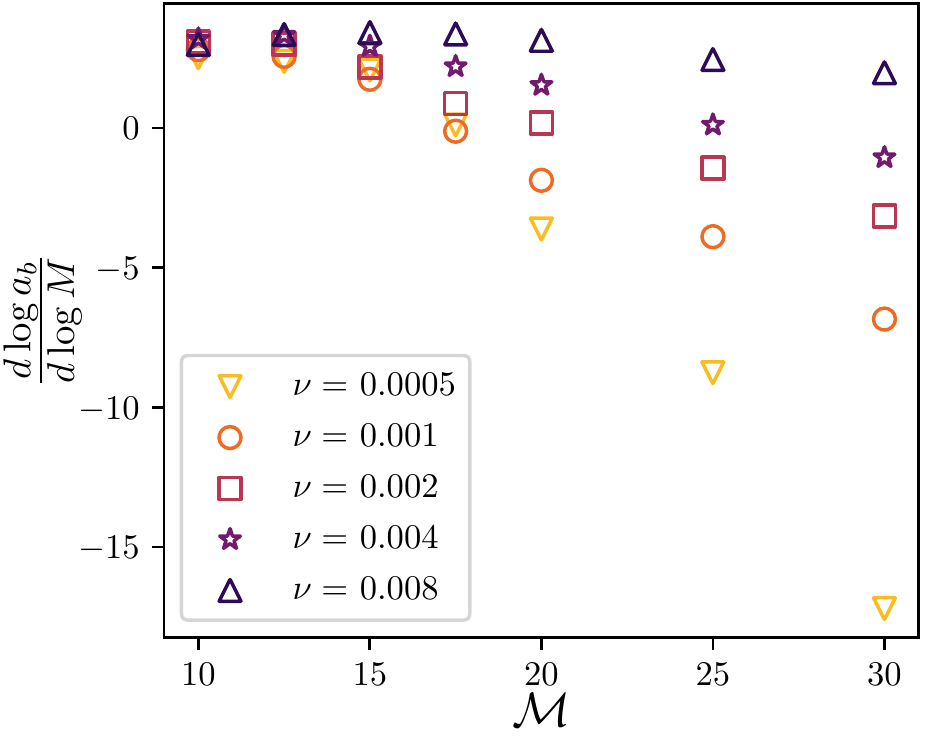}
\caption{Evolution of the binary semi-major axis for different Mach numbers and viscosities. Symbol shape and colour indicate viscosity, with lighter colours at low-$\nu$ and darker colours at high-$\nu$.}
\label{fig:qdlogs}
\end{figure}

\begin{figure}
\centering
\includegraphics[width=\linewidth]{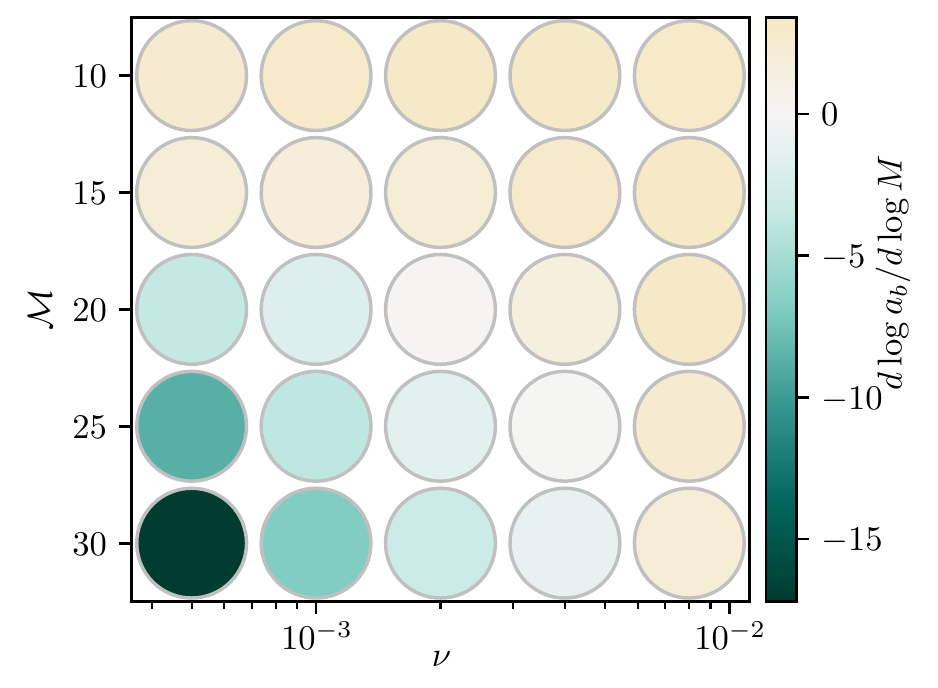}
\caption{Binary semi-major axis evolution in the $\nu-\mathcal{M}$ plane, where greens points indicate inspirals and and orange points indicate outspirals, with darker colours indicating faster evolution.}
\label{fig:spiralGrid}
\end{figure}

With the aforementioned general picture of circumbinary discs in mind, we investigate the orbital evolution of binaries. 
We measured the torques on the binaries as described in Section \ref{sec:diag} averaged over the final $500$ binary orbits of each simulation in order to determine $d\log{a}_b/d\log{M}$.
By this time each simulation had reached a quasi-steady-state in both accretion rate and torque on the binary.
As shown in Figure \ref{fig:qdlogs}, binaries inspiral \textit{very} rapidly at high Mach number compared to the timescale for their mass to grow, e.g. $|\dot{a}_b/a_b| \gg |\dot{M}/M|$. Another view of this is displayed in Figure \ref{fig:spiralGrid}, mapping out the $\nu-\mathcal{M}$ plane in terms of inspiral and outspiral. At low Mach numbers, $\mathcal{M}\sim10-15$, the evolution of the binary is almost independent of viscosity, even though the size and morphology of the cavity changes. This result was identified previously at $\mathcal{M}=10$ by \citet{2020ApJ...889..114M} and \citet{2020ApJ...901...25D}.
On the other hand, at higher Mach numbers, changes in viscosity by a factor of 2 can result in even larger changes in the rate of binary evolution. Clearly, knowledge gained from studies at $\mathcal{M}=10$ does not always translate to higher Mach numbers. 

\begin{figure}
\centering
\includegraphics[width=\linewidth]{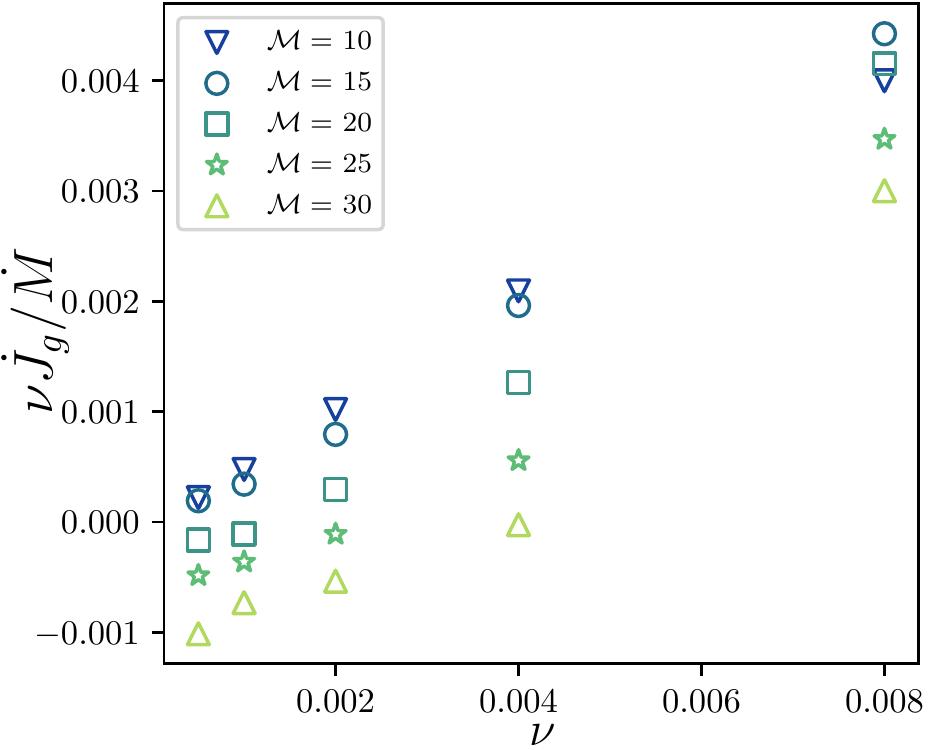}
\caption{The gravitational torque on the binary, scaled by a factor of $\dot{M}/\nu$ as a function of viscosity for various Mach numbers. This can be thought of as $\dot{J}_g/\Sigma_a$, where $\Sigma_a$ is a characteristic surface density of accreting material, thus directly probing how the geometry of the accretion flow influences the torque on the binary. Symbol shape and colour indicate Mach number, with lighter colours at high-$\mathcal{M}$ and darker colours at low-$\mathcal{M}$.}
\label{fig:nuTorque}
\end{figure}
Part of the variation of $d\log{a_b}/d\log{M}$ and $\dot{J}/\dot{M}$ with viscosity is independent of viscous modifications to disc morphology.
For example, consider circumbinary discs with identical morphologies, flow patterns, and accretion rates but different $\nu$: the accreting material at lower $\nu$ must have a higher surface density, and thus exert commensurately stronger gravitational torques. Alternatively, one can imagine two discs with the same surface density but different viscosities, in which case the viscous timescale is longer at low-viscosity, so gas can gravitationally interact with the binary for more dynamical times before being accreted. Along these lines, the dependence of $d\log{a_b}/d\log{M}$ on $\nu$ at higher Mach numbers is a less surprising result than the lack of $\nu-$dependence at lower Mach numbers. Indeed, the lack of $\nu-$dependence at higher aspect ratios suggests that viscosity \textit{strongly} affects the flow, changing the dynamics and geometry of the flow enough to counteract the effective rescaling of the surface density or timescales.

The dependence of $d\log{a_b}/d\log{M}$ on $\nu$ is also due to changes in the cavity dynamics and geometry. We can see this by examining the quantity $\dot{J}_g\nu/\dot{M}$, which scales out the $\nu$-dependence of $\dot{M}$. These results are displayed in Figure \ref{fig:nuTorque} as a function of $\nu$. This procedure effectively normalises the gravitational torque by the characteristic surface density of accreting material, and thus probes how the geometry of the flow changes the evolution of the binary as it varies with $\nu$ and $\mathcal{M}$. When using torque-free sinks, the average accretion torque is known analytically for equal-mass circular binaries, as accretion is divided equally between each binary member, so $d\log{a_b}/d\log{M}$ is a function of only $\dot{J}_g/\dot{M}$, leading to Equation (\ref{eq:orbJg}).  At low Mach numbers, we see that the slope of $\dot{J}_g\nu/\dot{M}$ remains roughly constant as a function of $\nu$, leading to the approximate viscosity-independence of $d\log{a_b}/d\log{M}$ often quoted in the literature. However, at higher Mach numbers we observe that the slope of $\dot{J}_g\nu/\dot{M}$ varies more significantly as a function of $\nu$, typically becoming shallower at low viscositites and leading thus to the corresponding dependence of $d\log{a}_b/d\log{M}$ in the high-$\mathcal{M}$ low-$\nu$ regime.

\subsection{Time-averaged profiles}

\begin{figure}
\centering
\includegraphics[width=\linewidth]{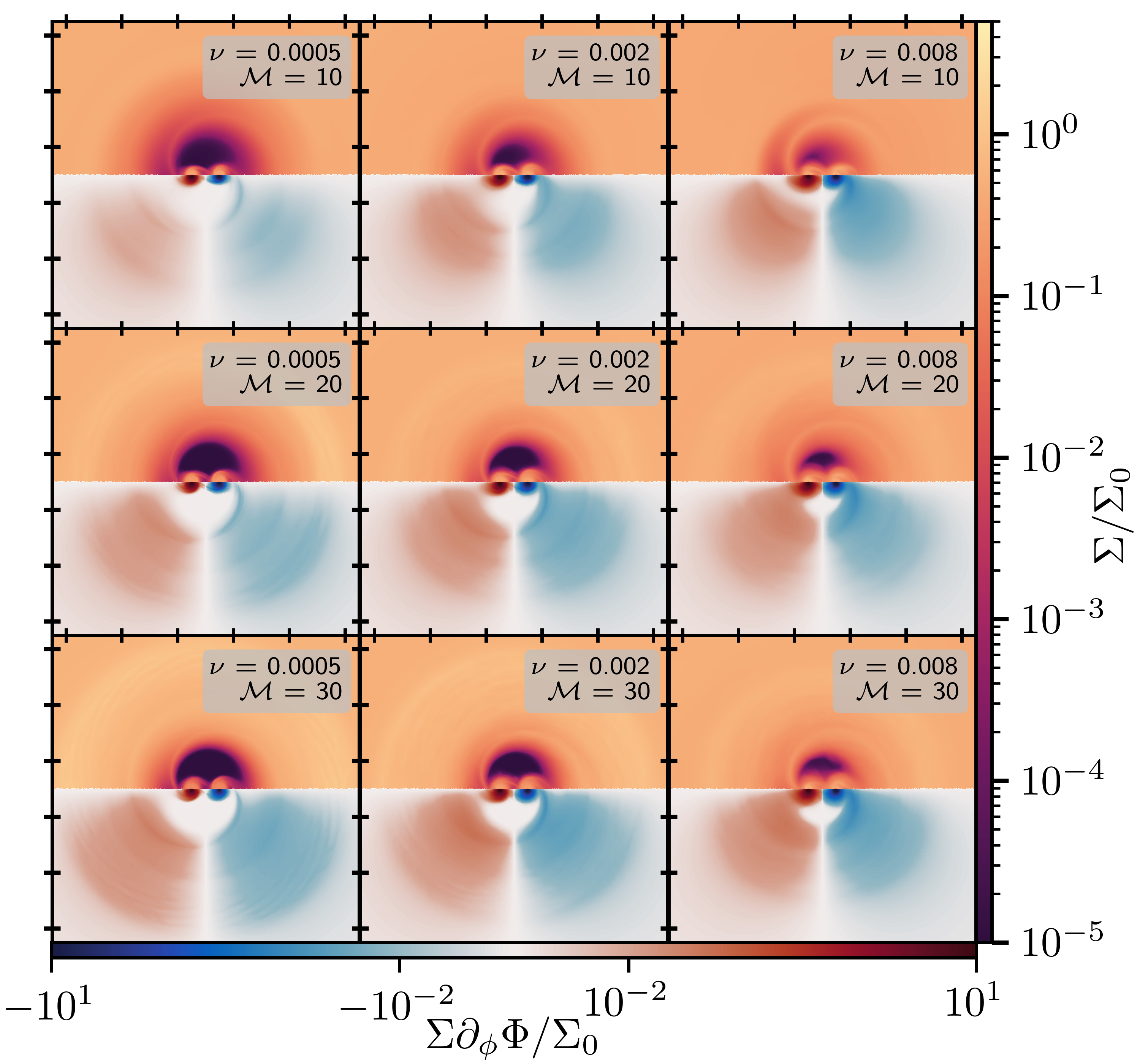}
\caption{Plots of the surface density and gravitational torque density in the frame of the binary, time-averaged over 500 orbits. Plots are colored according to the gravitational torque density for $\pi<\phi\leq2\pi$ and according to the surface density for $0<\phi\leq\pi$. Axis ticks separated by $2a_b$.}
\label{fig:averages}
\end{figure}

Even with the natural dependence of $d\log{a_b}/d\log{M}$ on $\nu$ scaled out, both $\nu$ and $\mathcal{M}$ significantly change the orbital evolution of the binary. In order to gain a better understanding of the origins of these dependencies, we investigate the time-averaged properties of the accretion flow.  We plot the surface density in Figures \ref{fig:averages} and \ref{fig:sigAvg}, average radial velocity in Figure \ref{fig:vrAvg}, and gravitational torque density on the binary in Figures \ref{fig:jgAvg}, for $\nu=\{0.0005, 0.002, 0.008\}$ and $\mathcal{M}=\{10,20,30\}$. 

In these plots, we have stacked snapshots from the last 500 orbits of each simulation, sampled once per orbit. Because each annulus in our mesh rotates with the average fluid velocity in that annulus, this procedure involves interpolating fluid quantities onto a single grid for plotting. Because we average over a longer timescale than the precession periods of the cavity in each simulation $\sim400-500$ orbits (see Section \ref{sec:diskOrb}), the resulting profiles are approximately $m=2$ rotationally symmetric. We show examples of the profiles in standard ($r-\phi$) coordinates in Figure \ref{fig:averages}, but otherwise present only the profiles from $0\leq\phi<\pi$, and `unroll' the $\phi$-coordinate of these averaged 2D profiles into a rectilinear grid to make comparisons with the average 1D gravitational torque and surface density profiles. The 1D profiles in Figures \ref{fig:sigAvg}, 
\ref{fig:vrAvg}, and \ref{fig:jgAvg} are averaged over the same period of 500 orbits, but the time averaging was sampled once per timestep rather than once per orbit. The x-axis in these unrolled plots is begins at $0.71a_b$, such that the plots include the outermost parts of the binary Roche lobes at $\sim0.88a_b$ \citep{1977MNRAS.181..441P}.

\subsubsection{Gas Morphology}

\begin{figure}
\centering
\includegraphics[width=\linewidth]{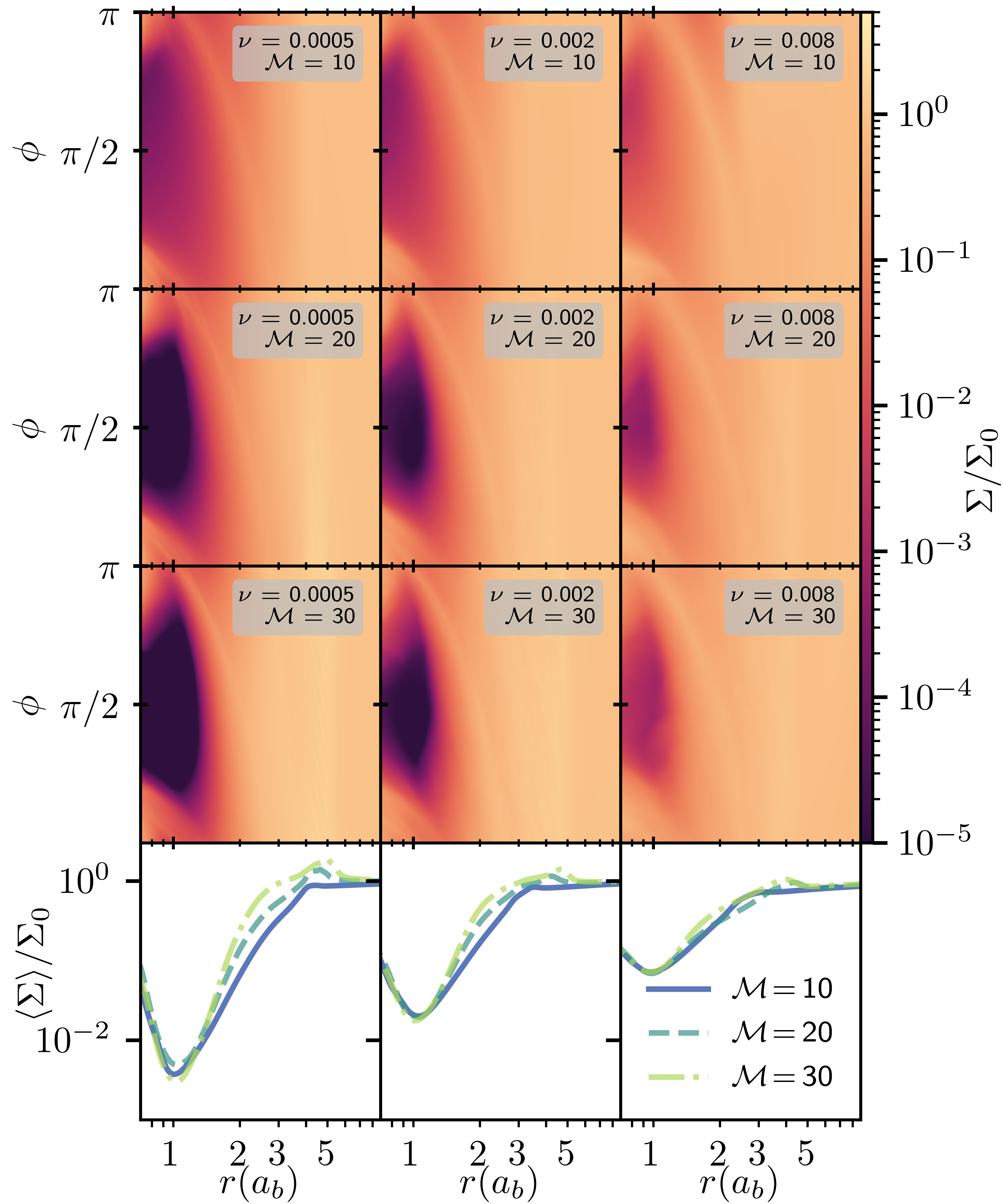}
\caption{The surface density distributions in a representative set of our simulations, normalised by $\Sigma_0$ with a logarithmic color scale. Note the overdense lumps, which become more prominent at higher Mach numbers, and how the cavity becomes more evacuated at high Mach number and low viscosity. The first column presents results at $\nu=0.0005$, the second column at $\nu=0.002$, and the third column at $\nu=0.008$. The first, second, and third rows present results for $\mathcal{M}=\{10,20,30\}$ respectively, while in the bottom row Mach 10 results are plotted using solid blue lines, Mach 30 results plotted using dash-dotted yellow-green, and Mach 20 results are plotted using dashed green lines.}
\label{fig:sigAvg}
\end{figure}

\begin{figure}
\centering
\includegraphics[width=\linewidth]{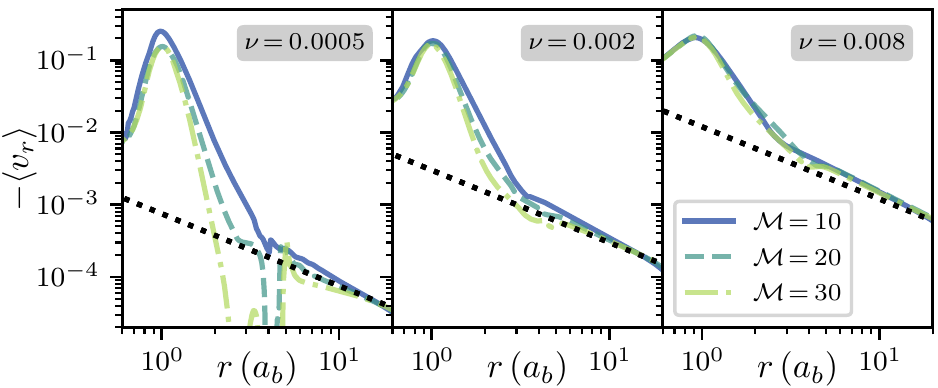}
\caption{The mass-weighted average radial velocity $\langle v_r\rangle$ for a subset of our simulations.  Mach 10 results are plotted using solid blue lines, Mach 20 results are plotted using dashed green lines, and Mach 30 results plotted using dash-dotted yellow-green lines. Dotted black lines indicate the viscous inflow rate, $v_r=-2\nu/3r$.}
\label{fig:vrAvg}
\end{figure}

Turning first to the time-averaged surface density profiles in Figures \ref{fig:averages} and \ref{fig:sigAvg}, we observe that in a time-averaged sense the cavity becomes deeper as $\nu$ decreases and $\mathcal{M}$ increases, the same pattern shown by the individual snapshots in Figure \ref{fig:sigGrid}. Furthermore, based on the $\phi-$averaged surface density profiles in the bottom row of Figure \ref{fig:sigAvg}, the average mass in the shallowest part of the cavity, around $r=a_b$, hardly changes as a function of Mach number despite the large differences shown in the upper panels of Figure \ref{fig:sigAvg}, particularly comparing the $\mathcal{M}=10$ and $\mathcal{M}=20$ profiles.

We have also measured that ${\rm min}\!\left(\langle\Sigma\rangle/\Sigma_0\right))\propto\nu$, or equivalently that ${\rm min}\!\left(\langle\Sigma\rangle\right)$ is independent of $\nu$ at constant $\dot{M}$ to within a few per cent. 
This suggests that the accretion rate through the cavity is roughly constant, but at a velocity that is governed by the binary rather than the viscous properties of the accretion disc. Examining the mass-weighted average radial velocity profiles in Figure \ref{fig:vrAvg}, we see that far from the binary ($r\gg a_b$) the radial velocity is given by the viscous rate shown by the dotted black line, but diverges from the viscous rate near the binary. The peak in $-\langle v_r\rangle$, around $r=a_b$, is nearly independent of both viscosity and Mach number. However, $\langle v_r\rangle$ clearly depends on Mach number for $a_b \lesssim r \lesssim 4\,a_b$, where the lower values of $\langle v_r\rangle$ at high Mach number agree with the higher surface densities shown in Figure \ref{fig:sigAvg}.

\begin{figure}
\centering
\includegraphics[width=\linewidth]{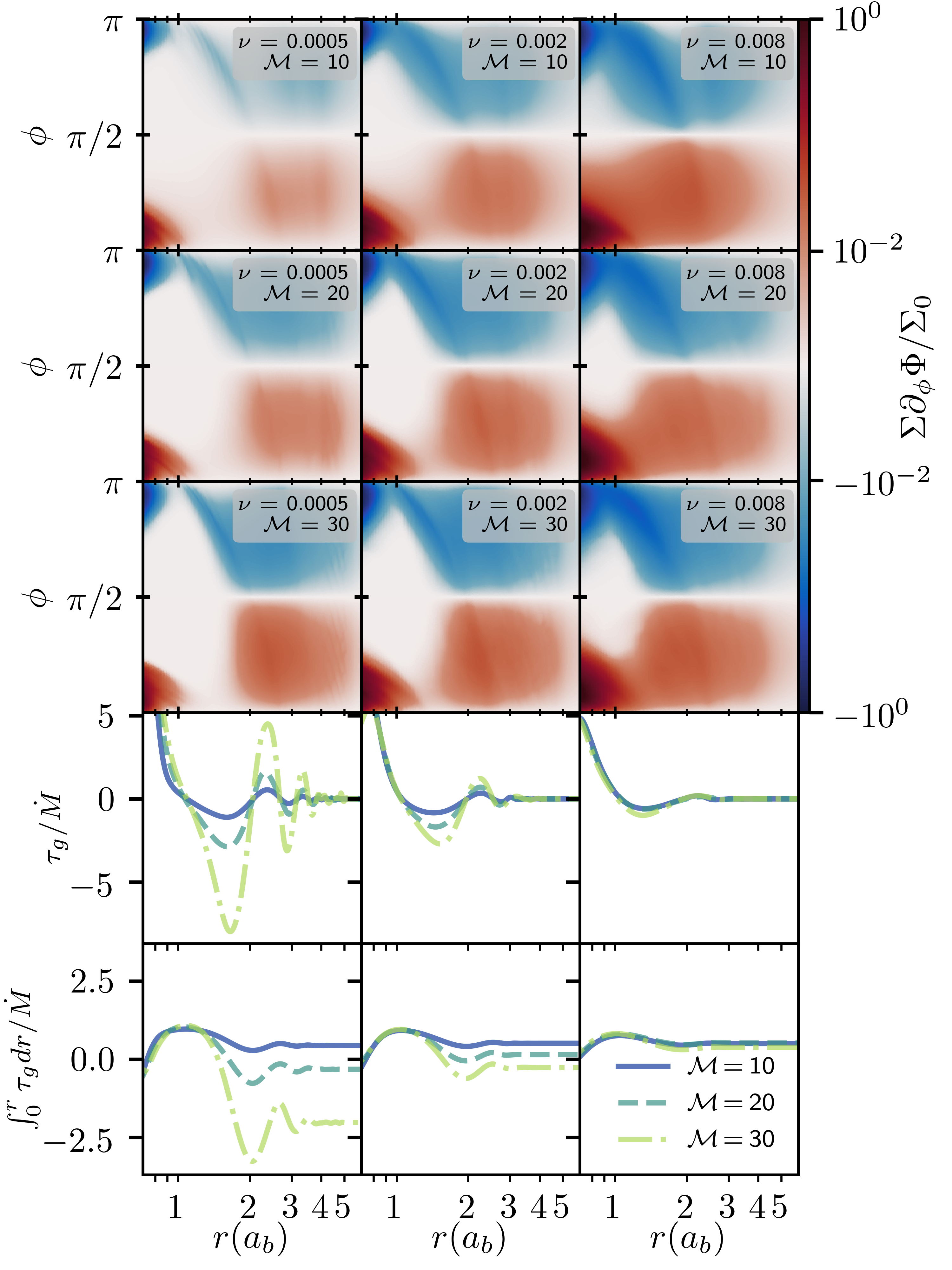}
\caption{Gravitational torques exerted by accereting gas on the binary. Areal torque densities, $\dot{J}_g=\Sigma\partial_\phi\Phi/\Sigma_0$ are presented in the first three rows and normalised by the surface density at the outer boundary condition. Linear torque densities, $\tau_g/\dot{M}_0=\pi r\langle \Sigma\partial_\phi\Phi \rangle/\dot{M}_0$ are presented in the fourth row, and integrated gravitational torque density from $0$ to $r$ is presented in the fifth row. 
The first column presents results at $\nu=0.0005$, the second column at $\nu=0.002$, and the third column at $\nu=0.008$. The first, second, and third rows present results for $\mathcal{M}=\{10,20,30\}$, while in the fourth and fifth rows Mach 10 results are plotted using solid blue lines, Mach 30 results plotted using dash-dotted yellow-green, and Mach 20 results are plotted using dashed green lines.}
\label{fig:jgAvg}
\end{figure}

\begin{figure*}
\centering
\includegraphics[width=\linewidth]{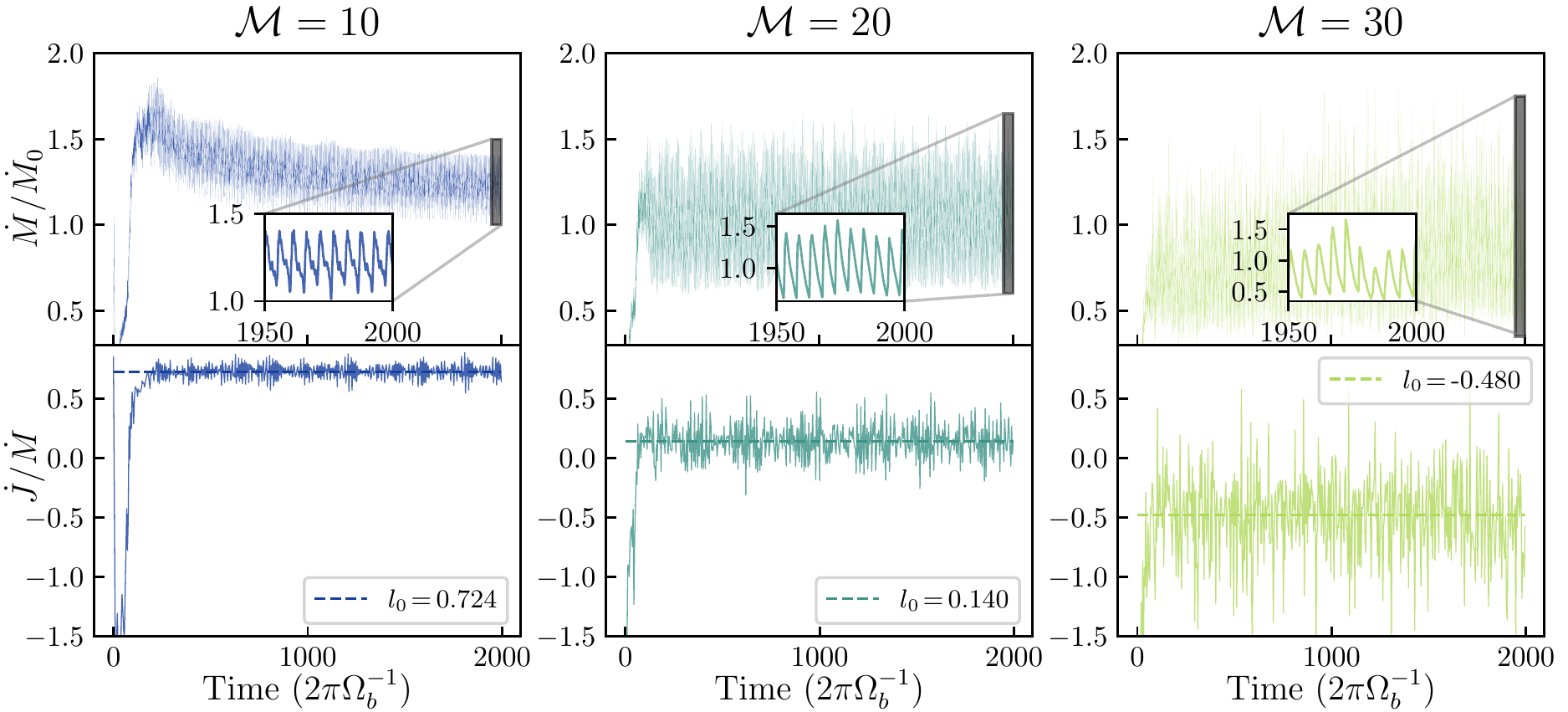}
\caption{Time series of the accretion rate (top row) and accreted specific angular momentum onto the binary (bottom row) over time for three $\nu=0.001$-simulations. The first, second, and third columns report data from the $\mathcal{M}=\{10,20,30\}$ simulations respectively. In the top row, the insets show the accretion rate over the last 50 orbits of each simulation.   In the bottom row, the dashed line indicates the average of $\dot{J}/\dot{M}$ over the last 500 orbits of each simulation, the average of which is reported in the legend. The average accretion rates over the last 500 orbits were $\dot{M}/\dot{M}_0=\{1.24, 1.03, 0.84\}$ for the $\mathcal{M}=\{10,20,30\}$ simulations respectively.} 
\label{fig:timeseries}
\end{figure*}

Focusing on the results displayed in the $\nu=0.0005$ columns of Figures \ref{fig:sigAvg} and \ref{fig:vrAvg}, we observe that $-\langle v_r\rangle$ decreases by more than an order of magnitude below the viscous rate from $4a_b \lesssim r\lesssim 5a_b$ at $\mathcal{M}=20$ and from $2a_b \lesssim r\lesssim 5a_b$ at $\mathcal{M}=30$, while in neither case the average surface density increases enough to keep $\dot{M}$ radially constant. Thus, these discs are not in a time-averaged steady state, although to a much lesser extent in the $\mathcal{M}=20$ case. In the $\mathcal{M}=30$ simulation, we measure the average accretion rate onto the binary over the last 500 orbits of the simulation to be $\sim 0.6\, \dot{M}_0$ (see Table \ref{tab:summary}), suggesting that the surface density is continuing to increase near the inner edge of the disc. We note that steady-state accretion disc profiles scale as $\Sigma(r)|_{\dot{J}/\dot{M}\neq 0}=\Sigma(r)|_{\dot{J}/\dot{M}=0}(1-\dot{J}\sqrt{r_0/r}/\dot{M})$ where $r_0$ is a characteristic length scale \citep{1974MNRAS.168..603L}. Thus, for $\dot{J}/\dot{M}<0$, the steady-state surface density at $r<\infty$ can be significantly larger than that for $\dot{J}/\dot{M}=0$. Accretion flows which initially have $\dot{J}/\dot{M}$ larger than their equilibrium value naturally experience decreases in $\dot{M}$ below the steady-state value as the discs approach a steady-state profile. This behaviour, along with $\dot{J}/\dot{M}$ at high Mach numbers, is likely at the heart of simulations which have indicated suppression of the accretion onto the binary in thinner disks \citep{2016MNRAS.460.1243R}. We explore this further in Section \ref{sec:acc}, showing that approximately steady accretion flows can be recovered even at high Mach numbers in simulations with initial $\dot{J}/\dot{M}$ closer to the equilibrium value.

\subsubsection{Torques}

With changes in cavity structure as a function of Mach number and viscosity in mind, we turn to the gravitational torque on the binary. The gravitational torque is essentially determined by the amount of mass leading the binary, exerting on it a positive torque, and material lagging behind the binary which exerts a negative torque. For the equal-mass binaries studied in this work, the contribution of the accretion torque to $d\log{a}/d\log{M}$ does not depend on Mach number or viscosity (e.g. Equation \ref{eq:orbJg}). We present maps of the time-averaged normalised torque surface density as a function of $\nu$ and $\mathcal{M}$ in the upper panels of Figure \ref{fig:jgAvg}. We show time-averaged plots of the linear gravitational torque density in the fourth row of Figure \ref{fig:jgAvg}, where we have integrated the torque surface density over each annulus, and the radially integrated torque in the fifth row, the total gravitational torque on the binary from material within $r$ from the binary barycentre. 

In general, the outer minidiscs are asymmetric and tend to exert a positive net torque around $r\lesssim a_b$. Especially at lower Mach number and higher viscosity, there can be an appreciable torque from material within the cavity itself. At lower viscosities, the gravitational torque from the accretion stream trailing the binary dominates from $a_b<r<2a_b$, although its contribution to the total torque becomes less pronounced at higher viscosities. The evolution of the binary is then determined by whether or not the negative contribution from trailing material is able to overwhelm the positive contributions from the minidiscs and cavity walls. We observe, because $\partial\Phi/\partial\phi$ is uniform between our simulations, that the streams of material shown in Figure \ref{fig:jgAvg} become more dense at high Mach numbers. We note that the wavelength of the torque density profiles shown in Figure \ref{fig:jgAvg} is independent of Mach number and viscosity, particularly at $r<3\,a_b$ where the negative torques which drive inspirals are excited. Thus, resonantly excited density waves are not responsible for the majority of the torque, as the wavelength of density waves should decrease at higher Mach numbers. Rather, as also shown by Figure \ref{fig:vrAvg}, the material responsible for the negative torque is predominantly ballistic with motion independent of viscosity and mach number.

The higher-density material in the accretion streams and inner disc at higher Mach numbers and lower viscosities directly leads to the rapid inspirals shown in Figures \ref{fig:qdlogs} and \ref{fig:spiralGrid}.
\citet{2020ApJ...900...43T} identified the same general trend: that as Mach number increases, the gravitational torque from both the disc and cavity region become increasingly negative, while the torque from the minidiscs remains positive and roughly constant \citep[e.g. Figures 7 and 8 of][]{2020ApJ...900...43T}. At lower Mach numbers and higher viscosities, we see that the higher ambient density material in the cavity can lead the binary, exerting a positive torque and diluting the influence of the streams.

\subsection{Accretion Time Series}\label{sec:acc}
As has been discussed previously \citep[e.g.][]{2017MNRAS.466.1170M,2020ApJ...889..114M,2020ApJ...900...43T}, it is crucial that simulations are run for long enough that initial transients to die out, which may have impaired earlier studies \citep[c.f.][]{2008ApJ...672...83M,2020ApJ...889..114M}.
In Figure \ref{fig:timeseries}, we show that even when the accretion rate onto the binary is changing adiabatically or is on average different than the accretion rate in our initial condition, the inferred evolution of the binary is minimally affected.
Specifically, after the first few hundred orbits, $\dot{J}/\dot{M}$ ceases to evolve on a timescale comparable to the binary orbital period, although we expect secular variation on much longer timescales, comparable to the viscous timescale at $r\gg a_b$, in cases where $\dot{M}/\dot{M}_0$ deviates significantly from unity.

In the $\mathcal{M}=10$ case, after an initial transient, the accretion rate onto the binary secularly decreases, averaging $\sim1.2$ times the value specified in our initial condition. However, after the initial transient dies out, the average angular momentum change of the binary per unit mass remains roughly constant. Higher Mach number simulations evolve similarly, but have seemingly shorter transient periods. Note that in all cases, the primary period of variability onto the binary is roughly $5$ orbits, and the amplitude of this accretion rate variability increases with Mach number, a trend which can also be seen in the time series presented in \citep{2020ApJ...900...43T}, and which parallels the higher surface density of the cavity walls (see, e.g. the 1D profiles in Figure \ref{fig:sigAvg}). The variability in the average binary angular momentum change per unit accreted mass also increases with Mach number. We note that \citet{2021MNRAS.501.3540D} also observed higher magnitude variability in simulations of a $q=10^{-3}$ binary as Mach number increased, and found that the variability was stronger at higher resolution and fixed Mach number.

We find that the average accretion rates in the simulations presented in Figure \ref{fig:timeseries} are anticorrelated with $\dot{J}/\dot{M}$. This trend was observed in \citet{2016MNRAS.460.1243R}, which found that the accretion rate onto binaries is suppressed at high Mach number. Similarly, \citet{2020A&A...641A..64H} conjectured that binaries inspiral at high Mach number \textit{because} accretion is suppressed. However, the accretion rate specified in our initial condition assumed a net angular momentum current of zero through the disc, which is inconsistent with $l_0\neq0$, so disagreement with the initial $\dot{M}_0$ and $\langle\dot{M}\rangle$ is natural if $|l_0|$ is large. 
To verify that our observed reduction in $\dot{M}$ is spurious, and test whether our conclusions about the orbital evolution of binaries are affected, we re-ran the $\nu=0.001,~\mathcal{M}=30$ simulation using the value of $l_0$ found in the initial simulation, similar to the procedures used in a number of other studies \citep[e.g.][]{2017MNRAS.466.1170M,2020ApJ...905..106M,2020ApJ...891..108D}. In the initial simulation, initialised using $l_i=0$, we found $\langle\dot{M}\rangle/\dot{M}_0=0.84,~l_0=-0.48,$ and $d\log{a_b}/d\log{M}=-6.84,$ while in the simulation initialised using $l_i=-0.48$, we found 
$\langle\dot{M}\rangle/\dot{M}_0=1.06, ~l_0=-0.37,$ and $d\log{a}/d\log{M}=-6.0,$
and remark that this value of $l_0$ is well within the scatter in the time series of $\dot{J}/\dot{M}$ in the right column of Figure \ref{fig:timeseries}. We have verified that this result holds for different choices of disc parameters, the most extreme case of which was a $\nu=0.0005,\,\mathcal{M}=50$ disc: using $l_i=0$, we found $l_0\approx-14.6$ and $\dot{M}/\dot{M}_0\approx0.14$, but after a single iteration using $l_i=-14.6$ we found $l_0\approx -11$ and $\dot{M}/\dot{M}\approx0.91$, and expect that another iteration would be sufficient for the disc to settle into an approximate viscous steady state. We note that this result is consistent with \citet{2017MNRAS.466.1170M}, which found that such iterations in $\mathcal{M}=10$ simulations reduced $\dot{M}/\dot{M}_0$ in cases where it was greater than unity, consistent with their $l_0>0$.

Thus, the decreased accretion rates in our initial simulations were indeed spurious, and the result of initial conditions that were too far from the eventual viscous steady state. Furthermore, binaries still rapidly inspiral without any suppression of the accretion rate. In general, we present results without any similar iterative procedure, so they are likely only precise to $\sim10\%$ at higher Mach numbers and lower viscosities, but are in general qualitatively sound. 

\subsection{$\alpha$-viscosity}\label{sec:alpha}
It is very popular to model angular momentum transport in astrophysical discs using an $\alpha$ viscosity model \citep{1973A&A....24..337S}, where angular momentum transport is assumed to take place through turbulence, which is related to the size and velocity of the largest eddies and thus governed by the disc scale height and disc sound speed through a dimensionless global constant $0<\alpha<1$ such that $\nu=\alpha c_s H$. Of course, a globally-constant $\alpha$ can not describe most AGN, where angular momentum transport is thought to be seeded by the magneto-rotational instability \citep{velikhov59,1960PNAS...46..253C,1998RvMP...70....1B}, although models where $\alpha$ varies as a function of time and space can provide a much-improved description of angular momentum transport \citep{1997MNRAS.292..679L,2016ApJ...826...40H}. \citet{2016ApJ...826...40H} found that the effective $\alpha$ in their magnetohydrodynamic simulations was log-normally distributed with values roughly bracketed by $0.02<\alpha<0.1$, which we use to inform our choices of $\alpha$. Many studies of circumbinary discs have used a constant-$\alpha$ viscosity \citep[e.g.][]{2017MNRAS.466.1170M,2014ApJ...783..134F,2020ApJ...889..114M,2019ApJ...875...66M,2019ApJ...871...84M,2020A&A...641A..64H}. However, in this formulation, $\nu\propto\alpha\mathcal{M}^{-2}$, which not only makes it more difficult to separate the effects of viscosity from pressure effects, but also makes carrying out converged high-Mach number simulations computationally prohibitive, as the viscous timescale at a given radius grows longer as the Mach number increases $t_\nu\propto r^2/\nu\propto\mathcal{M}^2$.

\begin{figure}
\centering
\includegraphics[width=\linewidth]{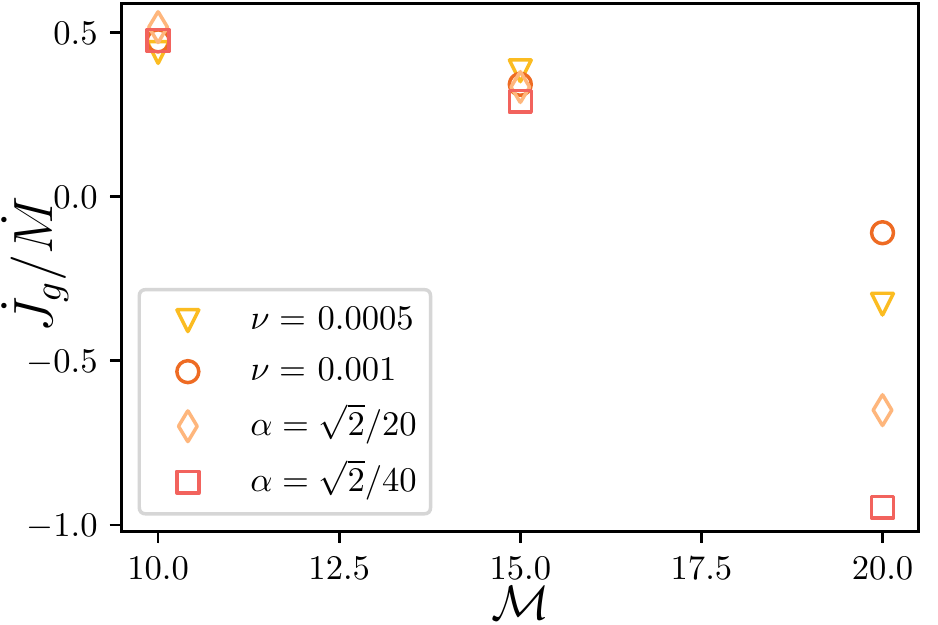}
\caption{Change in binary angular momentum, per unit accreted mass, due to gravitational interactions
for different Mach numbers, viscosities, and viscosity prescriptions. Symbol shape and colour indicate viscosity, with lighter colours at low-$\nu$ and darker colours at high-$\nu$. Recall that simulations using $\alpha$ viscosity, $\nu$ is additionally a function of radius and Mach number, and note that at $\mathcal{M}=10,$ $\alpha=\sqrt{2}/20$ corresponds to $\nu=0.001$ at $r=2a_b$. }
\label{fig:dlogAlpha}
\end{figure}

With these caveats in mind, we carried out a small suite of simulations using an $\alpha$-viscosity model, at $\alpha=\{\sqrt{2}/20, \sqrt{2}/40\}\sim\{0.0707,0.0354\}$, $\mathcal{M}=\{10,15,20\}$. Each of these was simulated for roughly $2\pi$ viscous timescales at $r=2a_b$, where we estimate the viscous timescale as $r^2/2\nu$, e.g. 9000 binary orbits for $\alpha=\sqrt{2}/40,~\mathcal{M}=15$, except for the $\alpha=\sqrt{2}/40,~\mathcal{M}=20$ simulation which was only run for 8000 binary orbits. Because we have found that decreasing $\nu$ leads to faster binary inspirals at higher Mach numbers, the use of an $\alpha$-viscosity should lead to even greater changes in orbital evolution as the Mach number increases. We present results in terms of the gravitational torque in Figure \ref{fig:dlogAlpha}, along with results from constant-$\nu$ simulations for comparison. 

As expected, at higher Mach numbers inspirals become more rapid at lower $\nu$ (at constant $\dot{M}$) as $\dot{J}_g/\dot{M}$ becomes more negative. Our result follows naturally from the fact that the accretion timescale becomes longer at low viscosities, and the structure of the accretion flows becoming less affected by viscosity at higher Mach numbers, as shown in Figure \ref{fig:nuTorque}. Thus, our results from constant$-\nu$ simulations are qualitatively consistent with those from simulations using an $\alpha-$viscosity model.\footnote{We note that \citet{2020ApJ...900...43T} found much smaller differences between their simulations using an $\alpha$-viscosity and holding $\nu$ constant (with the same proportionality at $\mathcal{M}=10$ as the viscosities tested here) further suggesting that numerical viscosity may have played a more significant role in their simulations.}
\begin{figure}
\centering
\includegraphics[width=\linewidth]{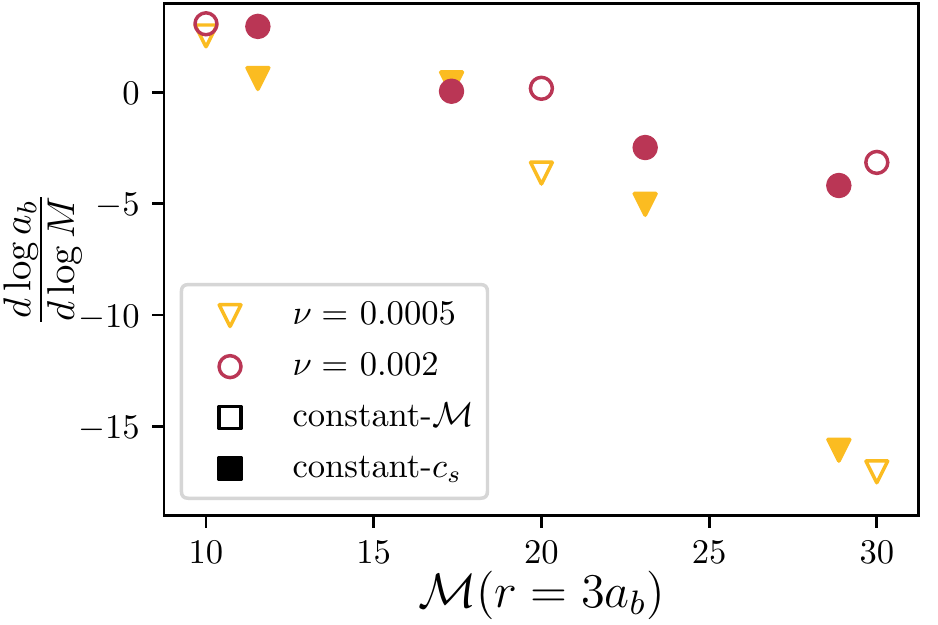}
\caption{Evolution of the binary semi-major axis for different Mach numbers, viscosities, and equations of state. Symbol shape and colour indicate viscosity, with lighter colours at low-$\nu$ and darker colours at high-$\nu$. Open symbols plot data from simulations using a locally isothermal equation of state, while filled symbols plot data from simulations using globally isothermal equations of state. Results are plotted as a function of the Mach number at $r=3a_b$.}
\label{fig:logsGlobal}
\end{figure}
\subsection{Equation of state}\label{sec:global}
One limitation of our study is its reliance on an isothermal equation of state. Our isothermal equation of state assumes that gas is heated or cooled extremely rapidly, potentially much faster than one would realistically expect in many systems. Additionally, as detailed by \citet{2020ApJ...892...65M,2019ApJ...878L...9M}, locally isothermal equations of state, where the sound speeds vary as a function of radius, lead to the angular momentum flux of density waves changing as they propagate radially. Neither discs with adiabatic indices not equal to 1 nor discs with globally constant sounds speeds display this behaviour. Because such waves play an important role in angular momentum transport through circumbinary discs, it is important to test whether or not this pathological property of locally isothermal discs has affected our conclusions. Although a thorough investigation of discs with realistic cooling and thermodynamics is necessary, it is beyond the scope of our present work. Instead, we performed a small set of exploratory simulations holding the sound speed \textit{globally} constant, as in Equation (\ref{globiso}).

\begin{figure*}
\centering
\includegraphics[width=\linewidth]{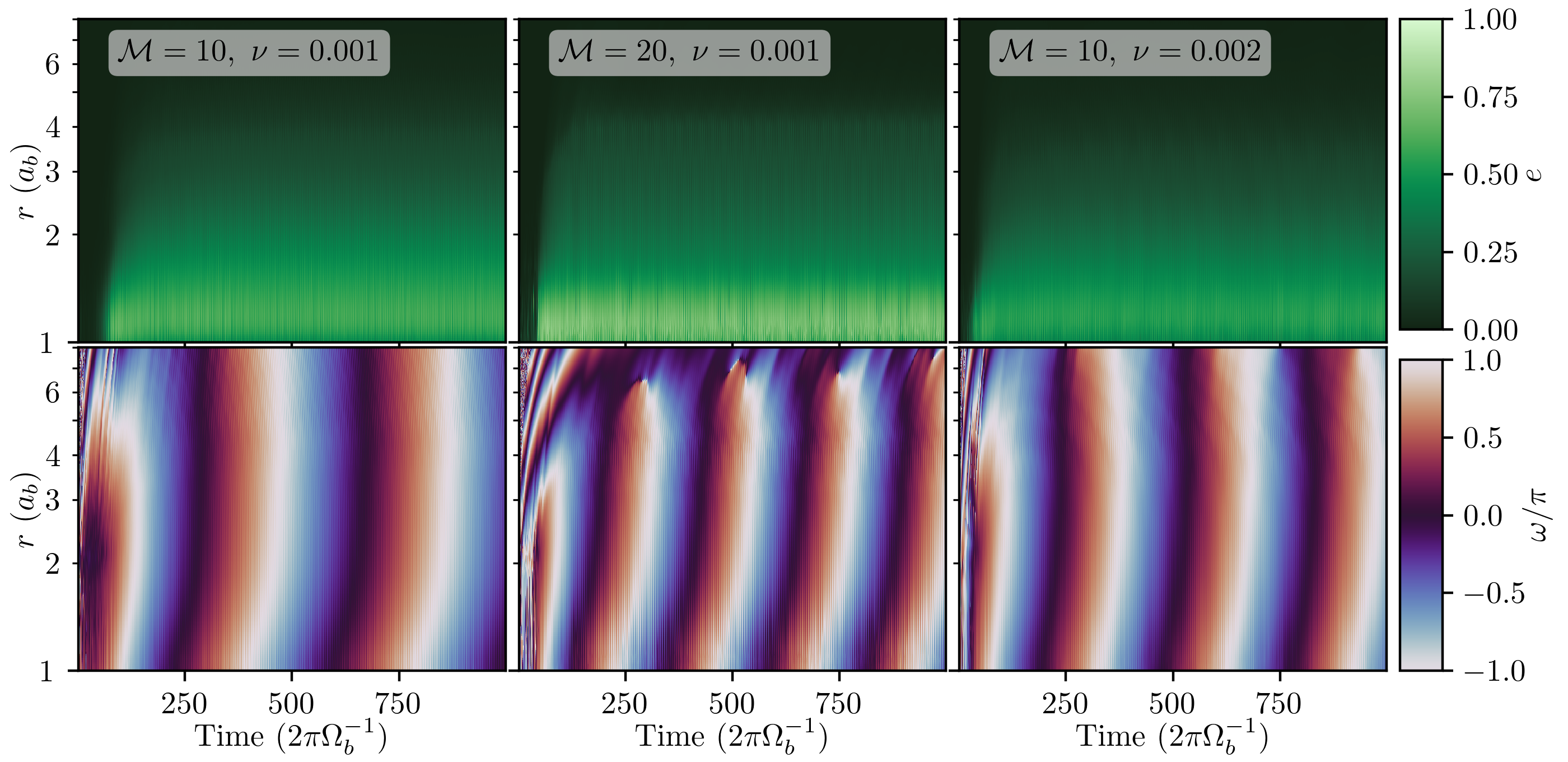}
\caption{The mass-weighted average eccentricity (top row) and argument of periapsis (bottom row) as a function of radius for the first 1000 orbits of a set of simulations. We plot averaged orbital elements once per binary orbit, in each case integrating over every timestep.}
\label{fig:3ecc}
\end{figure*}
We carried out a set of eight additional simulations using a globally isothermal equation of state at $\nu=\{0.0005, 0.002\},$ and $\mathcal{M}_*=\{20,30,40,50\}$ so that $\mathcal{M}\gtrsim10$ throughout the cavity. At a given Mach number, these simulations have cooler minidiscs but hotter circumbinary discs compared to our simulations using locally isothermal equations of state. 
We plot our results in terms of the Mach number at $r=3a_b$ in Figure \ref{fig:logsGlobal}. We find good agreement between globally and locally isothermal simulation results when parameterised in this manner, and therefore posit that the sound speed around the inner edge of the circumbinary disc plays a dominant role in binary evolution.
Due to their very different sound speed profiles $c_s(r)$, it is not surprising that the agreement is imperfect. 
Nevertheless, the same general trends occur: thinner and/or less viscous discs tend to reduce $d\log{a_b}/d\log{M}$ for their binaries. Thus, our results are not malignantly affected by the radial dependence of the wave angular momentum flux in locally isothermal discs. 

\section{Disc Morphology}\label{sec:diskOrb}
While interactions with the disc cause the binary to evolve, interactions with the binary also cause the disc to evolve, as shown by Figure \ref{fig:sigGrid}. Eccentric circumbinary discs are commonly observed in stellar binaries such as GG Tauri A \citep{1999A&A...348..570G,2002ApJ...575..974M,2014ApJ...787..148A,2020A&A...639A..62K}, and are commonly found in numerical simulations \citep[e.g.][]{2008ApJ...672...83M,2017MNRAS.466.1170M,2020MNRAS.499.3362R} of binaries with mass ratios $q\gtrsim0.04$ \citep{2016MNRAS.459.2379D,2020ApJ...901...25D}. The eccentric discs can precess coherently \citep{2017MNRAS.466.1170M,2020MNRAS.499.3362R}, which is consistent with the existence of trapped eccentric modes in circumbinary discs \citep{2020ApJ...905..106M}.
In unequal-mass binaries, odd$-m$ eccentric Lindblad resonances (ELRs) may excite eccentricity \citep{2003ApJ...585.1024G}, although equal-mass binaries for which odd$-m$ modes vanish are still able to excite eccentricity in the disc through even$-m$ ELRs \citep[e.g.][]{1991ApJ...381..259L,2017MNRAS.466.1170M}. In three-dimensional magnetohydrodynamic simulations of accretion onto equal-mass binaries, \citep{2012ApJ...749..118S} found that the growth rate of eccentricity in the circumbinary disc was consistent with excitation be streams of gas being flung away from the binary and shocking against the cavity wall.
Herein we examine how circumbinary disc eccentricity and cavity size vary as functions of Mach number and viscosity. 

We begin by examining the evolution of the disc's eccentricity vector in time and space for a few example Mach numbers and viscosities in Figure \ref{fig:3ecc}. We measure disc eccentricity $(e)$ and argument of periapsis $(\omega)$ as described in Section \ref{sec:diag}. It is clear in all cases that eccentricity quickly develops throughout the circumbinary disc out to distances of a few times the binary semi-major axis after a few tens of orbits. Additionally, it becomes immediately apparent that the $\mathcal{M}=20$ disc is more eccentric than either $\mathcal{M}=10$ disc, although differences in eccentricity along with viscosity are not as clear. 

Examining the average argument of periapsis over time, it is evident that the disc precesses at a roughly constant rate throughout, at least where $\omega$ is well-defined ($e>0$). At smaller radii where the cavity is present $(r \lesssim3a_b)$, the average arguments of periapse are out of phase with the bulk of the disc. The precession rate is thought to be dominated by the strength of the binary quadrupole moment \citep[e.g.][]{2004ApJ...609.1065M} evaluated at a characteristic cavity radius \citep[e.g.][]{2008ApJ...672...83M,2020ApJ...905..106M}, and accordingly may be used as an approximate and indirect probe of the cavity size. Thus, we can infer from Figure \ref{fig:3ecc} that both the $\mathcal{M}=20$ and $\nu=0.002$ discs have smaller cavities than the $\mathcal{M}=10,~\nu=0.001$ circumbinary disc. 

\begin{figure}
\centering
\includegraphics[width=\linewidth]{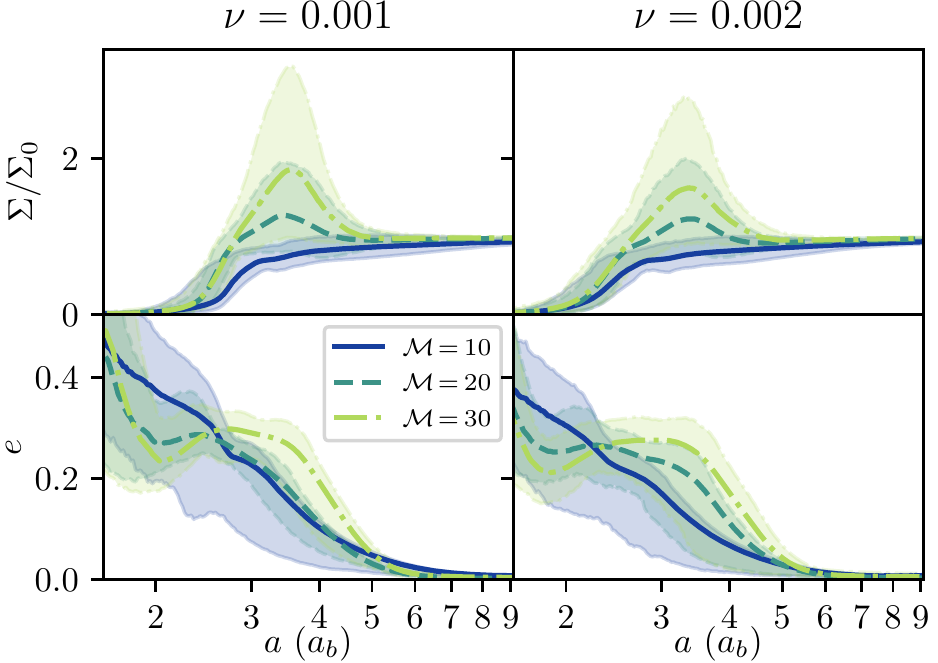}
\caption{Quantities related to the cavity as a function of fluid element semi-major axis, as a function of Mach number for $\nu=0.001$ and $\nu=0.002$. Top row: normalised surface density. Bottom row: fluid element eccentricity. 
Results for Mach 30 are shown using yellow-green dash-dotted lines, results for Mach 20 are shown using green dashed lines, and results for Mach 10 are shown using solid blue lines. Quantities have been averaged over the last 500 binary orbital periods of each simulation. Lines show the mean value over that period, while the shaded regions show the 90\% quantile symmetric about the median.}
\label{fig:aplots}
\end{figure}

To gain further insight, we investigate the disc surface density and eccentricity as functions of semi-major axis $a$: results for $\nu=\{0.001,0.002\}$, $\mathcal{M}=\{10,20,30\}$ are shown in figure \ref{fig:aplots}. We calculate $e$ and $a,$ for each fluid element, bin quantities according to their semi-major axis, and calculate mass-weighted averages in each semi-major axis bin. Averages were carried out over the last 500 orbits of each simulation, sampling once per binary orbit.

\begin{figure}
\centering
\includegraphics[width=\linewidth]{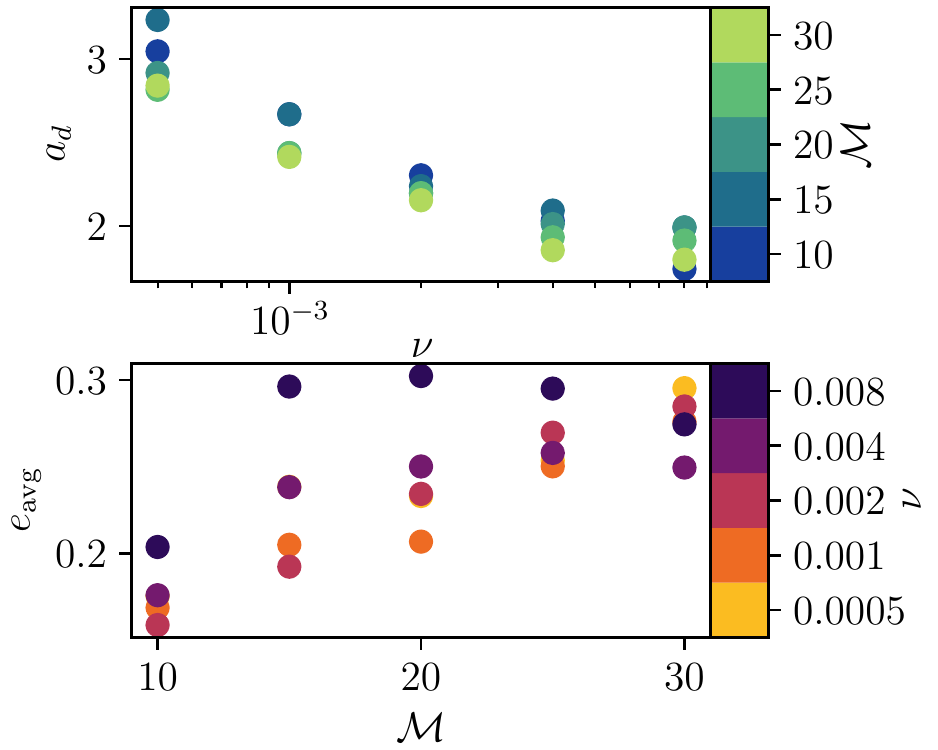}
\caption{Characteristic average circumbinary disc cavity semi-major axis $(a_d)$ and average disc eccentricity $e_{\rm avg}$ from simulations of $\nu=\{0.0005,0.001,0.002,0.004,0.008\}$ and $\mathcal{M}=\{10,15,20,25,30\}$ discs. The top panel plots $a_d$ as a function of $\nu,$ colouring points at each viscosity by their Mach number. The bottom panel plots $e_{\rm avg}$ as a function of Mach number, colouring points according to their viscosity. We note that in the bottom panel, the $\nu=0.0005$ and $\nu=0.004$ points overlap substantially at Mach 10 and Mach 15.}
\label{fig:orbitScatter}
\end{figure}

\begin{figure}
\centering
\includegraphics[width=\linewidth]{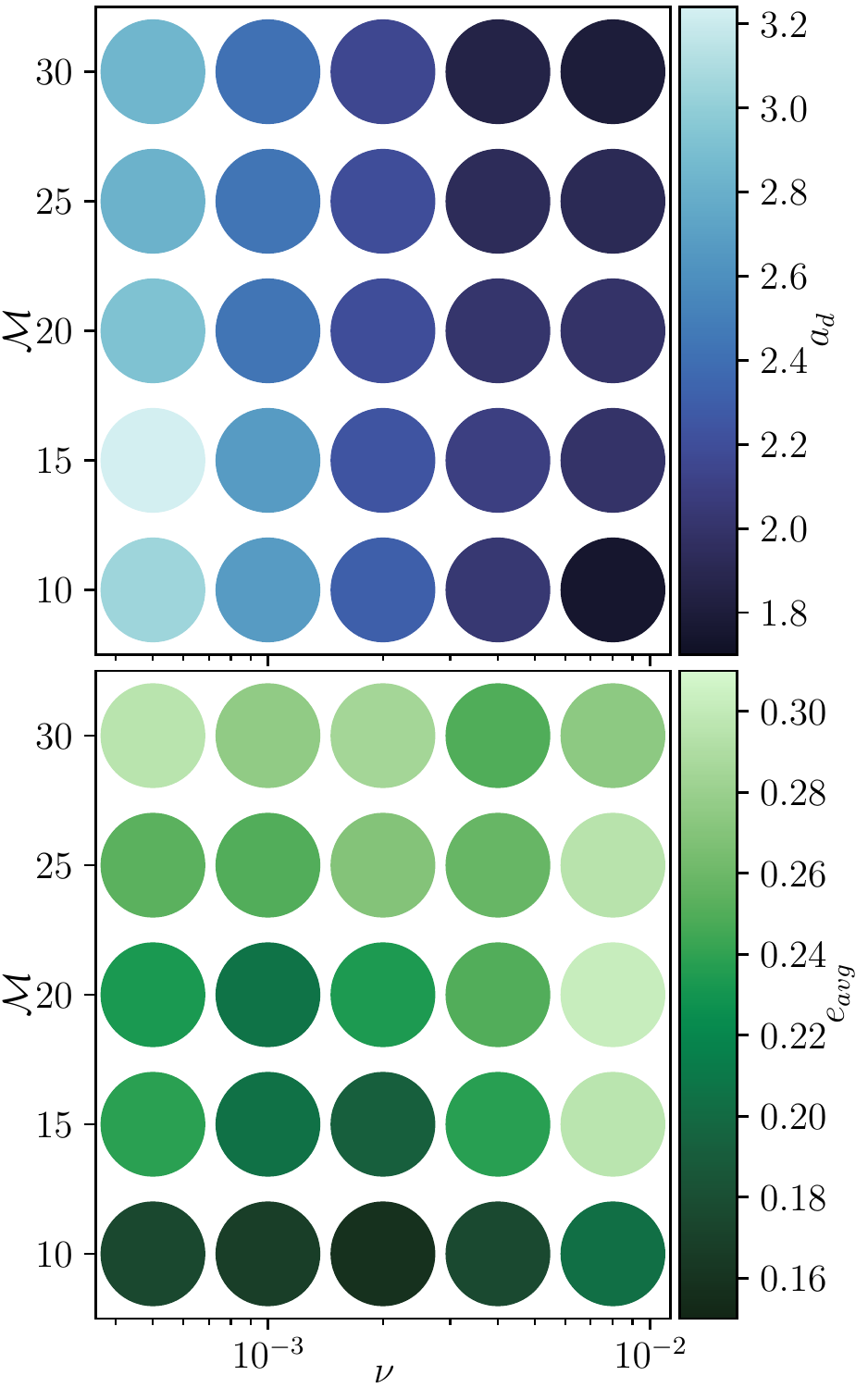}
\caption{Characteristic average circumbinary disc cavity semi-major axis ($a_d$, top panel) and average disc eccentricity ($e_{\rm avg}$, bottom panel) as functions of Mach number and viscosity from a subset of our simulations. Across Mach numbers, it is clear that $a_d$ tends to become smaller as viscosity increases. Similarly, $e_{\rm avg}$ tends to increase at higher Mach numbers, although there is scatter is both relations. There is some correlation between $a_d$ and $e_{\rm avg}$, which is at least in part due to fluid elements with smaller semi-major axes tending to have higher eccentricities.}
\label{fig:orbitDots}
\end{figure}

First, it is clear that at higher Mach numbers, typical surface densities at the cavity wall are much larger, the same trend which is shown in Figures \ref{fig:sigGrid} and \ref{fig:sigAvg}, and was also observed by \citet{2020ApJ...900...43T}. Additionally, in the higher-viscosity simulations, noticeable amounts of mass extend to lower semi-major axes. Out of the $\nu=0.001$ simulations, significant mass in the $\mathcal{M}=20$ simulation extends to lower semi-major axes than in the $\mathcal{M}=10$ simulation, confirming the difference in cavity size expected based on the rate of precession in the disc shown in Figure \ref{fig:3ecc}. Turning to eccentricity, it is clear that higher-Mach number discs tend to be more eccentric, at least where there is appreciable surface density $a \gtrsim 3 a_b$. We note that because plotting quantities at constant $a$ is equivalent to plotting them at constant specific orbital energy, and that the higher the eccentricity of a fluid element at a given semi-major axis, the lower its specific angular momentum, as $e=\sqrt{1-p/a}$ for Keplerian orbits, where $p=l^2/GM$ is the semilatus rectum, and $l$ is the specific angular momentum.

\begin{figure}
\centering
\includegraphics[width=\linewidth]{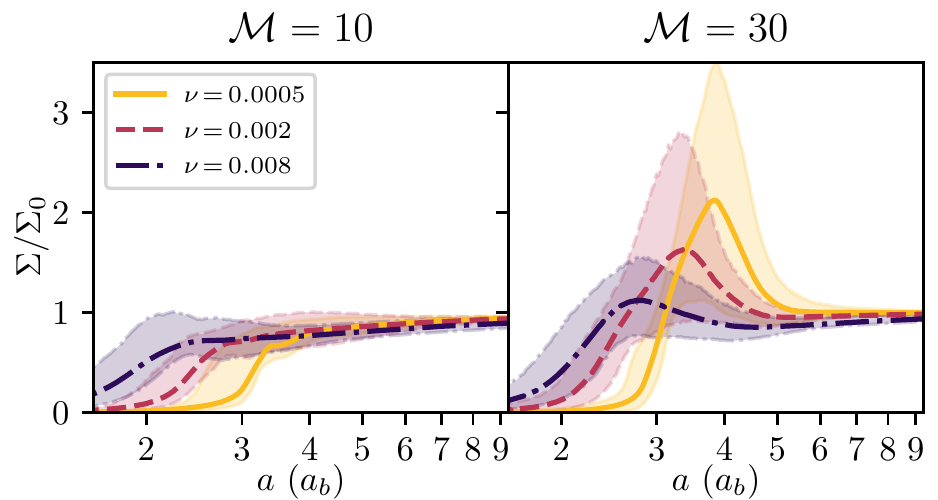}
\caption{Surface density as a function of binary semi-major axis for different viscosities, at $\mathcal{M}=\{10,30\}$, averaged over the last 500 binary orbital periods of each simulation, highlighting the dependence of cavity size on viscosity. Lines show the mean value, while the shaded regions show the 90\% quantile symmetric about the median. Solid yellow lines plot results for $\nu=0.0005$ simulations, dashed red lines plot results for $\nu=0.002$ simulations, and dash-dotted purple lines plot results for $\nu=0.008$ simulations.}
\label{fig:sigmaProfiles}
\end{figure}

With the surface density and eccentricity profiles presented in Figure \ref{fig:aplots} in mind, we turn to a larger set of simulations, at $\nu=\{0.0005,0.001,0.002,0.004,0.008\}$ and $\mathcal{M}=\{10,15,20,25,30\}$. For each, we calculate surface density and eccentricity profiles as in Figure \ref{fig:aplots}. We calculate a characteristic cavity semi-major axis $(a_d)$ by finding the point at which satisfies $a_d>1.5a_b$ and $\Sigma(a_d)=\Sigma_0/5$. We report average disc eccentricities $(e_{\rm avg}$), calculated by a mass-weighted average of the eccentricity from $a_d$ to $a_d+1.5a_b$. We have found that the precise values of $e_{\rm avg}$ depend on the choices made when averaging (e.g. including gas at larger semi-major axes naturally leads to lower average values) but find that the overall trends in $e_{\rm avg}$ as a function of $\nu$ and $\mathcal{M}$ are insensitive to our method of averaging. We present collections of these results in Figures \ref{fig:orbitScatter} and \ref{fig:orbitDots}.

We observe a strong correlation between disc eccentricity and Mach number, which is illustrated in Figure \ref{fig:orbitScatter}. We also observe correlations between the disc eccentricity and viscosity, which are likely due to the tendency of fluid elements to have higher eccentricity at smaller semi-major axes, which can be seen in Figure \ref{fig:aplots}, and the anticorrelation between viscosity and cavity size also illustrated in Figure \ref{fig:orbitScatter}. The anticorrelation between viscosity and cavity size follows from the reasoning that cavities are opened when gravitational torques are able to overcome viscous torques. We discuss potential reasons for the correlation between $e_{\rm avg}$ and $\mathcal{M}$ in Section \ref{sec:eccEv}.
We find no significant correlations between $e_{\rm avg}$ and $a_d$. This may seem contrary to the results presented by \citet{2020MNRAS.499.3362R}, which found a correlation between disc eccentricity and cavity size. However, \citet{2020MNRAS.499.3362R} primarily varied the binary mass ratio, where smaller mass ratios naturally lead to lower disc eccentricity and smaller cavity sizes if those quantities are controlled by the time-variable potential of the binary.

\subsection{Eccentricity evolution}\label{sec:eccEv}

Although Figure \ref{fig:3ecc} demonstrates the precession of the disc and the development of eccentricity throughout the disc, it still leaves the question of how eccentricity is excited and why higher Mach number discs are more eccentric. Furthermore, Figure \ref{fig:3ecc} illustrates that the $\mathcal{M}=20,~\nu=0.001$ disc precesses more quickly than the $\mathcal{M}=10,~\nu=0.002$ disc, while Figure \ref{fig:orbitScatter} shows that the latter disc has a smaller cavity. To gain further insight into these observations, we quantitatively analyse the disc precession rate, and investigate the process of eccentricity excitation. 

\begin{figure}
\centering
\includegraphics[width=\linewidth]{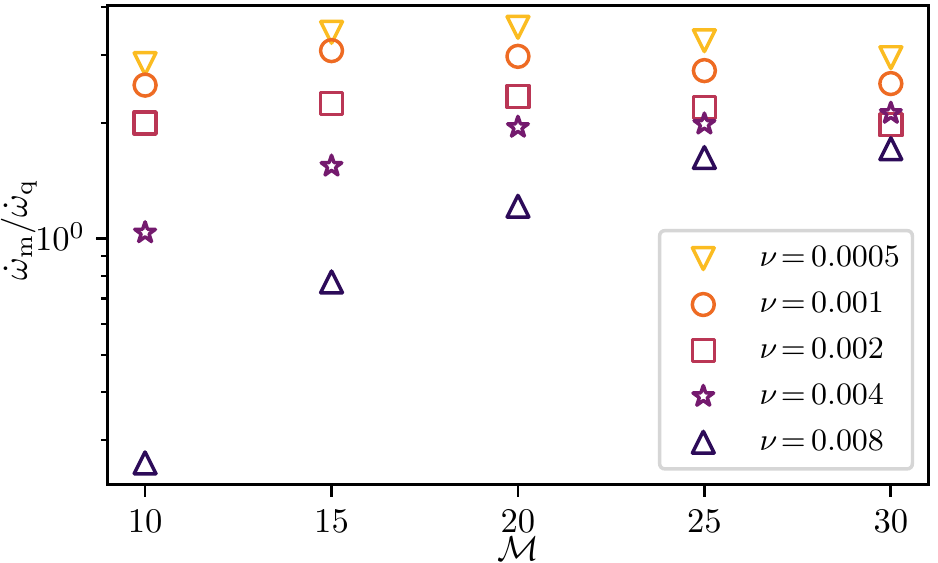}
\caption{The ratio of the measured precession rate $\dot{\omega}_m$ to the leading-order estimate for a quadrupole potential $\dot{\omega}_q$ for different Mach numbers and viscosities. Symbol shape and colour indicate viscosity, with lighter colours at low-$\nu$ and darker colours at high-$\nu$.}
\label{fig:preccomp}
\end{figure}

The leading order precession rate $\dot{\omega}$ of an orbit of semi-major axis $a$ and eccentricity $0\leq e<1$ due to the quadrupole potential of an equal-mass binary with semi-major axis $a_b$ and orbital frequency $\Omega_b$ is 
\begin{equation}\label{eq:preclin}
\dot{\omega}(a,e) = \frac{3}{16}\left(\frac{a_b}{a}\right)^{7/2}\left(1-e^2\right)^{-2}\Omega_b.
\end{equation}
As a prediction from linear theory, we calculate quadrupole precession rates $\dot{\omega}_q(a_{\rm d},e_{\rm avg})$ using our measurements of cavity semi-major axis and disc eccentricity. We measure the precession rate in our simulations using the time-variation of the mass-weighted average x and y components of the eccentricity vector, binned in semi-major axis as described in Section \ref{sec:diskOrb}. After averaging components of the eccentricity vector over 5 orbital periods to remove high-frequency variation on the orbital period of the cavity, we perform nonlinear least-squares fits of sine functions to the eccentricity time series for each semi-major axis bin from $a_d$ to $a_d+1.5a_b$, the range used to compute $e_{\rm avg}$. We then take the median frequency of the best-fitting models for each bin and eccentricity component as our measured precision rate $\dot{\omega}_m$, although typically variations between bins are less than one per cent. The precession rates measured in our simulations are compared with the expectation from the linear quadrupole contribution in Figure \ref{fig:preccomp}.

\begin{figure}
\centering
\includegraphics[width=\linewidth]{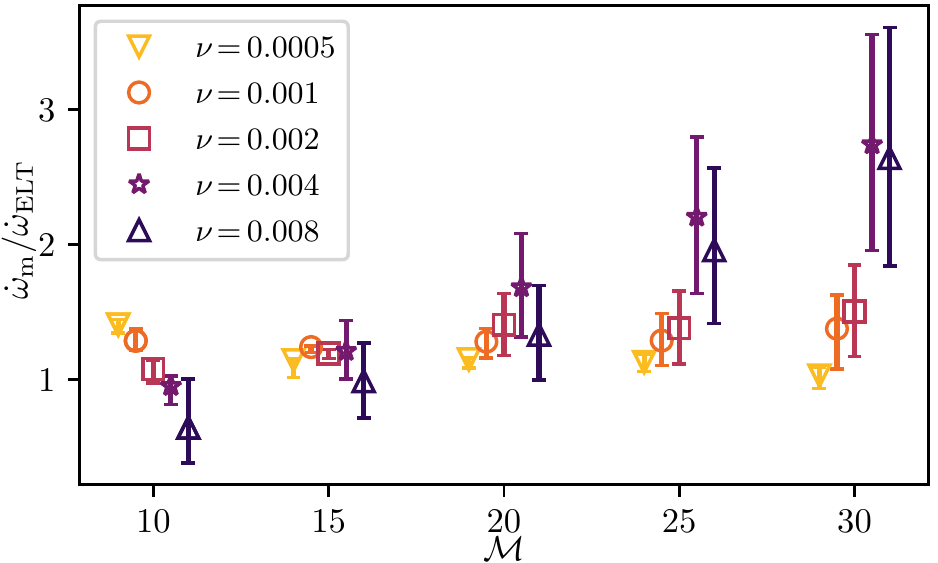}
\caption{The ratio of the measured precession rate $\dot{\omega}_m$ to a prediction based on the linear theory of eccentric discs $\dot{\omega}_{\rm ELT}$ \citep[e.g.][]{2006MNRAS.368.1123G,2016MNRAS.458.3221T} for different Mach numbers and viscosities, where a small horizontal offset has been added for visibility. Symbols indicate the median value for the range up upper and lower limits when calculating averages of $\dot{\omega}_{\rm ELT}$ while the upper and lower bars correspond to the 75th and 25th percentiles. Symbol shape and colour indicate viscosity, with lighter colours at low-$\nu$ and darker colours at high-$\nu$.}
\label{fig:precEccTheory}
\end{figure}

\begin{figure*}
\centering
\includegraphics[width=\linewidth]{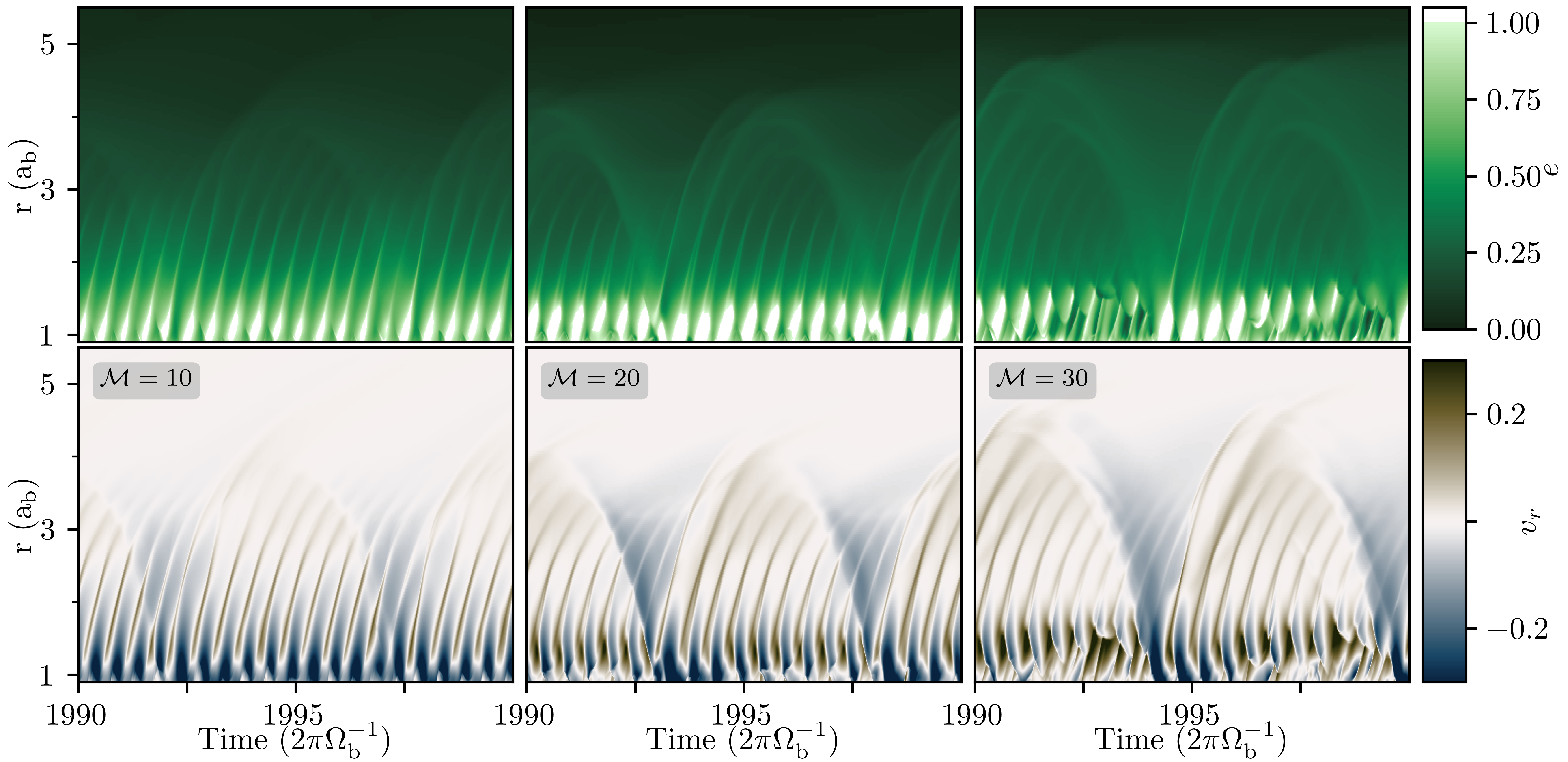}
\caption{Space-time plots of the azimuthally averaged and mass-weighted eccentricity (top row) and radial velocity (bottom row) for $\nu=0.001$ simulations at $\mathcal{M}=\{10,20,30\}$ (\{left, center, right\} columns). The color scale for the eccentricity panels is such that radii at which the average fluid element is on a hyperbolic or parabolic trajectory $(e\geq1)$ are coloured white.}
\label{fig:eccvr}
\end{figure*}

Overall, apart from discs with high viscosities $(\nu\gtrsim0.004)$ and low Mach numbers $(\mathcal{M}\lesssim20)$, the circumbinary disc tends to precess more quickly than predicted by the leading-order quadrupole estimate, by factors ranging from $\sim 1.5$ to $\sim 3$, although the measurements and quadrupole estimate could be brought into agreement by a mild redefinition of $a_d$. For lower viscosities, the ratio $\dot{\omega}_q/\dot{\omega}_m$ is roughly constant, suggesting that the precession rate still scales with eccentricity and semi-major axis with the same proportionality as in Equation (\ref{eq:preclin}). On average, higher-viscosity discs precess more slowly at constant Mach number, suggesting that viscosity plays a key role in reducing the rate of precession, which is neglected in Equation (\ref{eq:preclin}). Our results hold some indication that the precession rate decreases in thicker discs, although this is tenuous at best for $\nu\lesssim0.002$. 

We note that a previous investigation of the eccentric eigenfrequency of idealised discs found that thicker discs precess more slowly than thinner discs \citep[e.g. Figure 2 of][]{2020ApJ...905..106M}, approaching the quadrupole rate for $\mathcal{M}\gtrsim100$. It should be noted that Figures 2 and 3 of \citet{2020ApJ...905..106M} assumed a constant $\dot{J}/\dot{M}$ as a function of Mach number: the fact that we observe a similar trend only at higher viscosities may be a result of the weak dependence of $\dot{J}/\dot{M}$ on $\mathcal{M}$ at $\nu\gtrsim0.004$ (c.f. Figure \ref{fig:qdlogs} and equation \ref{eq:orbJg}), whereas in general $\dot{J}/\dot{M}$ depends strongly on Mach number. Other deviations from the predictions of \citet{2020ApJ...905..106M} may be due to their axisymmetric disc model, an increasingly poor assumption for higher Mach number discs, or assumption of an $\alpha$ viscosity (rather than our constant-$\nu$ model) when modelling surface density profiles. These differences may contribute to differences in $\dot{\omega}_m/\dot{\omega}_q$ between our study and \citet{2020ApJ...905..106M}: while we typically observe faster disc precession than the linear quadrupole estimate, \citet{2020ApJ...905..106M} found that the discs in their hydrodynamic simulations typically precessed more slowly than the linear quadrupole estimate.\footnote{\citet{2020ApJ...905..106M} did not directly report $\dot{\omega}_m/\dot{\omega}_q$, but $\dot{\omega}_m/\dot{\omega}_q<1$ can be inferred from their Figures 3 and 10 combined with the text of their Section 3.3.}
Our simulations were not run for long enough durations to identify any of the lower frequency harmonics predicted by \citet{2020ApJ...905..106M}.

More thorough predictions for the rate of precession of an eccentric disc explicitly take into account additional effects, such as pressure gradients and non-adiabatic effects related to locally-isothermal equations of state. Assuming that the eccentricity of a disc can be described by $E(r,t)=e(r)\exp{(it\dot{\omega}_{\rm ELT})}$, with magnitude as a function of radius $e(r)$ and uniform precession rate $\dot{\omega}_{\rm ELT}$, the evolution of the disc eccentricity can be described in linear theory by \citep{2006MNRAS.368.1123G,2016MNRAS.458.3221T}
\begin{equation}\label{eq:eccevol}
-i\Sigma r^2\Omega \frac{\partial E}{\partial t}=\frac{1}{r}\frac{\partial}{\partial r}\left(F\frac{\partial}{\partial r}\left(\frac{E}{c_s^2}\right)\right)+GE,
\end{equation}
where $F = \Sigma c_s^4 r^3/2,$
\begin{equation}
G = \frac{r}{2}\frac{\partial}{\partial r}\left(\Sigma c_s^2\right)+\Sigma\frac{\Omega^2 r^4}{a_b^2}b^{(1)}_{3/2}(2r/a_b),
\end{equation}
and 
\begin{equation}
b_s^{(j)}=\frac{1}{\pi}\int_0^{2\pi}\frac{\cos{(j\psi)d\psi}}{(1+\alpha^2-2\alpha\cos{(\psi)})^s}.
\end{equation}

The innermost radii of the circumbinary disk appear to precess coherently (e.g. Figure \ref{fig:3ecc}). In this region $\dot{\omega}_{\rm ELT}$ will be near constant with $r$ and we can derive an approximate expression:
\begin{equation}\label{eq:precelt}
\dot{\omega}_{\rm ELT}(r)=\frac{1}{r^3\Sigma\Omega e}\left[ \frac{\partial }{\partial r}\left(F\frac{\partial}{\partial r}\left(\frac{e}{c_s^2}\right)\right)+reG\right].
\end{equation}

To calculate these values of $\dot{\omega}_{\rm ELT}$, we have used time-averaged azimuthally integrated disc profiles, averaged once per orbit over the final five hundred orbits of each simulation. We calculate values of $\dot{\omega}_{\rm ELT}$ averaged over a radial range $r_{\rm min}$ to $r_{\rm max}$, where we survey 20 values of $r_{\rm min}$ from $a_d(1-e_{\rm avg})$, the approximate pericentre of the circumbinary disc, to $a_d(1-e_{\rm avg})+a_b$, and 20 values of $r_{\rm max}$ from $6a_b$ to $8a_b$, and report the 25th, 50th, and 75th percentiles of the resulting $\dot{\omega}_{\rm ELT}$ measurements compared with the measured precession frequencies $\dot{\omega}_m$ from our simulation in Figure \ref{fig:precEccTheory}. The predicted and measured values of the precession rate are almost always in very close agreement. Measured and predicted precession rates disagree most severely for simulations with both high Mach numbers and high viscosities, which may be be because the perturbative analysis in \citet{2006MNRAS.368.1123G,2016MNRAS.458.3221T} assumes that the disc viscosity and aspect ratio are both small quantities of similar order, which breaks down in the high-viscosity high-Mach number regime. Additionally, we find more mild disagreement between measured precession rates and the linear theory prediction at $\mathcal{M}=10$, where the thin disc approximation is weakest.

To investigate the excitation of eccentricity in the disc, we examine the azimuthally-averaged mass-weighted eccentricity and radial velocity in Figure \ref{fig:eccvr}, where we have recorded samples three hundred times per orbit over the final ten orbits of three $\nu=0.001$ simulations. First, we see from the shading in of the $e(r,t)$ plot that the $\mathcal{M}=\{20,30\}$ discs have a slightly higher average eccentricity than the $\mathcal{M}=10$ disc, confirming our expectations based on Figures \ref{fig:3ecc}, \ref{fig:orbitDots}, and \ref{fig:aplots}.

On timescales shorter than disc precession period, the two key timescales are the orbital period of the binary and the orbital period of the disc cavity, both of which can be identified in Figure \ref{fig:eccvr}. Specifically, the `streaks' through both the $e(r,t)$ and $v_r(r,t)$ diagrams occur roughly $\sim20$ times, which corresponds to the number of times one of the binary members has a close approach with the periapse of the circumbinary disc. An `envelope' can also be seen on a period of $\sim 4-5$ binary orbits, the orbital period of the disc cavity walls. The accretion of material from the `lump' onto the binary once per cavity orbital period is particularly clear in the plots of $v_r(r,t)$, which show that the accretion rate onto the binary temporarily increases dramatically every $\sim4-5$ binary orbits. 

Strong correlations between $v_r(r,t)$ and $e(r_t)$ are apparent in Figure \ref{fig:eccvr} within the cavity. Specifically, eccentricity in the disc is strongly correlated with \textit{positive} velocities - the streams of material travelling away from the binary. At larger radii, we see that eccentricity remains excited in the disc even after the net positive velocity vanishes. This suggests that eccentricity is primarily excited by shocks as high-eccentricity streams of material strike the cavity walls, as suggested by the 2D hydrodynamic simulation presented in \citet{2008ApJ...672...83M} the 3D magnetohydrodynamic simulations presented in \citet{2012ApJ...749..118S}, for example, rather than coherent eccentric resonances \citep[e.g.][]{1991ApJ...381..259L,2004AJ....128.1418P,2006ApJ...652.1698D}, although the latter may play a significant role in initially seeding eccentricity in the disc. In the case of non-resonant eccentricity excitation due to fluid elements which receive kicks upon pericentre passage shocking against the disc at apocentre, the role of pressure is clear as stronger shocks occur in colder discs. We note that in this way, the disc eccentricities observed in our simulations may be larger than in realistic (non-isothermal) discs at the similar average aspect ratios, which would heat up during shocks, limiting shock strengths. 

\section{Binary supermassive black holes}\label{sec:SMBHB}
Although the accretion discs around binary supermassive black holes cannot be resolved in images, circumbinary accretion may leave an observable imprint in both the lightcurves and gravitational wave signals of accreting SMBH binaries.

\subsection{Disc-driven inspirals}
Our results, shown in Figures \ref{fig:qdlogs} and \ref{fig:spiralGrid} suggest that the evolution of binary SMBHs can be significantly modified by circumbinary accretion. We begin by revisiting the competition between gas- and GW-driven binary coalescence. One common model for circumbinary evolution \citep{2009ApJ...700.1952H} follows from \citet{1999MNRAS.307...79I}, suggesting that for equal-mass binaries the gravitational wave frequency at which the gas and GW contributions to $da_b/dt$ are equal is $\sim3\times10^{-8}~\rm{Hz}$ for $10^7~\rm M_\odot$ binaries, or $\sim9\times10^{-10}~\rm{Hz}$ for $10^9~\rm M_\odot$ binaries \citep{2013CQGra..30x4009S}.

For simplicity, we denote the value of $d\log{a_b}/d\log{M}$ measured from our simulations as $-\xi$, such that the rate of binary semi-major axis evolution due to interaction with the circumbinary disc is
\begin{equation}\label{eq:daCB}
\left.\frac{da_b}{dt}\right|_{\rm CB}=-\xi a_b \frac{\dot{M}}{M}.
\end{equation}
The rate of semi-major axis evolution due to gravitational radiation for a circular equal-mass binary is \citep{1964PhRv..136.1224P}
\begin{equation}\label{eq:daGW}
\left.\frac{da_b}{dt}\right|_{\rm GW}=-\frac{16}{5}\frac{G^3M^3}{c^5a_b^3}.
\end{equation}
We parameterise the accretion rate onto the binary in terms of the Eddington-limited accretion rate, such that $\dot{M}/M=4\eta\pi G/\kappa c \epsilon_\bullet$, where $\eta$ is the ratio of the Eddington-limited rate at which the binary accretes, $\epsilon_\bullet$ is the fraction of rest mass radiated during accretion onto the black holes, and $\kappa$ is the gas opacity, for which we assume the electron-scattering value $\kappa\sim0.4~\rm{cm^2~g^{-1}}$.

Setting Equations (\ref{eq:daCB}) and (\ref{eq:daGW}) equal, and expressing the result in terms of the gravitational wave frequency $f_{\rm GW}=\sqrt{GM/a_b^3}/\pi$, we find that the frequency at which equality occurs is 
\begin{equation}
f_{eq}=2.1\times10^{-9}\xi^{3/8}\eta^{3/8}\left(\frac{\epsilon_\bullet}{0.1}\right)^{-3/8}\left(\frac{M}{10^9~\rm{M_\odot}}\right)^{-5/8}~\rm Hz.
\end{equation}
As an example, we consider the case of Eddington-limited accretion with $\eta=1$. Taking values of $\xi$ from our simulations, e.g. $\xi\sim15$, we find that for $10^9~\rm M_\odot$ binaries gravitational waves begin dominating the inspiral process near $\sim5.8\times10^{-9}$ Hz, while for $10^7~\rm M_\odot$ binaries gravitational waves dominate the inspiral near $\sim10^{-7}$ Hz. However, we only carried out simulations up to $\mathcal{M}=30$, whereas AGN are expected to have accretion discs with $\mathcal{M}\sim10^2-10^3$ \citep[e.g.][]{1999agnc.book.....K, 2001ApJ...559..680H}. 

Based on extrapolation of our results shown in Figure \ref{fig:qdlogs}, we expect much larger values of $\xi$ in realistic binary AGN circumbinary discs. For example, fixing $\nu=0.001$ and linearly extrapolating $\xi$ using the slope between our $\mathcal{M}=25$ and $\mathcal{M}=30$ results, we would expect $\xi\sim50$ at $\mathcal{M}\sim10^2$, and $\xi\sim600$ at $\mathcal{M}\sim10^3$.
Similarly, if we assume $\dot{J}_g/\dot{M}\propto \nu^{-1}\propto\mathcal{M}^2$, following the $\nu\propto\mathcal{M}^{-2}$ $\alpha-$viscosity scaling, and extrapolate the value of $\xi$ measured at $\nu=0.001,$ $\mathcal{M}=30$, we would expect $\xi\sim700$ at $\mathcal{M}=300$, even ignoring the observed increase of $\xi$ with $\mathcal{M}$ at constant $\nu$.  Because these scaling arguments are not guaranteed to hold at regimes of such high Mach number and low viscosity, or in the case of magnetohydrodynamic turbulence rather than Navier-Stokes viscosity, we take lower values of $\xi$ an example.

Using $\xi=250$, we expect gravitational waves to begin dominating the inpspiral at $\sim1.7\times10^{-8}$ Hz for $10^9~\rm M_\odot$ binaries, or at $\sim3\times10^{-7}$ Hz for $10^7~\rm M_\odot$ binaries. Thus, compared to earlier studies \citep[e.g.][]{2013CQGra..30x4009S,2018MNRAS.477..964K}, we expect interaction with circumbinary discs to dominate the evolution of binary SMBH separations until higher frequencies later in the inspiral, extending the range of frequencies over which interactions with circumbinary discs suppress the stochastic gravitational wave background \citep{2011MNRAS.411.1467K}. The difference is larger for more massive black holes, e.g. a factor of ~20 in frequency for $10^9~\rm M_\odot$ SMBHs, which are expected to dominate the signal in Pulsar Timing arrays \citep{2013CQGra..30x4009S}. Interactions with circumbinary discs can also remain the dominant driver of inspirals for binary candidates in optical surveys, with estimated orbital periods on the order of years \citep[e.g.][]{2015MNRAS.453.1562G,2020MNRAS.499.2245C}.

Considering SMBH binaries with mass $10^9~\rm M_\odot$, the inspirals of which become dominated by gravitational waves at $a_b\sim0.024(0.012)$ pc for $\xi=15(250)$, using the above figures above which assumed Eddington-limited accretion. Assuming circumbinary discs become the dominant driver of binary orbital evolution at $a_0\sim2$ pc, the semi-major axis must decrease by a factors of $\sim 83.6(172)$. Then, using $(a_b/a_0)^{-1/\xi}=(M/M_0),$ where $M_0$ is the initial binary mass, by the time the inspiral becomes dominated by gravitational waves $M\approx1.34(1.02)M_0$. As we have assumed Eddington-limited accretion, the related $e$-folding timescale for accretion \citep{1964ApJ...140..796S} implies timescales of $\sim1.3\times10^7$ years for $\xi=15$ or $\sim9\times10^5$ years for $\xi=250$ for the inspiral to become dominated by gravitational waves. 

Observations indicate that individual AGN accretion episodes may be as short as $\sim10^5$ years \citep{2015MNRAS.451.2517S,10.1093/mnrasl/slv098}, although galactic nuclei are expected to be active for total durations on the order of $\sim10^8$ years \citep[e.g.][]{2004cbhg.symp..169M}. Using the estimates of $\xi$ considered thus far, binaries could be driven to the GW-dominated regime over the course of a few $\sim10^5-$year accretion episodes, or possibly in just a single episode for larger values of $\xi$. However, our analysis is limited because in general SMBH binaries are not generally expected to be equal-mass, initially in circular orbits, and aligned with the larger-scale accretion flow, so our results for SMBH binary evolution must be viewed with caution. Additionally, our simulations assumed a fixed binary orbit, whereas for sufficiently large $\xi$ the orbit of the binary may evolve on hundreds to thousands of dynamical timescales, comparable to our simulation duration. In such cases, the evolution may be limited by the viscous spreading of the disc, and proceed with virtually no accretion similar to early models of circumbinary evolution \citep{1991MNRAS.248..754P,1991ApJ...370L..35A}.

\subsection{Variability}

Because the accretion rate onto the binary is heavily modulated at the orbital period of the inner edge of the circumbinary disc, the time series of accretion onto the binary is strongly affected by the structure of the circumbinary disc. Although simulations of 2D viscous isothermal hydrodynamics can only probe the observable characteristics of these systems to a limited extent, the accretion rate onto the binary can serve as a proxy for luminosity variability  (\citet{2014ApJ...783..134F,2015MNRAS.446L..36F,2018MNRAS.476.2249T}, but see also \citet{2016ApJ...832...22S}). 
We note that general relativistic magnetohydrodynamic simulations of black hole binaries approaching merger have found that the variability of jet Poynting fluxes matches that of the accretion rate \citep{2021arXiv210901307C}. 
Emission from the disc as a whole is less correlated with the instantaneous accretion rate onto the binary, although the variability shows statistical similarities \citep[e.g.][]{2012ApJ...755...51N,2021arXiv210312100N,2021arXiv211209773G}.
In binaries with larger mass ratios $q\gtrsim0.2$, accretion primarily occurs on timescales of $\sim4-5$ times that of the binary orbits, which is linked to the formation of the high-density clump near the inner edge of the circumbinary disc (see Figure \ref{fig:sigGrid}), which then feeds the binary at the clump's periapse \citep[e.g.][]{2008ApJ...672...83M,2012ApJ...749..118S,2012ApJ...755...51N,2013MNRAS.436.2997D,2015ApJ...807..131S,2017MNRAS.466.1170M, 2019ApJ...879...76B,2020ApJ...889..114M,2021ApJ...921...71D,2021arXiv210312100N}.

To characterise accretion rate variability, we begin by integrating the accretion rate onto the sink particles in our simulations down to a cadence of $\sim50$ samples per orbit. We searched for variability on time scales from 0.1 binary orbits to 200 orbital periods, using data from the final 500 orbits of each simulation. We quantified the strength of variability at a given frequency by constructing Lomb-Scargle periodograms \citep{{1976Ap&SS..39..447L},{1982ApJ...263..835S},{2010ApJS..191..247T}} after rescaling the accretion rate time series to have a mean of zero and variance of unity.

\begin{figure}
\centering
\includegraphics[width=\linewidth]{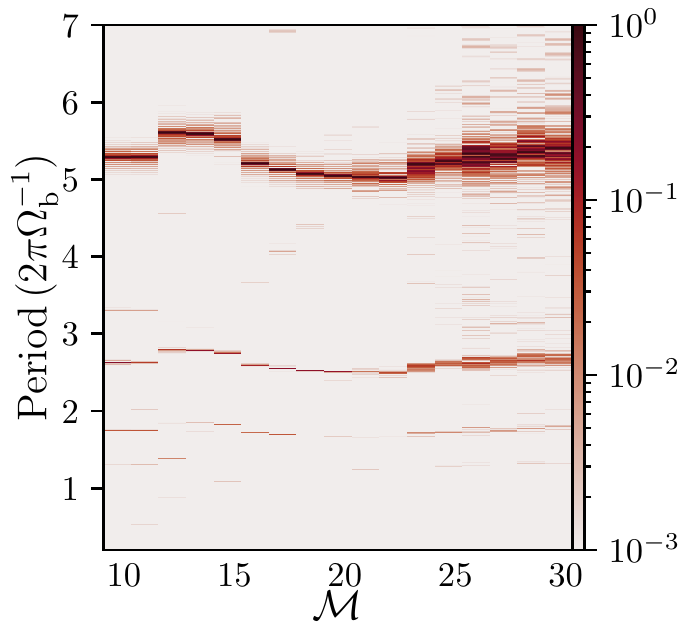}
\caption{Periodograms of the accretion rate onto the binary in a series of simulations at $\nu=0.001$ over a range of Mach numbers. The colour scale, indicating periodogram amplitudes, is normalised by the largest amplitude measured in each simulation.}
\label{fig:machVar}
\end{figure}

We examine variability as a function of Mach number at constant viscosity $(\nu=0.001)$ in Figure \ref{fig:machVar}, sampling Mach numbers between 10 and 30 in increments of 1.25. The largest variability amplitudes are typically occur at the orbital period of the cavity, roughly $\sim5$ times the binary orbital period, as well as the next-highest associated harmonic. Although the dominant variability period changes from one Mach number to another, it does not change monotonically, and period at which the peak amplitude occurs at $\mathcal{M}=25$ is almost the same as that at $\mathcal{M}=10$. As expected from the shorter precession period at $\mathcal{M}=20$ than at $\mathcal{M}=10$ seen in Figure \ref{fig:3ecc} and smaller cavity sizes shown in Figure \ref{fig:aplots}, we see that the dominant variability period is shorter at $\mathcal{M}=20$ than at $\mathcal{M}=10$ as well. The more chaotic nature of the accretion shown in Figure \ref{fig:timeseries} at high Mach numbers is visible in the periodograms, as peaks become scattered and more broad. It has been recently suggested, based on analysis of light curves in non-isothermal circumbinary disc simulations, that the chaotic nature of the accretion rate at realistically high Mach numbers for binary SMBHs may drown out any periodicity other than Doppler modulation \citep{2021arXiv211106882W}.

The lack of a correlation between the dominant periodicity of accretion and Mach number observed in Figure \ref{fig:machVar} also reaffirms the result shown in Figure \ref{fig:orbitScatter} that the cavity size does not vary secularly with Mach number at constant kinematic viscosity.\footnote{When using an $\alpha$-viscosity, secular increases in the dominant period of variability would be expected at higher Mach numbers due to the larger cavities following from $\nu\propto\mathcal{M}^{-2}$.} Based on these (albeit locally-isothermal 2D) simulations, it appears that the picture of accretion rate periodicity gleaned from lower Mach number simulations may extend to higher Mach numbers without modifications too severe, although a more realistic equation of state, probing more directly the disk luminosity, may change this picture. 

\section{Conclusions}\label{sec:conclude}
We have conducted a series of simulations of circumbinary accretion, surveying both viscosity and disc aspect ratio, and assessed their implications for the orbital evolution of binaries and their circumbinary discs. Although our simulations were limited in the sense of being vertically integrated, isothermal, and viscous rather than magnetohydrodynamic, our results do not depend on our parameterization of viscosity (constant$-\nu$ vs. constant$-\alpha$), and only weakly depend on our choice of locally- vs. globally-isothermal equations of state. 

At both higher Mach numbers and lower viscosities, binaries inspiral increasingly rapidly due to strong negative torques on the binary from trailing accretion streams. The dominant gravitational torques arise from interaction with streams of material in disc cavity, rather than resonances in the circumbinary disc. We find that no SMBH `stalling' should occur at large separations due to interactions with circumbinary discs as suggested in
\citep{2020ApJ...889..114M}, as AGN discs are expected to have high Mach numbers. For example, in our $\mathcal{M}=30$, $\nu=0.001$ simulation, over a single mass-doubling timescale for the binary, the binary semi-major axis would shrink by a factor of $\sim64$. At higher Mach numbers, even a single AGN accretion episode may be sufficient to drive SMBH binaries into the GW-driven regime, in a small fraction of a Salpeter time. Furthermore, we have shown that interactions with the circumbinary disc may dominate the orbital evolution of $\sim10^9\,M_\odot$ SMBH binaries up to gravitational wave frequencies of $>10^{-8}$ Hz, potentially limiting the number of systems which can contribute to the stochastic gravitational wave background as probed by pulsar timing arrays.

We have also shown that when the angular momentum current through the disc is sufficiently negative, simulation initial conditions specifying $\dot{J}/\dot{M}=0$ can lead to the spurious impression that accretion is suppressed at high Mach number, but this is remedied when simulations are initialised with a value of $\dot{J}/\dot{M}$ closer to the eventual steady-state $\dot{J}/\dot{M}$ of the disc. Thus, our analysis suggests that the accretion rate reductions which we observe are due to initial simulation conditions which deviate strongly from the eventual quasi-steady state. Such deviations become more common at high Mach numbers, but are not universal features of high-Mach number circumbinary accretion \citep[c.f.][]{2016MNRAS.460.1243R}.

We have found that at smaller disc aspect ratios, the circumbinary disc is driven to higher eccentricities. We have also found that as viscosity decreases, the size of the cavity increases. The latter follows naturally from the competition between gravitational and viscous torques, while the former is due to the excitation of eccentricity by spiral shocks which naturally become stronger in colder discs.  
It will be important in future studies of high-Mach number discs to investigate circumbinary disc eccentricity as a function of disc aspect ratio using more realistic thermodynamics, especially due to the role of shocks. Additionally, it will be important to study higher Mach number discs in a magnetohydrodynamic framework because of the more rich phenomenology of the MRI in eccentric discs \citep{2018ApJ...856...12C}, and it will be important to assess the extent to which our viscous models are applicable. 

\section*{Acknowledgements}
We are grateful to Paul Duffell, Zoltan Haiman, and Cole Miller for their comments and feedback on an earlier draft of this work.  We thank Chris Tiede for providing the data used to construct Figure \ref{fig:comparison}. We thank the anonymous referee for their insightful comments and suggestions.

The simulations presented in this paper were conducted in part on the Rusty Cluster at the Flatiron Institute. The authors acknowledge the University of Maryland supercomputing resources (http://hpcc.umd.edu) that were made available for conducting the research reported in this paper, and the YORP cluster administered by the Center for Theory and Computation within the University of Maryland Department of Astronomy. AJD is supported by NASA ADAP grant 80NSSC21K0649.  Research at Perimeter Institute is supported in part by the Government of Canada through the Department of Innovation, Science and Economic Development and by the Province of Ontario through the Ministry of Colleges and Universities.

\section*{Data Availability}
\Disco{} can be found on GitHub.\footnote{e.g. \href{https://github.com/NYU-CAL/Disco}{https://github.com/NYU-CAL/Disco}} The data generated in this work will be made available upon reasonable request to the corresponding author. 
\bibliographystyle{mnras}
\bibliography{references.bib}

\begin{thebibliography}{}
\makeatletter
\relax
\def\mn@urlcharsother{\let\do\@makeother \do\$\do\&\do\#\do\^\do\_\do\%\do\~}
\def\mn@doi{\begingroup\mn@urlcharsother \@ifnextchar [ {\mn@doi@}
  {\mn@doi@[]}}
\def\mn@doi@[#1]#2{\def\@tempa{#1}\ifx\@tempa\@empty \href
  {http://dx.doi.org/#2} {doi:#2}\else \href {http://dx.doi.org/#2} {#1}\fi
  \endgroup}
\def\mn@eprint#1#2{\mn@eprint@#1:#2::\@nil}
\def\mn@eprint@arXiv#1{\href {http://arxiv.org/abs/#1} {{\tt arXiv:#1}}}
\def\mn@eprint@dblp#1{\href {http://dblp.uni-trier.de/rec/bibtex/#1.xml}
  {dblp:#1}}
\def\mn@eprint@#1:#2:#3:#4\@nil{\def\@tempa {#1}\def\@tempb {#2}\def\@tempc
  {#3}\ifx \@tempc \@empty \let \@tempc \@tempb \let \@tempb \@tempa \fi \ifx
  \@tempb \@empty \def\@tempb {arXiv}\fi \@ifundefined
  {mn@eprint@\@tempb}{\@tempb:\@tempc}{\expandafter \expandafter \csname
  mn@eprint@\@tempb\endcsname \expandafter{\@tempc}}}

\bibitem[\protect\citeauthoryear{{Andrews} et~al.,}{{Andrews}
  et~al.}{2014}]{2014ApJ...787..148A}
{Andrews} S.~M.,  et~al., 2014, \mn@doi [\apj] {10.1088/0004-637X/787/2/148},
  \href {https://ui.adsabs.harvard.edu/abs/2014ApJ...787..148A} {787, 148}

\bibitem[\protect\citeauthoryear{{Artymowicz} \& {Lubow}}{{Artymowicz} \&
  {Lubow}}{1994}]{1994ApJ...421..651A}
{Artymowicz} P.,  {Lubow} S.~H.,  1994, \mn@doi [\apj] {10.1086/173679}, \href
  {https://ui.adsabs.harvard.edu/abs/1994ApJ...421..651A} {421, 651}

\bibitem[\protect\citeauthoryear{{Artymowicz}, {Clarke}, {Lubow}  \&
  {Pringle}}{{Artymowicz} et~al.}{1991}]{1991ApJ...370L..35A}
{Artymowicz} P.,  {Clarke} C.~J.,  {Lubow} S.~H.,   {Pringle} J.~E.,  1991,
  \mn@doi [\apjl] {10.1086/185971}, \href
  {https://ui.adsabs.harvard.edu/abs/1991ApJ...370L..35A} {370, L35}

\bibitem[\protect\citeauthoryear{{Arzoumanian} et~al.,}{{Arzoumanian}
  et~al.}{2020}]{2020ApJ...905L..34A}
{Arzoumanian} Z.,  et~al., 2020, \mn@doi [\apjl] {10.3847/2041-8213/abd401},
  \href {https://ui.adsabs.harvard.edu/abs/2020ApJ...905L..34A} {905, L34}

\bibitem[\protect\citeauthoryear{{Balbus} \& {Hawley}}{{Balbus} \&
  {Hawley}}{1998}]{1998RvMP...70....1B}
{Balbus} S.~A.,  {Hawley} J.~F.,  1998, \mn@doi [Reviews of Modern Physics]
  {10.1103/RevModPhys.70.1}, \href
  {https://ui.adsabs.harvard.edu/abs/1998RvMP...70....1B} {70, 1}

\bibitem[\protect\citeauthoryear{{Begelman}, {Blandford}  \& {Rees}}{{Begelman}
  et~al.}{1980}]{1980Natur.287..307B}
{Begelman} M.~C.,  {Blandford} R.~D.,   {Rees} M.~J.,  1980, \mn@doi [\nat]
  {10.1038/287307a0}, \href
  {https://ui.adsabs.harvard.edu/abs/1980Natur.287..307B} {287, 307}

\bibitem[\protect\citeauthoryear{{Boss}}{{Boss}}{1986}]{1986ApJS...62..519B}
{Boss} A.~P.,  1986, \mn@doi [\apjs] {10.1086/191150}, \href
  {https://ui.adsabs.harvard.edu/abs/1986ApJS...62..519B} {62, 519}

\bibitem[\protect\citeauthoryear{{Bowen}, {Mewes}, {Noble}, {Avara},
  {Campanelli}  \& {Krolik}}{{Bowen} et~al.}{2019}]{2019ApJ...879...76B}
{Bowen} D.~B.,  {Mewes} V.,  {Noble} S.~C.,  {Avara} M.,  {Campanelli} M.,
  {Krolik} J.~H.,  2019, \mn@doi [\apj] {10.3847/1538-4357/ab2453}, \href
  {https://ui.adsabs.harvard.edu/abs/2019ApJ...879...76B} {879, 76}

\bibitem[\protect\citeauthoryear{{Chan}, {Krolik}  \& {Piran}}{{Chan}
  et~al.}{2018}]{2018ApJ...856...12C}
{Chan} C.-H.,  {Krolik} J.~H.,   {Piran} T.,  2018, \mn@doi [\apj]
  {10.3847/1538-4357/aab15c}, \href
  {https://ui.adsabs.harvard.edu/abs/2018ApJ...856...12C} {856, 12}

\bibitem[\protect\citeauthoryear{{Chandrasekhar}}{{Chandrasekhar}}{1960}]{1960PNAS...46..253C}
{Chandrasekhar} S.,  1960, \mn@doi [Proceedings of the National Academy of
  Science] {10.1073/pnas.46.2.253}, \href
  {https://ui.adsabs.harvard.edu/abs/1960PNAS...46..253C} {46, 253}

\bibitem[\protect\citeauthoryear{{Charisi}, {Bartos}, {Haiman}, {Price-Whelan},
  {Graham}, {Bellm}, {Laher}  \& {M{\'a}rka}}{{Charisi}
  et~al.}{2016}]{2016MNRAS.463.2145C}
{Charisi} M.,  {Bartos} I.,  {Haiman} Z.,  {Price-Whelan} A.~M.,  {Graham}
  M.~J.,  {Bellm} E.~C.,  {Laher} R.~R.,   {M{\'a}rka} S.,  2016, \mn@doi
  [\mnras] {10.1093/mnras/stw1838}, \href
  {https://ui.adsabs.harvard.edu/abs/2016MNRAS.463.2145C} {463, 2145}

\bibitem[\protect\citeauthoryear{{Chen} et~al.,}{{Chen}
  et~al.}{2020}]{2020MNRAS.499.2245C}
{Chen} Y.-C.,  et~al., 2020, \mn@doi [\mnras] {10.1093/mnras/staa2957}, \href
  {https://ui.adsabs.harvard.edu/abs/2020MNRAS.499.2245C} {499, 2245}

\bibitem[\protect\citeauthoryear{{Combi}, {Lopez Armengol}, {Campanelli},
  {Noble}, {Avara}, {Krolik}  \& {Bowen}}{{Combi}
  et~al.}{2021}]{2021arXiv210901307C}
{Combi} L.,  {Lopez Armengol} F.~G.,  {Campanelli} M.,  {Noble} S.~C.,  {Avara}
  M.,  {Krolik} J.~H.,   {Bowen} D.,  2021, arXiv e-prints, \href
  {https://ui.adsabs.harvard.edu/abs/2021arXiv210901307C} {p. arXiv:2109.01307}

\bibitem[\protect\citeauthoryear{{D'Angelo}, {Lubow}  \& {Bate}}{{D'Angelo}
  et~al.}{2006}]{2006ApJ...652.1698D}
{D'Angelo} G.,  {Lubow} S.~H.,   {Bate} M.~R.,  2006, \mn@doi [\apj]
  {10.1086/508451}, \href
  {https://ui.adsabs.harvard.edu/abs/2006ApJ...652.1698D} {652, 1698}

\bibitem[\protect\citeauthoryear{{D'Orazio} \& {Duffell}}{{D'Orazio} \&
  {Duffell}}{2021}]{2021arXiv210309251D}
{D'Orazio} D.~J.,  {Duffell} P.~C.,  2021, arXiv e-prints, \href
  {https://ui.adsabs.harvard.edu/abs/2021arXiv210309251D} {p. arXiv:2103.09251}

\bibitem[\protect\citeauthoryear{{D'Orazio}, {Haiman}  \&
  {MacFadyen}}{{D'Orazio} et~al.}{2013}]{2013MNRAS.436.2997D}
{D'Orazio} D.~J.,  {Haiman} Z.,   {MacFadyen} A.,  2013, \mn@doi [\mnras]
  {10.1093/mnras/stt1787}, \href
  {https://ui.adsabs.harvard.edu/abs/2013MNRAS.436.2997D} {436, 2997}

\bibitem[\protect\citeauthoryear{{D'Orazio}, {Haiman}, {Duffell}, {MacFadyen}
  \& {Farris}}{{D'Orazio} et~al.}{2016}]{2016MNRAS.459.2379D}
{D'Orazio} D.~J.,  {Haiman} Z.,  {Duffell} P.,  {MacFadyen} A.,   {Farris} B.,
  2016, \mn@doi [\mnras] {10.1093/mnras/stw792}, \href
  {https://ui.adsabs.harvard.edu/abs/2016MNRAS.459.2379D} {459, 2379}

\bibitem[\protect\citeauthoryear{{Dempsey}, {Lee}  \& {Lithwick}}{{Dempsey}
  et~al.}{2020a}]{2020ApJ...891..108D}
{Dempsey} A.~M.,  {Lee} W.-K.,   {Lithwick} Y.,  2020a, \mn@doi [\apj]
  {10.3847/1538-4357/ab723c}, \href
  {https://ui.adsabs.harvard.edu/abs/2020ApJ...891..108D} {891, 108}

\bibitem[\protect\citeauthoryear{{Dempsey}, {Mu{\~n}oz}  \&
  {Lithwick}}{{Dempsey} et~al.}{2020b}]{2020ApJ...892L..29D}
{Dempsey} A.~M.,  {Mu{\~n}oz} D.,   {Lithwick} Y.,  2020b, \mn@doi [\apjl]
  {10.3847/2041-8213/ab800e}, \href
  {https://ui.adsabs.harvard.edu/abs/2020ApJ...892L..29D} {892, L29}

\bibitem[\protect\citeauthoryear{{Derdzinski}, {D'Orazio}, {Duffell}, {Haiman}
  \& {MacFadyen}}{{Derdzinski} et~al.}{2021}]{2021MNRAS.501.3540D}
{Derdzinski} A.,  {D'Orazio} D.,  {Duffell} P.,  {Haiman} Z.,   {MacFadyen} A.,
   2021, \mn@doi [\mnras] {10.1093/mnras/staa3976}, \href
  {https://ui.adsabs.harvard.edu/abs/2021MNRAS.501.3540D} {501, 3540}

\bibitem[\protect\citeauthoryear{{Dittmann} \& {Miller}}{{Dittmann} \&
  {Miller}}{2020}]{2020MNRAS.493.3732D}
{Dittmann} A.~J.,  {Miller} M.~C.,  2020, \mn@doi [\mnras]
  {10.1093/mnras/staa463}, \href
  {https://ui.adsabs.harvard.edu/abs/2020MNRAS.493.3732D} {493, 3732}

\bibitem[\protect\citeauthoryear{{Dittmann} \& {Ryan}}{{Dittmann} \&
  {Ryan}}{2021}]{2021ApJ...921...71D}
{Dittmann} A.~J.,  {Ryan} G.,  2021, \mn@doi [\apj] {10.3847/1538-4357/ac1bbd},
  \href {https://ui.adsabs.harvard.edu/abs/2021ApJ...921...71D} {921, 71}

\bibitem[\protect\citeauthoryear{{Duffell}}{{Duffell}}{2016}]{2016ApJS..226....2D}
{Duffell} P.~C.,  2016, \mn@doi [\apjs] {10.3847/0067-0049/226/1/2}, \href
  {https://ui.adsabs.harvard.edu/abs/2016ApJS..226....2D} {226, 2}

\bibitem[\protect\citeauthoryear{{Duffell}, {D'Orazio}, {Derdzinski}, {Haiman},
  {MacFadyen}, {Rosen}  \& {Zrake}}{{Duffell}
  et~al.}{2020}]{2020ApJ...901...25D}
{Duffell} P.~C.,  {D'Orazio} D.,  {Derdzinski} A.,  {Haiman} Z.,  {MacFadyen}
  A.,  {Rosen} A.~L.,   {Zrake} J.,  2020, \mn@doi [\apj]
  {10.3847/1538-4357/abab95}, \href
  {https://ui.adsabs.harvard.edu/abs/2020ApJ...901...25D} {901, 25}

\bibitem[\protect\citeauthoryear{{Farris}, {Duffell}, {MacFadyen}  \&
  {Haiman}}{{Farris} et~al.}{2014}]{2014ApJ...783..134F}
{Farris} B.~D.,  {Duffell} P.,  {MacFadyen} A.~I.,   {Haiman} Z.,  2014,
  \mn@doi [\apj] {10.1088/0004-637X/783/2/134}, \href
  {https://ui.adsabs.harvard.edu/abs/2014ApJ...783..134F} {783, 134}

\bibitem[\protect\citeauthoryear{{Farris}, {Duffell}, {MacFadyen}  \&
  {Haiman}}{{Farris} et~al.}{2015}]{2015MNRAS.446L..36F}
{Farris} B.~D.,  {Duffell} P.,  {MacFadyen} A.~I.,   {Haiman} Z.,  2015,
  \mn@doi [\mnras] {10.1093/mnrasl/slu160}, \href
  {https://ui.adsabs.harvard.edu/abs/2015MNRAS.446L..36F} {446, L36}

\bibitem[\protect\citeauthoryear{{Foord}, {Liu}, {G{\"u}ltekin}, {Whitley},
  {Shi}  \& {Chen}}{{Foord} et~al.}{2021}]{2021arXiv211002982F}
{Foord} A.,  {Liu} X.,  {G{\"u}ltekin} K.,  {Whitley} K.,  {Shi} F.,   {Chen}
  Y.-C.,  2021, arXiv e-prints, \href
  {https://ui.adsabs.harvard.edu/abs/2021arXiv211002982F} {p. arXiv:2110.02982}

\bibitem[\protect\citeauthoryear{{Goldreich} \& {Sari}}{{Goldreich} \&
  {Sari}}{2003}]{2003ApJ...585.1024G}
{Goldreich} P.,  {Sari} R.,  2003, \mn@doi [\apj] {10.1086/346202}, \href
  {https://ui.adsabs.harvard.edu/abs/2003ApJ...585.1024G} {585, 1024}

\bibitem[\protect\citeauthoryear{{Goodchild} \& {Ogilvie}}{{Goodchild} \&
  {Ogilvie}}{2006}]{2006MNRAS.368.1123G}
{Goodchild} S.,  {Ogilvie} G.,  2006, \mn@doi [\mnras]
  {10.1111/j.1365-2966.2006.10197.x}, \href
  {https://ui.adsabs.harvard.edu/abs/2006MNRAS.368.1123G} {368, 1123}

\bibitem[\protect\citeauthoryear{{Gottlieb} \& {Shu}}{{Gottlieb} \&
  {Shu}}{1998}]{1998MaCom..67...73G}
{Gottlieb} S.,  {Shu} C.~W.,  1998, Mathematics of Computation, \href
  {https://ui.adsabs.harvard.edu/abs/1998MaCom..67...73G} {67, 73}

\bibitem[\protect\citeauthoryear{{Gould} \& {Rix}}{{Gould} \&
  {Rix}}{2000}]{2000ApJ...532L..29G}
{Gould} A.,  {Rix} H.-W.,  2000, \mn@doi [\apjl] {10.1086/312562}, \href
  {https://ui.adsabs.harvard.edu/abs/2000ApJ...532L..29G} {532, L29}

\bibitem[\protect\citeauthoryear{{Graham} et~al.,}{{Graham}
  et~al.}{2015}]{2015MNRAS.453.1562G}
{Graham} M.~J.,  et~al., 2015, \mn@doi [\mnras] {10.1093/mnras/stv1726}, \href
  {https://ui.adsabs.harvard.edu/abs/2015MNRAS.453.1562G} {453, 1562}

\bibitem[\protect\citeauthoryear{{Guilloteau}, {Dutrey}  \&
  {Simon}}{{Guilloteau} et~al.}{1999}]{1999A&A...348..570G}
{Guilloteau} S.,  {Dutrey} A.,   {Simon} M.,  1999, \aap, \href
  {https://ui.adsabs.harvard.edu/abs/1999A&A...348..570G} {348, 570}

\bibitem[\protect\citeauthoryear{{Guti{\'e}rrez}, {Combi}, {Noble},
  {Campanelli}, {Krolik}, {L{\'o}pez Armengol}  \&
  {Garc{\'\i}a}}{{Guti{\'e}rrez} et~al.}{2021}]{2021arXiv211209773G}
{Guti{\'e}rrez} E.~M.,  {Combi} L.,  {Noble} S.~C.,  {Campanelli} M.,  {Krolik}
  J.~H.,  {L{\'o}pez Armengol} F.~G.,   {Garc{\'\i}a} F.,  2021, arXiv
  e-prints, \href {https://ui.adsabs.harvard.edu/abs/2021arXiv211209773G} {p.
  arXiv:2112.09773}

\bibitem[\protect\citeauthoryear{{Haiman}, {Kocsis}  \& {Menou}}{{Haiman}
  et~al.}{2009}]{2009ApJ...700.1952H}
{Haiman} Z.,  {Kocsis} B.,   {Menou} K.,  2009, \mn@doi [\apj]
  {10.1088/0004-637X/700/2/1952}, \href
  {https://ui.adsabs.harvard.edu/abs/2009ApJ...700.1952H} {700, 1952}

\bibitem[\protect\citeauthoryear{{Heath} \& {Nixon}}{{Heath} \&
  {Nixon}}{2020}]{2020A&A...641A..64H}
{Heath} R.~M.,  {Nixon} C.~J.,  2020, \mn@doi [\aap]
  {10.1051/0004-6361/202038548}, \href
  {https://ui.adsabs.harvard.edu/abs/2020A&A...641A..64H} {641, A64}

\bibitem[\protect\citeauthoryear{{Hogg} \& {Reynolds}}{{Hogg} \&
  {Reynolds}}{2016}]{2016ApJ...826...40H}
{Hogg} J.~D.,  {Reynolds} C.~S.,  2016, \mn@doi [\apj]
  {10.3847/0004-637X/826/1/40}, \href
  {https://ui.adsabs.harvard.edu/abs/2016ApJ...826...40H} {826, 40}

\bibitem[\protect\citeauthoryear{{Hubeny}, {Blaes}, {Krolik}  \&
  {Agol}}{{Hubeny} et~al.}{2001}]{2001ApJ...559..680H}
{Hubeny} I.,  {Blaes} O.,  {Krolik} J.~H.,   {Agol} E.,  2001, \mn@doi [\apj]
  {10.1086/322344}, \href
  {https://ui.adsabs.harvard.edu/abs/2001ApJ...559..680H} {559, 680}

\bibitem[\protect\citeauthoryear{{Ivanov}, {Papaloizou}  \&
  {Polnarev}}{{Ivanov} et~al.}{1999}]{1999MNRAS.307...79I}
{Ivanov} P.~B.,  {Papaloizou} J.~C.~B.,   {Polnarev} A.~G.,  1999, \mn@doi
  [\mnras] {10.1046/j.1365-8711.1999.02623.x}, \href
  {https://ui.adsabs.harvard.edu/abs/1999MNRAS.307...79I} {307, 79}

\bibitem[\protect\citeauthoryear{{Kashi} \& {Soker}}{{Kashi} \&
  {Soker}}{2011}]{2011MNRAS.417.1466K}
{Kashi} A.,  {Soker} N.,  2011, \mn@doi [\mnras]
  {10.1111/j.1365-2966.2011.19361.x}, \href
  {https://ui.adsabs.harvard.edu/abs/2011MNRAS.417.1466K} {417, 1466}

\bibitem[\protect\citeauthoryear{{Kelley}, {Blecha}, {Hernquist}, {Sesana}  \&
  {Taylor}}{{Kelley} et~al.}{2018}]{2018MNRAS.477..964K}
{Kelley} L.~Z.,  {Blecha} L.,  {Hernquist} L.,  {Sesana} A.,   {Taylor} S.~R.,
  2018, \mn@doi [\mnras] {10.1093/mnras/sty689}, \href
  {https://ui.adsabs.harvard.edu/abs/2018MNRAS.477..964K} {477, 964}

\bibitem[\protect\citeauthoryear{{Keppler} et~al.,}{{Keppler}
  et~al.}{2020}]{2020A&A...639A..62K}
{Keppler} M.,  et~al., 2020, \mn@doi [\aap] {10.1051/0004-6361/202038032},
  \href {https://ui.adsabs.harvard.edu/abs/2020A&A...639A..62K} {639, A62}

\bibitem[\protect\citeauthoryear{{Khan}, {Holley-Bockelmann}, {Berczik}  \&
  {Just}}{{Khan} et~al.}{2013}]{2013ApJ...773..100K}
{Khan} F.~M.,  {Holley-Bockelmann} K.,  {Berczik} P.,   {Just} A.,  2013,
  \mn@doi [\apj] {10.1088/0004-637X/773/2/100}, \href
  {https://ui.adsabs.harvard.edu/abs/2013ApJ...773..100K} {773, 100}

\bibitem[\protect\citeauthoryear{King \& Nixon}{King \&
  Nixon}{2015}]{10.1093/mnrasl/slv098}
King A.,  Nixon C.,  2015, \mn@doi [Monthly Notices of the Royal Astronomical
  Society: Letters] {10.1093/mnrasl/slv098}, 453, L46

\bibitem[\protect\citeauthoryear{{Kocsis} \& {Sesana}}{{Kocsis} \&
  {Sesana}}{2011}]{2011MNRAS.411.1467K}
{Kocsis} B.,  {Sesana} A.,  2011, \mn@doi [\mnras]
  {10.1111/j.1365-2966.2010.17782.x}, \href
  {https://ui.adsabs.harvard.edu/abs/2011MNRAS.411.1467K} {411, 1467}

\bibitem[\protect\citeauthoryear{{Kolykhalov} \& {Syunyaev}}{{Kolykhalov} \&
  {Syunyaev}}{1980}]{1980SvAL....6..357K}
{Kolykhalov} P.~I.,  {Syunyaev} R.~A.,  1980, Soviet Astronomy Letters, \href
  {https://ui.adsabs.harvard.edu/abs/1980SvAL....6..357K} {6, 357}

\bibitem[\protect\citeauthoryear{{Kormendy} \& {Ho}}{{Kormendy} \&
  {Ho}}{2013}]{2013ARA&A..51..511K}
{Kormendy} J.,  {Ho} L.~C.,  2013, \mn@doi [\araa]
  {10.1146/annurev-astro-082708-101811}, \href
  {https://ui.adsabs.harvard.edu/abs/2013ARA&A..51..511K} {51, 511}

\bibitem[\protect\citeauthoryear{{Kratter}, {Murray-Clay}  \&
  {Youdin}}{{Kratter} et~al.}{2010}]{2010ApJ...710.1375K}
{Kratter} K.~M.,  {Murray-Clay} R.~A.,   {Youdin} A.~N.,  2010, \mn@doi [\apj]
  {10.1088/0004-637X/710/2/1375}, \href
  {https://ui.adsabs.harvard.edu/abs/2010ApJ...710.1375K} {710, 1375}

\bibitem[\protect\citeauthoryear{{Krolik}}{{Krolik}}{1999}]{1999agnc.book.....K}
{Krolik} J.~H.,  1999, {Active galactic nuclei : from the central black hole to
  the galactic environment}.
Princeton University Press

\bibitem[\protect\citeauthoryear{{Kurganov} \& {Tadmor}}{{Kurganov} \&
  {Tadmor}}{2000}]{2000JCoPh.160..241K}
{Kurganov} A.,  {Tadmor} E.,  2000, \mn@doi [Journal of Computational Physics]
  {10.1006/jcph.2000.6459}, \href
  {https://ui.adsabs.harvard.edu/abs/2000JCoPh.160..241K} {160, 241}

\bibitem[\protect\citeauthoryear{{Liao} et~al.,}{{Liao}
  et~al.}{2021}]{2021MNRAS.500.4025L}
{Liao} W.-T.,  et~al., 2021, \mn@doi [\mnras] {10.1093/mnras/staa3055}, \href
  {https://ui.adsabs.harvard.edu/abs/2021MNRAS.500.4025L} {500, 4025}

\bibitem[\protect\citeauthoryear{{Liu}, {Gezari}  \& {Miller}}{{Liu}
  et~al.}{2018}]{2018ApJ...859L..12L}
{Liu} T.,  {Gezari} S.,   {Miller} M.~C.,  2018, \mn@doi [\apjl]
  {10.3847/2041-8213/aac2ed}, \href
  {https://ui.adsabs.harvard.edu/abs/2018ApJ...859L..12L} {859, L12}

\bibitem[\protect\citeauthoryear{{Lomb}}{{Lomb}}{1976}]{1976Ap&SS..39..447L}
{Lomb} N.~R.,  1976, \mn@doi [\apss] {10.1007/BF00648343}, \href
  {https://ui.adsabs.harvard.edu/abs/1976Ap&SS..39..447L} {39, 447}

\bibitem[\protect\citeauthoryear{{Lubow}}{{Lubow}}{1991}]{1991ApJ...381..259L}
{Lubow} S.~H.,  1991, \mn@doi [\apj] {10.1086/170647}, \href
  {https://ui.adsabs.harvard.edu/abs/1991ApJ...381..259L} {381, 259}

\bibitem[\protect\citeauthoryear{{Lubow}, {Martin}  \& {Nixon}}{{Lubow}
  et~al.}{2015}]{2015ApJ...800...96L}
{Lubow} S.~H.,  {Martin} R.~G.,   {Nixon} C.,  2015, \mn@doi [\apj]
  {10.1088/0004-637X/800/2/96}, \href
  {https://ui.adsabs.harvard.edu/abs/2015ApJ...800...96L} {800, 96}

\bibitem[\protect\citeauthoryear{{Lynden-Bell} \& {Pringle}}{{Lynden-Bell} \&
  {Pringle}}{1974}]{1974MNRAS.168..603L}
{Lynden-Bell} D.,  {Pringle} J.~E.,  1974, \mn@doi [\mnras]
  {10.1093/mnras/168.3.603}, \href
  {https://ui.adsabs.harvard.edu/abs/1974MNRAS.168..603L} {168, 603}

\bibitem[\protect\citeauthoryear{{Lyubarskii}}{{Lyubarskii}}{1997}]{1997MNRAS.292..679L}
{Lyubarskii} Y.~E.,  1997, \mn@doi [\mnras] {10.1093/mnras/292.3.679}, \href
  {https://ui.adsabs.harvard.edu/abs/1997MNRAS.292..679L} {292, 679}

\bibitem[\protect\citeauthoryear{{MacFadyen} \&
  {Milosavljevi{\'c}}}{{MacFadyen} \&
  {Milosavljevi{\'c}}}{2008}]{2008ApJ...672...83M}
{MacFadyen} A.~I.,  {Milosavljevi{\'c}} M.,  2008, \mn@doi [\apj]
  {10.1086/523869}, \href
  {https://ui.adsabs.harvard.edu/abs/2008ApJ...672...83M} {672, 83}

\bibitem[\protect\citeauthoryear{{Martini}}{{Martini}}{2004}]{2004cbhg.symp..169M}
{Martini} P.,  2004, in {Ho} L.~C.,  ed., Coevolution of Black Holes and
  Galaxies. p.~169 (\mn@eprint {arXiv} {astro-ph/0304009})

\bibitem[\protect\citeauthoryear{{McCabe}, {Duch{\^e}ne}  \& {Ghez}}{{McCabe}
  et~al.}{2002}]{2002ApJ...575..974M}
{McCabe} C.,  {Duch{\^e}ne} G.,   {Ghez} A.~M.,  2002, \mn@doi [\apj]
  {10.1086/341479}, \href
  {https://ui.adsabs.harvard.edu/abs/2002ApJ...575..974M} {575, 974}

\bibitem[\protect\citeauthoryear{{Milosavljevi{\'c}} \&
  {Merritt}}{{Milosavljevi{\'c}} \& {Merritt}}{2003a}]{2003ApJ...596..860M}
{Milosavljevi{\'c}} M.,  {Merritt} D.,  2003a, \mn@doi [\apj] {10.1086/378086},
  \href {https://ui.adsabs.harvard.edu/abs/2003ApJ...596..860M} {596, 860}

\bibitem[\protect\citeauthoryear{{Milosavljevi{\'c}} \&
  {Merritt}}{{Milosavljevi{\'c}} \& {Merritt}}{2003b}]{2003AIPC..686..201M}
{Milosavljevi{\'c}} M.,  {Merritt} D.,  2003b, in {Centrella} J.~M.,  ed.,
  American Institute of Physics Conference Series Vol. 686, The Astrophysics of
  Gravitational Wave Sources. pp 201--210 (\mn@eprint {arXiv}
  {astro-ph/0212270}), \mn@doi{10.1063/1.1629432}

\bibitem[\protect\citeauthoryear{{Miranda} \& {Lai}}{{Miranda} \&
  {Lai}}{2015}]{2015MNRAS.452.2396M}
{Miranda} R.,  {Lai} D.,  2015, \mn@doi [\mnras] {10.1093/mnras/stv1450}, \href
  {https://ui.adsabs.harvard.edu/abs/2015MNRAS.452.2396M} {452, 2396}

\bibitem[\protect\citeauthoryear{{Miranda} \& {Rafikov}}{{Miranda} \&
  {Rafikov}}{2019}]{2019ApJ...878L...9M}
{Miranda} R.,  {Rafikov} R.~R.,  2019, \mn@doi [\apjl]
  {10.3847/2041-8213/ab22a7}, \href
  {https://ui.adsabs.harvard.edu/abs/2019ApJ...878L...9M} {878, L9}

\bibitem[\protect\citeauthoryear{{Miranda} \& {Rafikov}}{{Miranda} \&
  {Rafikov}}{2020}]{2020ApJ...892...65M}
{Miranda} R.,  {Rafikov} R.~R.,  2020, \mn@doi [\apj]
  {10.3847/1538-4357/ab791a}, \href
  {https://ui.adsabs.harvard.edu/abs/2020ApJ...892...65M} {892, 65}

\bibitem[\protect\citeauthoryear{{Miranda}, {Mu{\~n}oz}  \& {Lai}}{{Miranda}
  et~al.}{2017}]{2017MNRAS.466.1170M}
{Miranda} R.,  {Mu{\~n}oz} D.~J.,   {Lai} D.,  2017, \mn@doi [\mnras]
  {10.1093/mnras/stw3189}, \href
  {https://ui.adsabs.harvard.edu/abs/2017MNRAS.466.1170M} {466, 1170}

\bibitem[\protect\citeauthoryear{{Moody}, {Shi}  \& {Stone}}{{Moody}
  et~al.}{2019}]{2019ApJ...875...66M}
{Moody} M. S.~L.,  {Shi} J.-M.,   {Stone} J.~M.,  2019, \mn@doi [\apj]
  {10.3847/1538-4357/ab09ee}, \href
  {https://ui.adsabs.harvard.edu/abs/2019ApJ...875...66M} {875, 66}

\bibitem[\protect\citeauthoryear{{Moriwaki} \& {Nakagawa}}{{Moriwaki} \&
  {Nakagawa}}{2004}]{2004ApJ...609.1065M}
{Moriwaki} K.,  {Nakagawa} Y.,  2004, \mn@doi [\apj] {10.1086/421342}, \href
  {https://ui.adsabs.harvard.edu/abs/2004ApJ...609.1065M} {609, 1065}

\bibitem[\protect\citeauthoryear{{Mu{\~n}oz} \& {Lithwick}}{{Mu{\~n}oz} \&
  {Lithwick}}{2020}]{2020ApJ...905..106M}
{Mu{\~n}oz} D.~J.,  {Lithwick} Y.,  2020, \mn@doi [\apj]
  {10.3847/1538-4357/abc74c}, \href
  {https://ui.adsabs.harvard.edu/abs/2020ApJ...905..106M} {905, 106}

\bibitem[\protect\citeauthoryear{{Mu{\~n}oz}, {Miranda}  \& {Lai}}{{Mu{\~n}oz}
  et~al.}{2019}]{2019ApJ...871...84M}
{Mu{\~n}oz} D.~J.,  {Miranda} R.,   {Lai} D.,  2019, \mn@doi [\apj]
  {10.3847/1538-4357/aaf867}, \href
  {https://ui.adsabs.harvard.edu/abs/2019ApJ...871...84M} {871, 84}

\bibitem[\protect\citeauthoryear{{Mu{\~n}oz}, {Lai}, {Kratter}  \&
  {Miranda}}{{Mu{\~n}oz} et~al.}{2020}]{2020ApJ...889..114M}
{Mu{\~n}oz} D.~J.,  {Lai} D.,  {Kratter} K.,   {Miranda} R.,  2020, \mn@doi
  [\apj] {10.3847/1538-4357/ab5d33}, \href
  {https://ui.adsabs.harvard.edu/abs/2020ApJ...889..114M} {889, 114}

\bibitem[\protect\citeauthoryear{{Noble}, {Mundim}, {Nakano}, {Krolik},
  {Campanelli}, {Zlochower}  \& {Yunes}}{{Noble}
  et~al.}{2012}]{2012ApJ...755...51N}
{Noble} S.~C.,  {Mundim} B.~C.,  {Nakano} H.,  {Krolik} J.~H.,  {Campanelli}
  M.,  {Zlochower} Y.,   {Yunes} N.,  2012, \mn@doi [\apj]
  {10.1088/0004-637X/755/1/51}, \href
  {https://ui.adsabs.harvard.edu/abs/2012ApJ...755...51N} {755, 51}

\bibitem[\protect\citeauthoryear{{Noble}, {Krolik}, {Campanelli}, {Zlochower},
  {Mundim}, {Nakano}  \& {Zilh{\~a}o}}{{Noble}
  et~al.}{2021}]{2021arXiv210312100N}
{Noble} S.~C.,  {Krolik} J.~H.,  {Campanelli} M.,  {Zlochower} Y.,  {Mundim}
  B.~C.,  {Nakano} H.,   {Zilh{\~a}o} M.,  2021, \mn@doi [\apj]
  {10.3847/1538-4357/ac2229}, \href
  {https://ui.adsabs.harvard.edu/abs/2021ApJ...922..175N} {922, 175}

\bibitem[\protect\citeauthoryear{{Pan} \& {Sari}}{{Pan} \&
  {Sari}}{2004}]{2004AJ....128.1418P}
{Pan} M.,  {Sari} R.,  2004, \mn@doi [\aj] {10.1086/423214}, \href
  {https://ui.adsabs.harvard.edu/abs/2004AJ....128.1418P} {128, 1418}

\bibitem[\protect\citeauthoryear{{Papaloizou} \& {Pringle}}{{Papaloizou} \&
  {Pringle}}{1977}]{1977MNRAS.181..441P}
{Papaloizou} J.,  {Pringle} J.~E.,  1977, \mn@doi [\mnras]
  {10.1093/mnras/181.3.441}, \href
  {https://ui.adsabs.harvard.edu/abs/1977MNRAS.181..441P} {181, 441}

\bibitem[\protect\citeauthoryear{{Peters}}{{Peters}}{1964}]{1964PhRv..136.1224P}
{Peters} P.~C.,  1964, \mn@doi [Physical Review] {10.1103/PhysRev.136.B1224},
  \href {https://ui.adsabs.harvard.edu/abs/1964PhRv..136.1224P} {136, 1224}

\bibitem[\protect\citeauthoryear{{Price} et~al.,}{{Price}
  et~al.}{2018}]{2018PASA...35...31P}
{Price} D.~J.,  et~al., 2018, \mn@doi [\pasa] {10.1017/pasa.2018.25}, \href
  {https://ui.adsabs.harvard.edu/abs/2018PASA...35...31P} {35, e031}

\bibitem[\protect\citeauthoryear{{Pringle}}{{Pringle}}{1991}]{1991MNRAS.248..754P}
{Pringle} J.~E.,  1991, \mn@doi [\mnras] {10.1093/mnras/248.4.754}, \href
  {https://ui.adsabs.harvard.edu/abs/1991MNRAS.248..754P} {248, 754}

\bibitem[\protect\citeauthoryear{{Ragusa}, {Lodato}  \& {Price}}{{Ragusa}
  et~al.}{2016}]{2016MNRAS.460.1243R}
{Ragusa} E.,  {Lodato} G.,   {Price} D.~J.,  2016, \mn@doi [\mnras]
  {10.1093/mnras/stw1081}, \href
  {https://ui.adsabs.harvard.edu/abs/2016MNRAS.460.1243R} {460, 1243}

\bibitem[\protect\citeauthoryear{{Ragusa}, {Alexander}, {Calcino}, {Hirsh}  \&
  {Price}}{{Ragusa} et~al.}{2020}]{2020MNRAS.499.3362R}
{Ragusa} E.,  {Alexander} R.,  {Calcino} J.,  {Hirsh} K.,   {Price} D.~J.,
  2020, \mn@doi [\mnras] {10.1093/mnras/staa2954}, \href
  {https://ui.adsabs.harvard.edu/abs/2020MNRAS.499.3362R} {499, 3362}

\bibitem[\protect\citeauthoryear{{Roedig}, {Sesana}, {Dotti}, {Cuadra},
  {Amaro-Seoane}  \& {Haardt}}{{Roedig} et~al.}{2012}]{2012A&A...545A.127R}
{Roedig} C.,  {Sesana} A.,  {Dotti} M.,  {Cuadra} J.,  {Amaro-Seoane} P.,
  {Haardt} F.,  2012, \mn@doi [\aap] {10.1051/0004-6361/201219986}, \href
  {https://ui.adsabs.harvard.edu/abs/2012A&A...545A.127R} {545, A127}

\bibitem[\protect\citeauthoryear{{Salpeter}}{{Salpeter}}{1964}]{1964ApJ...140..796S}
{Salpeter} E.~E.,  1964, \mn@doi [\apj] {10.1086/147973}, \href
  {https://ui.adsabs.harvard.edu/abs/1964ApJ...140..796S} {140, 796}

\bibitem[\protect\citeauthoryear{{Scargle}}{{Scargle}}{1982}]{1982ApJ...263..835S}
{Scargle} J.~D.,  1982, \mn@doi [\apj] {10.1086/160554}, \href
  {https://ui.adsabs.harvard.edu/abs/1982ApJ...263..835S} {263, 835}

\bibitem[\protect\citeauthoryear{{Schawinski}, {Koss}, {Berney}  \&
  {Sartori}}{{Schawinski} et~al.}{2015}]{2015MNRAS.451.2517S}
{Schawinski} K.,  {Koss} M.,  {Berney} S.,   {Sartori} L.~F.,  2015, \mn@doi
  [\mnras] {10.1093/mnras/stv1136}, \href
  {https://ui.adsabs.harvard.edu/abs/2015MNRAS.451.2517S} {451, 2517}

\bibitem[\protect\citeauthoryear{{Sesana}}{{Sesana}}{2013}]{2013CQGra..30x4009S}
{Sesana} A.,  2013, \mn@doi [Classical and Quantum Gravity]
  {10.1088/0264-9381/30/24/244009}, \href
  {https://ui.adsabs.harvard.edu/abs/2013CQGra..30x4009S} {30, 244009}

\bibitem[\protect\citeauthoryear{{Shakura} \& {Sunyaev}}{{Shakura} \&
  {Sunyaev}}{1973}]{1973A&A....24..337S}
{Shakura} N.~I.,  {Sunyaev} R.~A.,  1973, \aap, \href
  {https://ui.adsabs.harvard.edu/abs/1973A&A....24..337S} {500, 33}

\bibitem[\protect\citeauthoryear{{Shi} \& {Krolik}}{{Shi} \&
  {Krolik}}{2015}]{2015ApJ...807..131S}
{Shi} J.-M.,  {Krolik} J.~H.,  2015, \mn@doi [\apj]
  {10.1088/0004-637X/807/2/131}, \href
  {https://ui.adsabs.harvard.edu/abs/2015ApJ...807..131S} {807, 131}

\bibitem[\protect\citeauthoryear{{Shi} \& {Krolik}}{{Shi} \&
  {Krolik}}{2016}]{2016ApJ...832...22S}
{Shi} J.-M.,  {Krolik} J.~H.,  2016, \mn@doi [\apj]
  {10.3847/0004-637X/832/1/22}, \href
  {https://ui.adsabs.harvard.edu/abs/2016ApJ...832...22S} {832, 22}

\bibitem[\protect\citeauthoryear{{Shi}, {Krolik}, {Lubow}  \& {Hawley}}{{Shi}
  et~al.}{2012}]{2012ApJ...749..118S}
{Shi} J.-M.,  {Krolik} J.~H.,  {Lubow} S.~H.,   {Hawley} J.~F.,  2012, \mn@doi
  [\apj] {10.1088/0004-637X/749/2/118}, \href
  {https://ui.adsabs.harvard.edu/abs/2012ApJ...749..118S} {749, 118}

\bibitem[\protect\citeauthoryear{{Shlosman} \& {Begelman}}{{Shlosman} \&
  {Begelman}}{1989}]{1989ApJ...341..685S}
{Shlosman} I.,  {Begelman} M.~C.,  1989, \mn@doi [\apj] {10.1086/167526}, \href
  {https://ui.adsabs.harvard.edu/abs/1989ApJ...341..685S} {341, 685}

\bibitem[\protect\citeauthoryear{{Springel}}{{Springel}}{2010}]{2010MNRAS.401..791S}
{Springel} V.,  2010, \mn@doi [\mnras] {10.1111/j.1365-2966.2009.15715.x},
  \href {https://ui.adsabs.harvard.edu/abs/2010MNRAS.401..791S} {401, 791}

\bibitem[\protect\citeauthoryear{{Tang}, {MacFadyen}  \& {Haiman}}{{Tang}
  et~al.}{2017}]{2017MNRAS.469.4258T}
{Tang} Y.,  {MacFadyen} A.,   {Haiman} Z.,  2017, \mn@doi [\mnras]
  {10.1093/mnras/stx1130}, \href
  {https://ui.adsabs.harvard.edu/abs/2017MNRAS.469.4258T} {469, 4258}

\bibitem[\protect\citeauthoryear{{Tang}, {Haiman}  \& {MacFadyen}}{{Tang}
  et~al.}{2018}]{2018MNRAS.476.2249T}
{Tang} Y.,  {Haiman} Z.,   {MacFadyen} A.,  2018, \mn@doi [\mnras]
  {10.1093/mnras/sty423}, \href
  {https://ui.adsabs.harvard.edu/abs/2018MNRAS.476.2249T} {476, 2249}

\bibitem[\protect\citeauthoryear{{Teyssandier} \& {Ogilvie}}{{Teyssandier} \&
  {Ogilvie}}{2016}]{2016MNRAS.458.3221T}
{Teyssandier} J.,  {Ogilvie} G.~I.,  2016, \mn@doi [\mnras]
  {10.1093/mnras/stw521}, \href
  {https://ui.adsabs.harvard.edu/abs/2016MNRAS.458.3221T} {458, 3221}

\bibitem[\protect\citeauthoryear{{Tiede}, {Zrake}, {MacFadyen}  \&
  {Haiman}}{{Tiede} et~al.}{2020}]{2020ApJ...900...43T}
{Tiede} C.,  {Zrake} J.,  {MacFadyen} A.,   {Haiman} Z.,  2020, \mn@doi [\apj]
  {10.3847/1538-4357/aba432}, \href
  {https://ui.adsabs.harvard.edu/abs/2020ApJ...900...43T} {900, 43}

\bibitem[\protect\citeauthoryear{{Tiede}, {Zrake}, {MacFadyen}  \&
  {Haiman}}{{Tiede} et~al.}{2021}]{2021arXiv211104721T}
{Tiede} C.,  {Zrake} J.,  {MacFadyen} A.,   {Haiman} Z.,  2021, arXiv e-prints,
  \href {https://ui.adsabs.harvard.edu/abs/2021arXiv211104721T} {p.
  arXiv:2111.04721}

\bibitem[\protect\citeauthoryear{{Toro}, {Spruce}  \& {Speares}}{{Toro}
  et~al.}{1994}]{1994ShWav...4...25T}
{Toro} E.~F.,  {Spruce} M.,   {Speares} W.,  1994, \mn@doi [Shock Waves]
  {10.1007/BF01414629}, \href
  {https://ui.adsabs.harvard.edu/abs/1994ShWav...4...25T} {4, 25}

\bibitem[\protect\citeauthoryear{{Townsend}}{{Townsend}}{2010}]{2010ApJS..191..247T}
{Townsend} R.~H.~D.,  2010, \mn@doi [\apjs] {10.1088/0067-0049/191/2/247},
  \href {https://ui.adsabs.harvard.edu/abs/2010ApJS..191..247T} {191, 247}

\bibitem[\protect\citeauthoryear{{Vaughan}, {Uttley}, {Markowitz},
  {Huppenkothen}, {Middleton}, {Alston}, {Scargle}  \& {Farr}}{{Vaughan}
  et~al.}{2016}]{2016MNRAS.461.3145V}
{Vaughan} S.,  {Uttley} P.,  {Markowitz} A.~G.,  {Huppenkothen} D.,
  {Middleton} M.~J.,  {Alston} W.~N.,  {Scargle} J.~D.,   {Farr} W.~M.,  2016,
  \mn@doi [\mnras] {10.1093/mnras/stw1412}, \href
  {https://ui.adsabs.harvard.edu/abs/2016MNRAS.461.3145V} {461, 3145}

\bibitem[\protect\citeauthoryear{{Velikhov}}{{Velikhov}}{1959}]{velikhov59}
{Velikhov} E.,  1959, JETP, 36, 1398

\bibitem[\protect\citeauthoryear{{Westernacher-Schneider}, {Zrake}, {MacFadyen}
   \& {Haiman}}{{Westernacher-Schneider} et~al.}{2021}]{2021arXiv211106882W}
{Westernacher-Schneider} J.~R.,  {Zrake} J.,  {MacFadyen} A.,   {Haiman} Z.,
  2021, arXiv e-prints, \href
  {https://ui.adsabs.harvard.edu/abs/2021arXiv211106882W} {p. arXiv:2111.06882}

\bibitem[\protect\citeauthoryear{{Wood}, {Crosas}  \& {Ghez}}{{Wood}
  et~al.}{1999}]{1999ApJ...516..335W}
{Wood} K.,  {Crosas} M.,   {Ghez} A.,  1999, \mn@doi [\apj] {10.1086/307104},
  \href {https://ui.adsabs.harvard.edu/abs/1999ApJ...516..335W} {516, 335}

\bibitem[\protect\citeauthoryear{{Zrake} \& {MacFadyen}}{{Zrake} \&
  {MacFadyen}}{2012}]{2012ApJ...744...32Z}
{Zrake} J.,  {MacFadyen} A.~I.,  2012, \mn@doi [\apj]
  {10.1088/0004-637X/744/1/32}, \href
  {https://ui.adsabs.harvard.edu/abs/2012ApJ...744...32Z} {744, 32}

\bibitem[\protect\citeauthoryear{{Zrake}, {Tiede}, {MacFadyen}  \&
  {Haiman}}{{Zrake} et~al.}{2021}]{2020arXiv201009707Z}
{Zrake} J.,  {Tiede} C.,  {MacFadyen} A.,   {Haiman} Z.,  2021, \mn@doi [\apjl]
  {10.3847/2041-8213/abdd1c}, \href
  {https://ui.adsabs.harvard.edu/abs/2021ApJ...909L..13Z} {909, L13}

\bibitem[\protect\citeauthoryear{{van Leer}}{{van
  Leer}}{1979}]{1979JCoPh..32..101V}
{van Leer} B.,  1979, \mn@doi [Journal of Computational Physics]
  {10.1016/0021-9991(79)90145-1}, \href
  {https://ui.adsabs.harvard.edu/abs/1979JCoPh..32..101V} {32, 101}

\makeatother
\end{thebibliography}
\appendix
\section{Simulation Summaries}\label{app:summary}
We show in Table \ref{tab:summary} the results of various simulation outputs related to the orbital evolution of the binary and flow of mass and angular momentum between the binary and the disc presented in Section \ref{sec:fiducial}. Table \ref{tab:summary2} holds the results presented in Section \ref{sec:alpha}, and Table \ref{tab:summary3} holds the results presented in Section \ref{sec:global}.
\begin{table}
\begin{tabular}{cccccc}\hline\hline
$\mathcal{M}$ & $\nu$ & $\dot{M}/\dot{M}_0$ & $\dot{J}_g/\dot{M}$ & $\dot{J}/\dot{M}$ & $d\log{a}_b/d\log{M}$ \\\hline
10 & 0.0005 & 1.220 & 0.439 & 0.689 & 2.514 \\
12.5 & 0.0005 & 1.228 & 0.422 & 0.672 & 2.376 \\
15 & 0.0005 & 1.219 & 0.383 & 0.634 & 2.069 \\
17.5 & 0.0005 & 1.121 & 0.138 & 0.388 & 0.104 \\
20 & 0.0005 & 0.965 & -0.328 & -0.077 & -3.619 \\
25 & 0.0005 & 0.797 & -0.971 & -0.721 & -8.766 \\
30 & 0.0005 & 0.607 & -2.027 & -1.777 & -17.212 \\
10 & 0.001 & 1.238 & 0.475 & 0.725 & 2.799 \\
12.5 & 0.001 & 1.240 & 0.445 & 0.695 & 2.561 \\
15 & 0.001 & 1.202 & 0.342 & 0.593 & 1.741 \\
17.5 & 0.001 & 1.109 & 0.110 & 0.360 & -0.117 \\
20 & 0.001 & 1.030 & -0.110 & 0.141 & -1.875 \\
25 & 0.001 & 0.952 & -0.362 & -0.112 & -3.894 \\
30 & 0.001 & 0.841 & -0.731 & -0.480 & -6.844 \\
10 & 0.002 & 1.270 & 0.509 & 0.759 & 3.073 \\
12.5 & 0.002 & 1.277 & 0.500 & 0.751 & 3.005 \\
15 & 0.002 & 1.233 & 0.397 & 0.647 & 2.177 \\
17.5 & 0.002 & 1.158 & 0.233 & 0.484 & 0.868 \\
20 & 0.002 & 1.124 & 0.147 & 0.397 & 0.178 \\
25 & 0.002 & 1.037 & -0.055 & 0.195 & -1.442 \\
30 & 0.002 & 0.955 & -0.268 & -0.019 & -3.148 \\
10 & 0.004 & 1.215 & 0.523 & 0.773 & 3.187 \\
12.5 & 0.004 & 1.238 & 0.541 & 0.791 & 3.329 \\
15 & 0.004 & 1.233 & 0.490 & 0.740 & 2.924 \\
17.5 & 0.004 & 1.199 & 0.400 & 0.650 & 2.198 \\
20 & 0.004 & 1.170 & 0.316 & 0.566 & 1.529 \\
25 & 0.004 & 1.106 & 0.139 & 0.388 & 0.107 \\
30 & 0.004 & 1.065 & -0.006 & 0.242 & -1.061 \\
10 & 0.008 & 1.159 & 0.500 & 0.748 & 2.988 \\
12.5 & 0.008 & 1.182 & 0.544 & 0.793 & 3.347 \\
15 & 0.008 & 1.199 & 0.553 & 0.802 & 3.414 \\
17.5 & 0.008 & 1.202 & 0.548 & 0.796 & 3.367 \\
20 & 0.008 & 1.201 & 0.519 & 0.767 & 3.138 \\
25 & 0.008 & 1.185 & 0.434 & 0.681 & 2.450 \\
30 & 0.008 & 1.166 & 0.375 & 0.622 & 1.978 \\\hline
\end{tabular}\label{tab:summary2}
\caption{Simulation-derived measurements pertaining to binary evolution from our locally-isothermal simulations which used a constant-$\nu$ viscosity.}
\end{table}
\begin{table}
\begin{tabular}{cccccc}\hline
\hline
$\mathcal{M}$ & $ \alpha$ & $\dot{M}/\dot{M}_0$ & $\dot{J}_g/\dot{M}$ & $\dot{J}/\dot{M}$ & $d\log{a}_b/d\log{M}$ \\ \hline
10 & $\sqrt{2}/20$ & 1.180 & 0.517 & 0.767 & 3.132 \\
15 & $\sqrt{2}/20$ & 1.104 & 0.334 & 0.584 & 1.672 \\
20 & $\sqrt{2}/20$ & 0.879 & -0.650 & -0.400 & -6.201 \\
10 & $\sqrt{2}/40$ & 1.152 & 0.476 & 0.726 & 2.807 \\
15 & $\sqrt{2}/40$ & 1.127 & 0.291 & 0.541 & 1.326 \\
20 & $\sqrt{2}/40$ & 0.113 & -0.946 & -0.696 & -8.568 \\\hline
\end{tabular}\label{tab:summary3}
\caption{Simulation-derived measurements pertaining to binary evolution from our locally-isothermal simulations which used a constant-$\alpha$ viscosity.}
\end{table}
\begin{table}
\begin{tabular}{cccccc}\hline\hline
$\mathcal{M}_*$ & $ \nu$ & $\dot{M}/\dot{M}_0$ & $\dot{J}_g/\dot{M}$ & $\dot{J}/\dot{M}$ & $d\log{a}_b/d\log{M}$ \\
\hline
20 & 0.0005 & 1.136 & 0.201 & 0.452 & 0.614 \\
30 & 0.0005 & 1.171 & 0.177 & 0.427 & 0.418 \\
40 & 0.0005 & 0.931 & -0.505 & -0.254 & -5.035 \\
50 & 0.0005 & 0.640 & -1.883 & -1.633 & -16.066 \\
20 & 0.002 & 1.321 & 0.494 & 0.744 & 2.955 \\
30 & 0.002 & 1.136 & 0.130 & 0.381 & 0.046 \\
40 & 0.002 & 1.004 & -0.185 & 0.065 & -2.480 \\
50 & 0.002 & 0.929 & -0.398 & -0.148 & -4.185 \\\hline
\end{tabular}\label{tab:summary}
\caption{Simulation-derived measurements pertaining to binary evolution from our globally-isothermal simulations which used a constant-$\nu$ viscosity.}
\end{table}
$\\$
\bsp	
\label{lastpage}
\end{document}